# Attojoule Optoelectronics for Low-Energy Information Processing and Communications – a Tutorial Review

David A. B. Miller, *Fellow, IEEE*

*Abstract*—Optics offers unique opportunities for reducing energy in information processing and communications while simultaneously resolving the problem of interconnect bandwidth density inside machines. Such energy dissipation overall is now at environmentally significant levels; the source of that dissipation is progressively shifting from logic operations to interconnect energies. Without the prospect of substantial reduction in energy per bit communicated, we cannot continue the exponential growth of our use of information. The physics of optics and optoelectronics fundamentally addresses both interconnect energy and bandwidth density, and optics may be the only scalable solution to such problems. Here we summarize the corresponding background, status, opportunities, and research directions for optoelectronic technology and novel optics, including sub-femtojoule devices in waveguide and novel 2D array optical systems. We compare different approaches to low-energy optoelectronic output devices and their scaling, including lasers, modulators and LEDs, optical confinement approaches (such as resonators) to enhance effects, and the benefits of different material choices, including 2D materials and other quantum-confined structures. With such optoelectronic energy reductions, and the elimination of line charging dissipation by the use optical connections, the next major interconnect dissipations are in the electronic circuits for receiver amplifiers, timing recovery and multiplexing. We show we can address these through the integration of photodetectors to reduce or eliminate receiver circuit energies, free-space optics to eliminate the need for timing and multiplexing circuits (while also solving bandwidth density problems), and using optics generally to save power by running large synchronous systems. One target concept is interconnects from ~ 1 cm to ~ 10 m that have the same energy (~ 10fJ/bit) and simplicity as local electrical wires on chip.

*Index Terms*—Optical interconnections, optical communications, space-division multiplexing, wavelength-division multiplexing, integrated optoelectronics, quantum-confined Stark effect, optical resonators, optical arrays, optoelectronic devices, optical computing

## I. INTRODUCTION

ENERGY already limits our ability to process and communicate information. It constrains the design of information processing machines for simple reasons of power delivery, battery life, power dissipation and heat removal. The fraction of energy used for handling information has risen to a level that is environmentally significant [1], [2]. For these reasons, if we cannot continue reducing the energy required to handle each bit, then we cannot continue our exponential growth in the use of information.

In the early days of transistors and integrated circuits, much of the power was in the logic devices themselves. Over time, ever smaller transistors ("Moore's Law" [3]) reduced that logic energy per bit. That reduction is continuing, even if at a slower pace [4], [5], [6]. But, the energy to send information inside electronic machines does not scale down in the same way, especially for longer connections. As a result, most of the energy dissipated inside electronic machines is used to communicate; for example, even on silicon chips 50 – 80% of gates are for the "repeater" amplifiers in long interconnect lines on the chip [7], and information also has to be driven off chip [5], and over data links and networks [8], [9], at much greater energies per bit.

The remarkable and growing role of optics in the past few decades has enabled a continuing [10], [11] exponential growth of long-distance communications; the capacity of an individual optical fiber has grown at a rate comparable to Moore's law [12]. Increasingly, optics is allowing higher densities of communication inside large systems, as in optical data links in data centers [13], [14]. But, now we are facing a need to have optics help at shorter distances, and not just to enable higher interconnect densities. Now a key question is whether optics can reduce energy in interconnects inside cabinets, racks, and circuit boards, down at least to the edges of the chips themselves, and possibly even on the chip. This question is critical: if we cannot solve these problems with optics, it is not clear that we have any other way of tackling them.

### A. Goals for this review

At the time of writing this review, we are approximately at the point where, with current and emerging technology, optics is poised to provide at least modest energy reductions for data links compared to electrical approaches, even for relatively short links between cabinets and in backplanes or module connections [14], [15], [16], [17], [18], [19], [20].

The main point of this tutorial review is to expose the opportunities, requirements and challenges if we are to take such reduction in energy for communication substantially further, possibly by orders of magnitude. The main focus will be on potential applications within racks, or possibly local groups of racks, and down to the chip, or possibly for longer interconnects on chip – essentially, lengths from ~ 1cm to ~

This project was supported by Multidisciplinary University Research Initiative grant (Air Force Office of Scientific Research, FA9550-12-1-0024).

D. A. B. Miller is with the Ginzton Laboratory, Stanford University, Stanford, CA 94305-4088 (e-mail: dabm@ee.stanford.edu).

10m. For this article, we will call all such communications links "interconnects". Such interconnects may correspond to the majority of power dissipation in large information processing and communications systems today.

We can propose several key goals for such an approach.

1) We should move from energies for such ~1cm to ~10m interconnects that are currently in the range of picojoules or larger total energy per bit, down towards ~ 10 fJ or lower total energy per bit.

2) Such interconnects should look, both in use and in energy, as being as simple as a short electrical wire.

3) This interconnect approach should have sufficient density largely to eliminate the bandwidth bottleneck in current interconnect systems.

4) Such an optical technology should be one that can be the mainstream technology for all communications at distances from ~ 1 cm up to ~ 10 m, so we get these benefits for wide ranges of systems.

These goals are aggressive and even radical. Nonetheless, we argue here for how we could reach them, and we propose various promising research directions and opportunities. Such an optical approach would transform the power dissipation of modules, boards, server racks, internet routers, and supercomputers, while freeing the architectures from their current bandwidth constraints.

### B. How optics can reduce interconnect energy

There are two major ways in which optics can reduce energy for interconnects.

#### 1) Avoid charging electrical lines

The charging and discharging of electrical wires is the ultimate source of dissipation in simple electrical interconnects [2], [21], [22], [23]; optics can eliminate this through "quantum impedance conversion" [21].

Such optical interconnects become attractive energetically when the energy to run the optoelectronic devices – the photodetectors, modulators and/or lasers – becomes less than the charging energy of the corresponding electrical line. This requirement drives us to make low-energy, "attojoule" optoelectronic devices intimately integrated with electronics. Work towards such device technologies is under way and there are several promising directions.

#### 2) Eliminate electronic circuitry used to run links

This second way in which optics can reduce energy for interconnects is less obvious, but equally important: optics can eliminate the power dissipation of the electronic circuits used to operate data links. For links much longer than simple chip-to-chip lines, and possibly even at that level to some degree, both optical and electrical links currently have to add various other circuits to ensure reliable communications.

These circuits include receiver amplifiers, clock and data recovery (CDR) circuits, line coders (to allow AC coupled amplifiers and also CDR), and serialization and deserialization (SERDES) circuits (i.e., for time multiplexing and demultiplexing), in addition to the basic line or output driver circuits[1]. Such circuits together currently consume energies of the order of picojoules per bit (see, e.g., [24]). If we do not eliminate such energies, then we will see limited additional energy benefit from "attojoule" optoelectronics for any such longer links.

Fortunately, optics has additional features that can eliminate such circuits. In addition to saving energy, such approach can also help simplify the interconnect so that quite long interconnects look as simple as short local wires. The integration of low-capacitance photodetectors can largely eliminate the dissipation from electronic circuits used for receiver amplifiers, in what we will call "receiverless" or "near-receiverless" operation, and we will discuss this below. To eliminate the line coder, CDR and SERDES circuits, we can exploit two other features of optics, which we will also discuss below:

i. optics offers very large numbers and densities of physical channels for links of all lengths [25], including very large parallelism with free-space array optics, which means we can choose to avoid SERDES and line coding while simultaneously eliminating bandwidth density problems;

ii. optics offers the possibility of very large (e.g., ~ 10 m) synchronous zones [22] because of the timing precision and stability of optical channels, which means we can avoid CDR.

These latter two features of optics have not been part of much recent discussion, but they represent a substantial opportunity, at least as important as the reduction of energy in optoelectronic devices themselves[2].

### C. Organization of this paper

This paper is organized as follows. In Section II we will summarize some of the background in energy in information processing and communication systems. Section III examines some general aspects of energy dissipation in optoelectronic devices and their scaling to attojoule energy ranges. Section IV summarizes approaches and mechanisms for low-energy optoelectronic output devices, including modulators and light-emitters. Section V discusses photodetectors together with their receiver circuits. Section VI compares long, medium and short distance optical communication systems, showing in particular the different requirements for short distance interconnects. In Section VII, we concentrate on the specific issues and opportunities in optical systems themselves in short distance interconnects. Section VIII discusses the issues and power dissipation of circuits to deal with timing problems in links, and how to eliminate these using optics. Section IX gives a sketch of an example optical system approach for exploiting the many benefits of optics for reducing energy in information processing. Conclusions and recommendations for key research directions are summarized in Section X. To make the article easier to read, various detailed topics are covered in Appendices. Appendix A in particular is an extended discussion

---

[1] Electrical links may also have to add equalization and multilevel signaling circuits to allow sufficient data rates in the presence of signal distortion and loss on electrical lines.

[2] This article may represent the first substantial exposition of such ideas to use optics to eliminate line coding, SERDES and CDR circuits.

of physical mechanisms for optical modulators.

In giving numerical examples, for the sake of definiteness, we will typically calculate for devices and systems running at ∼ 1.5 µm wavelength. That wavelength is certainly consistent with many current technologies, such as silicon photonics and fiber telecommunications. This choice is not meant to be restrictive, however; for short interconnects especially, other wavelengths are possible, including near-infrared such as 850 nm wavelength, or even visible wavelengths, though in broad terms the choice of wavelength does not substantially change the arguments and conclusions here.

We should emphasize before going any further that this article is only intended to provide the context and overall background for research in this area. Because of its wide scope, it cannot review any area in great detail. As a result, the references and citations here can only be representative rather than comprehensive. Though we attempt to cite some seminal work, generally we reference just recent representative examples in many fields. This should allow readers themselves to trace backwards for more depth, but this author apologizes to the authors of the many worthy papers that are not credited appropriately.

## II. ENERGY IN INFORMATION PROCESSING AND COMMUNICATION

### A. Growth in information communications

Since the beginning of the internet, the bandwidth of information communicated over it has grown remarkably. The total bit rate for internet traffic surpassed that of telephone traffic approximately at the beginning of the 21st century [26], at which time internet traffic was growing at a rate of approximately a factor of 100 per decade [26]. Total internet traffic as of 2016 is estimated at ∼ 280 Tb/s ($280 \times 10^{12}$ b/s) [27]. To get some sense of scale, we can compare to voice data rates; at ∼ 32 kb/s for a voice channel, such a data rate corresponds approximately to everyone in the world talking at once. One current estimate [27] predicts a further factor of 3 increase in internet traffic over 5 years.

Though such an internet data rate may seem large, there is much more data sent over shorter links. One estimate is that ∼ $10^6$ bits are communicated inside a data center for every 1 bit that leaves it [28]. Already in 2012, a network communicating servers inside just one data center had a capacity of >1 Pb/s ($10^{15}$ b/s) [29]; such data center network traffic largely does not count the communication of information inside the racks of servers or within the servers themselves, which can only be larger.

To get a sense of interconnect traffic at shorter distances deeper inside information processing machines, we can look at the interconnect rates associated directly with silicon chips themselves. An example graphics processor chip [5] has a peak data rate on and off the chip of 1.4 Tb/s, so just 200 such chips are capable of generating as much information transmission as the entire global long-distance internet traffic. Another recent processor chip [30] has interconnections to off-chip memory with 1.28 Tb/s bandwidth, and other input and output (I/O) connections supporting more than 600 Gb/s, for a total of nearly 2 Tb/s off-chip bandwidth.

TABLE I
ENERGIES FOR COMMUNICATIONS AND COMPUTATIONS

| Operation | Energy per bit | References and notes |
|---|---|---|
| Wireless data | $10 - 30 \mu J$ | [31] |
| Internet: access | $40 - 80 nJ$ | [8],[8]; (a),(b) |
| Internet: routing | 20nJ | [9]; (c) |
| Internet: optical WDM links | 3nJ | [32]; (d) |
| Reading DRAM | 5pJ | [5]; (e) |
| Communicating off chip | $1 - 20$ pJ | [5], [15], [16] |
| Data link multiplexing and timing circuits | ∼ 2 pJ | [24] |
| Communicating across chip | 600 fJ | [5]; (f) |
| Floating point operation | 100fJ | [5]; (g) |
| Energy in DRAM cell | 10fJ | [33]; (h) |
| Switching CMOS gate | ∼50aJ − 3fJ | [4], [6], [34], [35]; (i) |
| 1 electron at 1V, or | 0.16aJ | |
| 1 photon @1eV | (160zJ) | |

WDM – wavelength division multiplexing
DRAM – dynamic random-access memory
CMOS – complementary metal-oxide-semiconductor transistor
(a) Uses projections to 2016 from [8]
(b) Presumes wired connections (optical or electrical) to the network
(c) Total for 20 "hops" per internet connection, and derating energies from the 2008 numbers in [9] using a factor of 0.74 per year (from [8]) for improved electronic energy efficiency.
(d) Total for 20 "hops" per internet connection, and using projections to 2016 in [32]
(e) Rounded sum of the DRAM access and interface energies as projected for 2017 in [5], for off-chip DRAM
(f) Based on 2017 projects in [5] for a 10mm line on the chip
(g) Double-precision fused multiply-add (floating-point) operation using the projection in [5] of ∼ 6.5 pJ in 2017 for this 64-bit operation to calculate energy per bit.
(h) Based on the relative constancy of DRAM cell capacitance at greater than ∼ 20fF, and a ∼ 1V charging voltage.
(i) We might estimate a lower limit ∼10aJ for switching a gate based on projected reductions in transistor capacitance, referenced in [34], and simulations of ∼ 20 aF gate capacitance in current technologies [35], but such an energy is just for charging the gate itself, and further parasitic capacitance of at least ∼ 40 aF is likely [35], even if we completely neglect other load capacitances and the fact that "complementary" electronic technology with two transistors per stage. On this basis, and allowing some room for continued improvement, we quote the minimum of ∼ 50 aJ. A projected overall energy per logic gate operation in an optimized processor core is ∼ 200 aJ [4], which includes leakage power dissipation and some local connection and other energies. Current logic gate operating energies in systems with a fan-out of 3 are ∼ 3 fJ [6].

The communications traffic inside the chip itself again can only be larger still. That same recent processor chip [30], for example, has an on-chip network supporting more than 4Tb/s of bisection bandwidth[3], and the total bandwidth in and out of the "level 3" (L3) cache memory on the chip is 12.8 Tb/s. We can generally expect yet more on-chip traffic into and out of lower level cache memory and within logic operations themselves.

---

[3] Bisection bandwidth is the amount of data traffic that we would find if we divided a data network into two parts, and counted the traffic passing from one part to the other; usually, this will refer to the largest possible number we would find from any such division into two parts.

As we look to reduce the energy in handling information, the energy in all such interconnects inside machines will be particularly critical.

### B.   Overall energy consumption

Information processing and computing, including data centers, personal computers and networks, were estimated to consume 4.6% of world electricity production in 2012 [1]; the growth rate of that consumption exceeds the growth rate in electricity generation capacity. If wireless communications and displays are included, the total rises to ~ 9 % of electricity consumption. With such growth in the amount of information we are handling, information processing and communications cannot continue to grow at their recent exponential rates without continued, major reductions in the energy per bit.

### C.   Energy per bit in communications and processing

To understand where the energy is consumed, we can look at the approximate energies per bit in various processing and communications operations in Table I. Actual numbers can vary considerably, of course, and they will change as technology advances, but the overall orders of magnitude here give us insight, nonetheless. We can examine these energies in a few categories, starting from the smaller energies at the bottom of the table and working up to the larger energies at the top.

#### 1)   Energies for logic operations

The energies for logic operations themselves are small, ranging from possibly as low as ~ 10-100 aJ per bit inside a given logic gate to ~ 100fJ per bit in a complicated operation such as floating point multiplication. Such energies have decreased over the decades as transistors have become smaller.

Note that even these small energies are much larger than the energy associated with one electron or one photon. Modern low-energy electronic devices work with relatively large numbers of electrons; even 10aJ corresponds to ~ 60 electrons at 1V. Changing to information processing systems that would use energies much smaller than 10 aJ would raise serious issues of statistical fluctuations; though we can consider reliable systems based on "unreliable" components[4], such systems would require a major change in the paradigm of digital information processing as we know it.

For much of the history of Moore's law, as the transistors became smaller, so also did the voltage to run them, following a rule known as Dennard scaling [36], [37]. The "dynamic" energy in operating a logic gate comes largely from charging and discharging the capacitances of the transistor itself and of the local wiring. (Logic gates can also dissipate "static" power even when they are not operating, such as through leakage currents.) The reduction in operating voltage meant the "dynamic" energy dissipation shrank even faster than the reduction in size would suggest.

More recently, however, the reduction of transistor operating voltage has largely stopped. This is because low gate voltages lead to a correspondingly smaller potential barrier between the source and the drain of a transistor in its "off" state, which leads to leakage current; the potential barrier height becomes too close to the average thermal energy of an electron at room temperature $T$ ($k_B T \simeq 25$meV where $k_B$ is Boltzmann's constant). So, to minimize the "static" power dissipation in chips, the operating voltage of logic gates is decreasing only slowly if at all [6], [38], [37]. Operating voltages of a substantial fraction of a volt (e.g., 0.8V) are typical [6]. This approximate constancy of transistor operating voltage has also meant that the voltages on the interconnect lines on chips have stopped decreasing, which influences the energy of electrical interconnections, as we discuss below.

Present CMOS technology is based on FinFET structures [19], [39], [40] or approaches like fully-depleted silicon on insulator (FDSOI) [19], but the scaling approach and arguments here are quite different from the Dennard scaling. One main point of such devices is to reduce drain-source leakage currents and related "short-channel" effects. The minimum dimension in such transistors is typically not now the gate or channel length, but rather the effective thickness of the channel; an effectively thinner channel allows it to be more fully depleted of carriers (electrons or holes) and reduces the drain-source leakage and "short-channel" effects.

Nonetheless, with smaller sizes in the devices the capacitances overall may still scale down [6], allowing correspondingly lower logic energies per bit. The combination of logic, local interconnection and leakage energies may, however, lead to a saturation in the total energy per bit in logic operations within a processing core [4], possibly in the range of ~100aJ/bit.

#### 2)   Clock speeds and power dissipation in electronic chips

We might think we run electronic processor chips at clock rates of ~ 2 – 3 GHz because the transistors are slow. In fact, the basic operation speed of an electronic gate, even when driving a standard "fan-out" load of 3 other gates, would be ~ 3 ps with current technology [6].

In modern electronic processor chips, we limit clock speeds for two main reasons related to power dissipation:

(1)   running transistors faster requires somewhat higher voltages [38] which means more energy per bit;

(2)   increasing clock speeds mean more switching transitions per second – so more power dissipation even for the same energy per bit – but chips are already limited by the ability to extract heat from them [5], [38].

Note that, as we scale down transistors and wires, the total capacitance per unit area of the chip in wires and logic gates does not decrease; in fact, device [6] and wiring capacitance per unit chip area can even increase somewhat[5]. So, for a given clock frequency, we could actually have more power dissipation per unit area as we charge and discharge device and wiring capacitance[6].

---

[4] The human brain is a good example of a system that can work well based on a somewhat statistical operation of potentially unreliable individual parts.

[5] For example, wires of smaller cross-section could lead to more total length of wiring in a given area; since wire capacitance per unit length is largely constant (see Section II D below), that would mean more wiring capacitance.

### 3) Energies for interconnects inside chips and off chips

As we move up Table I, we see that the energy to communicate bits across a chip (e.g., ~ 600 fJ/bit) can be larger than some quite substantial and complicated logical operation, like a floating point multiplication (e.g., ~ 100 fJ/bit), on those same bits. Similarly, the energy stored in a DRAM[7] cell itself is quite small, at ~ 10fJ. But, especially if the DRAM cell is on a different chip, the energy to read that cell becomes totally dominated by interconnection energy, and can be almost three orders of magnitude higher (e.g., 5 pJ).

In communicating off chip, interconnecting on short lines to adjacent chips may just involve charging line lengths of the order of centimeters to the logic voltage, but even that simple operation can lead to picojoule energies per bit communicated (see Section II D below).

Longer connections off chip may use lower voltage signaling or more sophisticated links, but the energy of these may not lower than the on-chip or local simple interconnections, leading to multiple picojoules of dissipation per bit, in part because of the more sophisticated receiver and transmitter circuits required (see Section II D below). A significant amount of energy per bit can also be used in the circuits that multiplex to higher bit rates per line or channel for what we can call data links; as mentioned above, such circuits perform functions like line coding, CDR, and SERDES, in addition to receiver amplification and line or output drivers. We will return to discuss such dissipations below (see Sections V and VIII).

### 4) Long-distance telecommunications

As we move to long distance, it might seem obvious that the majority of the energy for telecommunications networks for the internet that should be in the long-distance links themselves. Long distance optical links consume relatively low energy per bit, however, primarily because of the very low loss of optical fibers [9]. Because of switching of information, such as internet packets, in the many routers along the way, the larger part of that energy in the core of internet transmission is actually dissipated in the routers [9]. And, that energy is actually the energy dissipated inside electronic machines, which, as we will see, is predominantly interconnection energy at short distances.

### 5) Access networks and connections

The largest amounts of energy per bit in internet and telecommunications networks can be at associated with the last connections to the user (sometimes called "access" connections). Wireless connections, as in WiFi and mobile cellular connections, consume particularly large energies per bit [31]. For fixed connections over fiber or cable, the access network and any modem connecting the customer to the network tend to have a relatively fixed power, so the energy per bit depends on the bandwidth to the customer [8]. As that bandwidth rises, the energy cost for access reduces, possibly below the next largest energy cost, which is the energy dissipated inside the routers.

### 6) General conclusions on energies in information processing and communication

We conclude, first, that the majority of energy in information processing and communications is predominantly in sending the information from one point to another, not in the logical processing itself, and second, with the possible exception of wireless links, most of that energy is in local electrical interconnects inside information processing and switching systems. Hence, we should move to optics and optoelectronics for such local interconnects if we want to reduce energy per bit overall. At the present time, we appear to have no viable new approach other than optics for solving interconnect energy and density problems inside machines.

### D. Physics of electrical interconnect energies

The energy for communicating in electrical wires essentially is bounded by the energy required to charge up the appropriate line capacitance to the driving signal voltage. For short interconnections, that capacitance will be the total capacitance of the line, and the drive voltage will be essentially the same as the logic voltage; that is mostly the situation for interconnect lines on chips. For longer connections off chips, only the line length corresponding to one bit (that is, ~ one clock cycle) needs to be charged for each bit; but such lengths can be substantial (e.g., up to 30 cm at 1 GHz or 1 ns, and up to 3 cm at 10 GHz or 100 ps).

### 1) Capacitance of electrical lines

To understand the dissipation in electrical signaling, we need to understand the capacitance of lines. One key point is that the capacitance of electrical lines per unit length only depends on the relative geometry of the line, not the size scale. And, that dependence on geometry is predominantly logarithmic for lines where the size of the conductors is comparable to their separation [2], [21], [22], [23]. For example, the capacitance per unit length of a coaxial line only depends on the logarithm of the ratios of the inner and outer conductor radii, not on the actual cross-sectional size or overall diameter of the line.

$$C \, / \, L \approx 2 \text{ pF/cm} \equiv 200 \text{ aF/}\mu\text{m}$$

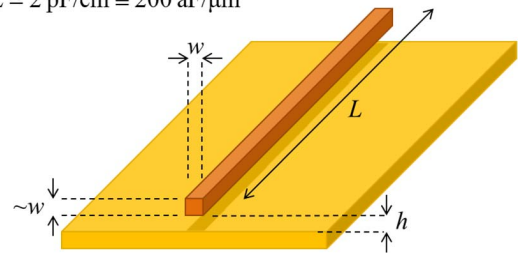

Fig. 1. A typical interconnection line will have cross-sectional dimensions that are similar in both directions, so ~ $w$ in the figure, and a separation $h$ between the two conductors that is also similar. This balances the need to keep the overall cross-section of the line relatively small so we can have high densities of interconnections, while avoiding large capacitances from conductors that are very close. A line above a ground plane is shown here, but because of the approximately logarithmic dependence of capacitance on geometry, the results are similar for all such lines. A typical value of capacitance per unit length is ~ 2pF/cm ≡ 200aF/μm.

When we are trying to get reasonably large densities of

---

interconnections, we do not want to waste cross-sectional area by separating the two conductors in a line by a large amount; anyway, doing so would only reduce the capacitance approximately logarithmically. As a result, lines typically have a separation between the conductors in the line that is comparable to the cross-sectional dimension of the smaller of the conductors. See Fig. 1.

Hence, the capacitances per unit length of all electrical transmission or interconnect lines are very similar, within factors of order unity. Typical $50\Omega$ coaxial cable with $\sim$ 1cm diameter has a capacitance of $\sim$ 1.5 pF/cm. Interconnect lines on chip with only 80nm center-to-center spacing (so $\sim \times 10^5$ smaller in linear size, and possibly $\sim \times 10^{10}$ smaller cross-sectional area, than the coaxial cable) also have a capacitance of $\sim$ 2 pF/cm ($\equiv$ 200aF/$\mu$m) [6]).

### 2)  Capacitive charging energies on lines

Because the voltages for logic operation on chips are not reducing substantially, as discussed above, and the capacitances per unit length of wires are relatively fixed and bounded by physical laws, the energies to communicate logic levels across the chip have not reduced significantly in recent years. Charging a capacitance to a voltage $V$ leads to an energy $(1/2)CV^2$ stored on the capacitor, with an equal energy dissipated in the series resistance through which the capacitor is charged (see, e.g., [41] for a discussion of capacitive charging energies). When the capacitor is discharged, this energy is dissipated into the discharging resistance, for a total dissipated energy[8] of $CV^2$. On this basis, we can see that charging a line of $\sim$ 1cm length to some fraction of a volt to send information down it leads to dissipated energies approaching or on the scale of a picojoule, which is the source of the 600fJ/bit energy for communicating across a chip[9] in Table I.

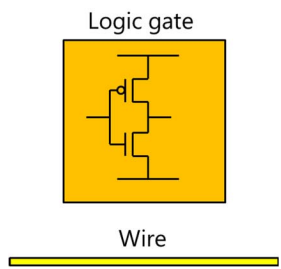

Logic gate

Wire

Fig. 2 The total capacitance of the transistors in a small logic gate is comparable to that of a wire to another nearby gate.

A key point in comparing interconnect and logic energies, however, is to note that the capacitance of the transistors in a logic gate is comparable to the capacitance of a wire from one logic gate to another that is relatively close by [34], [42]. For example, [42] estimates that the signal only has to go a distance $\sim$ 3 times the width of a transistor for the energy to charge the wire to become equal to the energy to switch the transistor (see Fig. 2). Because most signals go further than this and essentially all will go at least this far, it is simple to conclude that the majority of dynamic energy dissipation[10] in electronics is for communication, not logic.

### 3)  Energies for electrical off-chip communications

There is no simple answer for calculating the electrical energy per bit for connections off chip. Driving high-speed or high-data-rate signals on electrical lines over even 10's of centimeters can also be difficult because of loss and signal distortion on electrical lines [25]. As a result, such electrical connections may change to links where the format of the signaling can be quite different from simple "on-off" signals at the logic voltage.

In such links, it is possible to have lower voltage signaling or to allow complex modulation formats that can increase the number of bits per symbol sent, which would tend to reduce energy per bit, but that decrease can be more than offset by the necessity to run the required complex electronic circuitry to support the signaling. Typically, such links with more complex modulation formats are designed to increase the data capacity of lines, not to reduce energy per bit communicated. Additionally, such links often require clocking to establish the necessary timing for signals, and clock recovery circuitry can consume significant power (e.g., 50% of the total receiver power in a recent example [43]). Even on chips themselves, the power to run the clocking inside logic blocks can also be comparable to other power dissipations [44].

As a result of these various factors, energies per bit for off-chip electrical interconnects can typically range from picojoules per bit to significantly higher energies [5], [16]. This issue of off-chip connection energy and the difficulty in reducing it is well-known also in the context of supercomputers and their future scaling [15], for example.

### E.  Physics of optical interconnect energies

#### 1)  Quantum impedance conversion

The key reason why optics can save energy compared to electrical approaches in simple interconnects is that in optics we do not have to charge the line or other electromagnetic medium to the signal voltage; instead, we only have to charge or discharge the optoelectronic detector (or whatever is the equivalent load capacitance of the detector and the circuit to which it is connected).

Fig. 3(a) illustrates this point. The core physics is the photo-electric effect. The voltage that we can generate in a photodiode even in a simple photovoltaic mode is comparable to the photon energy in electron volts, and we can generate something close to one electron of current for each absorbed photon. The detection of light is a quantum-mechanical process of absorbing photons, not a classical process of measuring the voltage in the light beam itself (see, e.g., [84]).

---

[8] Incidentally, it is common to quote an energy of $(1/4)CV^2$ per bit for communications involving a capacitance $C$. This can be correct for the following reason. If the bit changes from one state to another, we dissipate $(1/2)CV^2$, either in the charging resistance or in the discharging resistance. On the average, for any two bit sequence, in an effectively random string of bits, half the time the next bit has the same value as the current one, so we change from one state to another every 2 bits, on the average; hence we dissipate $(1/2)CV^2$ on the average every two bits. So, we dissipate $(1/4)CV^2$ per bit, on the average.

[9] Long connections on chips are often broken up into shorter lengths of line with "repeater" amplifiers between these short lengths. This is to reduce delay. The capacitance and the resistance of the line are both proportional to length, so the overall charging time of the line grows as the square of the length; hence, breaking the line up into sections with intervening repeater amplifiers can reduce the overall delay. Even with repeaters, the effective signal propagation velocity on such lines can be, e.g., only $\sim$ 1/5 or less of the velocity of light (see, e.g., [23]), leading to significant "latency" or delay problems in systems.

[10] Dynamic energy is associated with the active processing of information, as opposed to static, background power dissipation.

In a classical electromagnetic beam of power $P$ propagating in free space, the power in the beam can be written as $P = V_{RMS}^2 / Z_o$; here $V_{RMS}$ is the root mean square ( RMS) voltage from one side of the (linearly polarized) beam to the other and $Z_o \simeq 377\Omega$ is the impedance of free space[11]. Then

$$V_{RMS} = \sqrt{PZ_o} \qquad (1)$$

For an example power of 1nW in a beam, the classical voltage would therefore be $V_{RMS} \sim 600\mu V$.

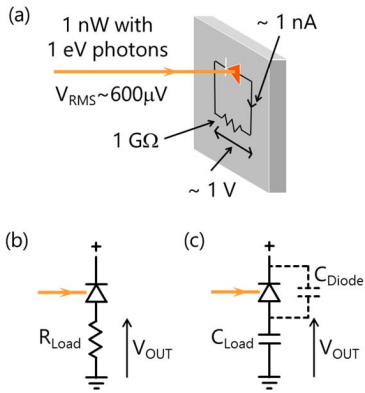

Fig. 3 (a) Illustration of quantum impedance conversion, in which a beam with a small classical voltage can generate a much larger voltage at the output of a photodiode. (b) A reverse-biased photodetector with a resistive load, and (c) with a capacitive load, including the diode's own capacitance.

The photodetector, however, does not measure classical voltage; it counts photons, and can give $\sim 1$ electron of current for each absorbed photon. For photon energy[12] $h\nu \equiv 1eV$ (where $h$ is Planck's constant and $\nu$ is the optical frequency), then absorbing this 1nW of power could give a current[13] of $\sim$ 1nA. The photodiode could also give a voltage[14] of up to 1V. 1nA in a 1G$\Omega$ load resistor[15] would correspond to 1V. The key point here is that the voltage in the load resistor can be much larger than any classical voltage in the light beam.

A simple photodiode therefore can transform power propagating in a low impedance medium into power in a high impedance load, and it can do so with some reasonable efficiency. This process can be called "quantum impedance conversion" [21], [22].

The circuit of Fig. 3(b) may be more practical, with the photodiode operated in reverse bias. In this case, the diode can likely generate $\sim 1$ electron per photon, giving a current $I_{PC} = Pe / h\nu$ (where $e$ is the magnitude of the electronic charge) for an absorbed optical power $P$, and an output voltage $V_{OUT} = R_{LOAD} Pe / h\nu$.

We can also think of this process in terms of optical energies, rather than powers; indeed, we may well be operating with a circuit more like that of Fig. 3(c), which has no load resistor[16]. Here, absorbing some amount of optical energy from a pulse can lead to an output voltage because the photogenerated charge can charge (or discharge) the capacitance $C_{Diode}$ of the diode and any load capacitance $C_{Load}$, such as an input to a transistor or logic gate, to change the output voltage $V_{OUT}$.

A given optical energy $E_A$ made up of photons of energy $h\nu$ corresponds to a number of photons $N_{ph} = E_A / h\nu$. If we absorb all that energy, generating $\sim 1$ electron per photon, then a total charge $Q_{PC} = E_A e / h\nu$ will flow in the circuit. This will lead to a change in output voltage $\Delta V_{OUT} = E_A e / h\nu C_{TOT}$ where $C_{TOT}$ is the total capacitance[17] $C_{TOT} = C_{Load} + C_{Diode}$. For $C_{TOT} = 1fF$, and an optical pulse of energy $E_A = 1fJ$, then $\Delta V_{OUT} = 1V$.

We will discuss device capacitances below in Section V; we might hypothesize a photodetector of $C_{Diode} \simeq 100$ aF , corresponding to a 1$\mu$m cube, connected to a transistor with input capacitance of 100aF through some wire of capacitance 100 aF, for a $C_{TOT} = 300$aF . Then $\sim$200aJ of optical energy in an optical pulse at 1.55$\mu$m wavelength (0.8eV photon energy) efficiently absorbed in such a detector would lead to $\Delta V_{OUT} \simeq 0.8V$ , which is more than the input voltage swing required to switch a logic gate[18].

Hence with low-capacitance photodetectors connected to a low-capacitance load, such as the input to a CMOS inverter circuit, even optical energies less than 1fJ could give to enough voltage change to switch a logic gate, without any electronic amplification (a so-called "receiverless" mode of operation [45], [46]). We discuss this benefit in detail in Section V.

### 2) Additional physical benefits of optics

Using optical interconnections brings many additional benefits. See also [2], [22], [23]. The interconnect bandwidth densities, especially for connections off chip, and the precision of timing possible with optics will turn out to be particularly interesting. We will come back in Section IX to discuss an example system that could simultaneously take maximum

---

[11] We could use other somewhat different impedances if the electromagnetic beam was propagating in a dielectric, such as glass, on in a transmission line, but the essence of this argument is not changed by that.

[12] Because the wavelength of light (in free space) $\lambda = c / \nu$ and the photon energy in electron volts, $h\nu_{eV}$ is the energy $h\nu$ in joules divided by the magnitude of the electronic charge $e$, then $h\nu_{eV} = hc/e\lambda \cong 1.24/\lambda_{microns}$, where $\lambda_{microns}$ is the wavelength in microns (micrometers). This relation, and the complementary one $\lambda_{microns} \cong 1.24/h\nu_{eV}$ are very convenient. So for $h\nu_{eV} = 1eV$, $\lambda \cong 1.24\mu m$, and for $\lambda = 1.55\mu m$, $h\nu_{eV} \cong 0.8eV$.

[13] This would be the so-called "short-circuit" current of the photodiode.

[14] This would be an "open-circuit" voltage under "flat-band" conditions.

[15] This example is somewhat simplified because we will not simultaneously obtain "short circuit" current and "open-circuit" conditions, and there are some other practical limits with diodes.

[16] Of course, such a simple circuit with no load resistor would have difficulty resetting itself; once triggered with an optical pulse, the resulting voltage change would remain unless some other leakage current discharged it. Later, in Section VIII B, we will discuss "dual-rail" operation with stacked pairs of diodes, which avoids this difficulty for circuits with no load resistor.

[17] The diode and load capacitances are effectively in parallel in a circuit like this. To change the voltage $V_{OUT}$ we have to charge or discharge both capacitances.

[18] Indeed 0.8V corresponds to a typical supply voltage for CMOS logic circuits [6]

advantage of all these benefits of optics to minimize energy dissipation overall.

### a) Density of interconnects

A major benefit of optics is that it allows very high densities of information to flow, in the sense of Gb/s per square millimeter of cable cross-section or Gb/s per linear millimeter of the edge of some card or board; this is one of the major reasons that optical interconnects are in current use for longer distances inside large machines. Optical fibers can carry high data rates over very thin (e.g., 125 μm diameter) "wires". Smaller waveguides (e.g., $\sim 0.2 - 3$ μm cross-sectional dimensions) are also possible on substrates, as in silicon photonics [47], [48], [49], [50], [51], [52], [53], [54], [55], [56], [57], [58], [59], [60] and integrated III-V photonics [61]. There are the additional opportunities of much larger amounts of information transmission using wavelength division multiplexing (WDM) (use of multiple different wavelengths as independent channels) or space-division multiplexing (SDM); SDM could use multiple spatial modes in a fiber (mode-division multiplexing) or free-space, two-dimensional interconnects off the surface of the chip (see Section VII below).

Electrical interconnects run into a basic limitation [2], [25] on bit rate $B$ that is proportional to the total cross sectional area $A$ of the wiring and inversely proportional to the square $l^2$ of the length $l$ of the wiring, i.e., $B = B_o A / l^2$ where the prefactor $B_o \sim 10^{15} - 10^{16}$ b/s . This limit, which results from the resistance and capacitance of electrical wires, applies to simple "on-off" signaling. It severely restricts the amount of information we can send through wiring systems[19].

This "aspect ratio limit" [25] is routinely encountered on chips, on boards, and transmission lines. It can be avoided to some degree by using sophisticated signaling techniques, such as equalization and/or multilevel signaling and modem technologies, so as to approach the Shannon limit for such electrical connections; that, however, requires more complex transmitter and receiver circuits, which in turn lead to increasing energy per bit. Since optical connections do not have the resistance of electrical wires, they completely avoid this particular limit, and can exceed it in practice by many orders of magnitude[20].

### b) Signal integrity

Another key benefit of optical connections is that they can avoid some of the problems of the propagation of high-frequency electrical signals. Over the scale of a machine, such as meters or 10's of meters, there can be negligible distortion of optical signals due to dispersion, even for picosecond pulses (see Section VIII B below for example calculations).

Electrical cables, by contrast, show very large pulse broadening even for much longer pulse widths [25]. Any crosstalk or loss in optical signals is also essentially independent of the signal bandwidth[21], so in general the optics itself in optical links can be designed to support very large signal bandwidths over the size scales of physical information processing systems.

Since optical signals operate by transmitting and detecting photons rather than measuring classical voltages, all optical connections intrinsically offer voltage isolation, just like inserting "optical isolators" in every link. This means ground voltage variations over systems do not matter in optically interconnected systems.

### c) Timing precision

As discussed above, optics can deliver even short pulses without significant distortion over quite large distances; that could allow electrical systems to be clocked optically with very little "jitter"[22], for example, into the sub-picosecond range [62]. So optics can be used for low-jitter clock distribution.

One additional aspect of optics that has not been substantially exploited is that such timing or clocking pulses can be delivered with a very well defined *absolute* delay [22]. Electrical wires have an effective delay that depends on the variation of the resistance in the wire with temperature because the slope of the rising or falling edge of electrical pulses depends on that resistance. As a practical matter, we typically do not rely on long electrical wires having any particular predictable delay, and we recover the clock phase (i.e., timing) using clock recovery circuitry and associated buffering.

The delay on optical fibers is, however, quite precisely predictable and substantially independent of temperature over the $1 - 10$ m distances involved inside a system (see Section VIII B); it could substantially reduce power dissipation in links because it could eliminate clock recovery circuitry entirely.

When using modulators as the output devices, we can also automatically retime the output signals by having the optical input to the modulators be such well-timed pulses [63], [64], as we also discuss in Section VIII B below.

### 3) Conclusions for physical benefits of optics for interconnects

In summary, optics offers various physical benefits compared to electrical lines:

- it can reduce interconnect energy by eliminating the charging of electrical lines;
- it can send information over large distances at high rates without additional loss or distortion;

---

[19] For example, a coaxial cable, 1cm$^2$ in cross-sectional area and 10 m long, would be able to carry $\sim$ 1 Gb/s in simple on/off signaling (the $\sim 10^{15}$ value of $B_o$ is appropriate for such an "LC" transmission line) [25]. A line on a chip with $\sim$ 1μm$^2$ cross-sectional area with a simple on/off signaling at 2Gb/s, could have a length up to $\sim$ 7 mm (the $\sim 10^{16}$ value of $B_o$ is appropriate for such an "RC" transmission line) [25]

[20] A hypothetical electrical line 125μm in diameter and 60km long would be able to carry about 0.03b/s with simple signaling (the capacitance of the wire would take $\sim$ 30 s to charge up through its own series resistance). An optical fiber of the same dimensions can carry bandwidths exceeding 10Gb/s with

simple on/off keying on one frequency channel [10], and may have many Tb/s of capacity with sophisticated signaling and multiplexing [11], [12].

[21] For modulation bandwidths (e.g., GHz to 100's of GHz) that are small compared to the carrier frequencies of optics (e.g., 200 THz), that modulation makes essentially no difference to the loss in propagating optical signals, nor to the cross-talk between adjacent waveguides or beams. If the system is running with wavelength-division multiplexing, of course the modulation can induce cross-talk between channels of different center wavelengths.

[22] Jitter is the pulse-to-pulse variation in the timing of a pulse in a pulse train, usually viewed as being from random or unpredictable causes.

- it can allow very high densities of high-bandwidth connections;
- it can offer very precise timing and retiming of signals.

We will discuss these various points below in more depth in Sections VII and VIII.

### III. Scaling optoelectronics into the attojoule range

The core energy benefit of optics in reducing the energy per bit for interconnects in simple connections requires that the energy to operate the optoelectronic device is itself lower than the energy required to charge an equivalent length of electrical line. Hence, the operating energy of optoelectronic devices is a very important consideration. Here we look at the prospects and approaches that could allow us to scale to very low operating energies in optoelectronics, ideally even into the sub-femtojoule or "attojoule" range.

The energy involved in operating optoelectronic devices themselves can be separated approximately into two parts:

(A) the electrostatic (capacitive) energies required to swing the necessary voltages across the device, as either a photodetector or an output device like a laser or modulator

(B) the other energies involved in running some devices, such as the energy to inject carriers in a light emitter or change carrier density in some modulators.

There will be yet other energies in operating a system, especially from optical losses; we will return to such energies later, however, concentrating here only on these specific energies involved in running the devices themselves.

#### A. Electrostatic energies

If our goal is make devices that operate with energies less than a femtojoule, then we must make sure that the capacitive charging and discharging energy for the devices themselves is less than this amount. To get a sense of scale, first we can look at the capacitance of a simple cube of semiconductor between two opposite surfaces, sketched in Fig. 4.

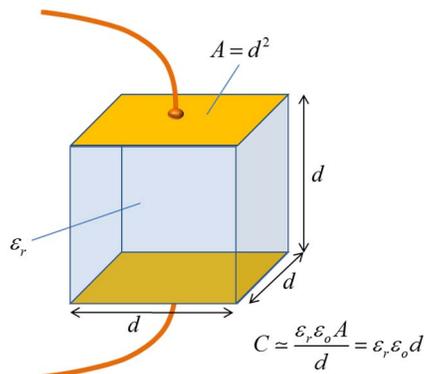

$$C \simeq \frac{\varepsilon_r \varepsilon_o A}{d} = \varepsilon_r \varepsilon_o d$$

Fig. 4. A cube of semiconductor material, with dielectric constant $\varepsilon_r$, and sides of length $d$, area $A$ of each cube surface, with capacitance between two opposing faces of $C \sim \varepsilon_r \varepsilon_o d$.

For the sake of definiteness, we will take the dielectric constant of semiconductor material to be $\varepsilon_r \sim 12$, which is a typical value. Because of this large dielectric constant, to a rough approximation, we will neglect the fringing fields, and treat this as a simple "plane-parallel capacitor" between opposing surfaces of this cube. With capacitor plate area $A$ and separation $d$, the capacitance would be

$$C = \varepsilon_r \varepsilon_o A / d \qquad (2)$$

where $\varepsilon_o \simeq 8.85 \times 10^{-12}$ F/m is the electric constant (permittivity) of free space). Hence for a cube of side $d$ (and hence of plate area $A = d^2$, the capacitance is

$$C = \varepsilon_r \varepsilon_o d \qquad (3)$$

So for a 1 μm cube, the capacitance is $\sim 100$ aF. If we presume that running some device requires charging or discharge the device capacitance by 1V, for example, then we can see that with such a micron size, the resulting total energy to charge the device would be $\sim CV^2 \sim 100$ aJ. For a 100 nm cube, the energy would be $\sim 10$ aJ. For some waveguide device that was, say, 300 nm wide, 200 nm thick, and 3 μm long, then the capacitance between the top and bottom faces (i.e., over the 200 nm thickness) would be $\sim 450$ aF for 1V drive, so the associated energy would be $\sim 450$ aJ.

Since integrated semiconductor devices are not likely to be more than $\sim 1$ μm thick[23], this simple approximation tells us that, for 1 V operation, or indeed operation at typical logic swing or supply voltages (e.g., 0.8V [6]), the optoelectronic devices have to be micron or sub-micron in size if we are to run at single femtojoule or sub-femtojoule operating energies[24].

Above in considering wires, we presumed $\sim 200$ aF/μm wire capacitance (see, e.g., [6]). So, if the capacitance of the wire that connects the device to its associated driver or receiver electronics is not to dominate the capacitance overall for such sub-femtojoule optoelectronics, then such connecting wiring also needs to be on a scale of no more than a few microns. Connection to photodetectors should use particularly short wires; any increase in overall input capacitance can cause the entire operating energy per bit to scale nearly in proportion, a point we discuss in greater detail in Sections V C and IX A.

The transistors to which the photodetector devices connect will have input capacitances in the range of ~20aF to ~ 100aF if they are near to minimum-size transistors (see [4], [6], [34], [35], and the discussion in footnote (i) of Table I), so their capacitance may not dominate overall, but should be included in the overall capacitance.

The simple overall conclusion here on electrostatic energies is that, if we are running optoelectronics at voltage swings comparable with the logic voltages, then the devices have to be micron or sub-micron in size and they have to be integrated right beside the associated electronics (e.g., within a few microns or less); otherwise electrostatic operating energies will raise the total energy out of the sub-femtojoule range. Hence the integration technology has to be a core part of any serious proposal for attojoule optoelectronic devices.

---

[23] Fabrication in substantially planar structures using lithography typically uses thickness of this scale or smaller, and layered growth techniques in general also use such thicknesses, for example.

[24] Note also that these capacitances and energies are slight under-estimates since they are neglecting fringing fields; the fact that these are under-estimates reinforces the need for small sizes.

## B.  Operating energies

To give some sense of energies for optical output devices (i.e., modulators and light emitters) and some of the requirements to achieve them, we can perform a simple scaling of two such devices that are each based on strong microscopic mechanisms, namely III-V semiconductor lasers and quantum-confined Stark effect (QCSE) electro-absorption modulators [65], [66], [67], [68].

<div style="text-align:center">

TABLE II

Example laser and modulator energy scaling

</div>

| Active device volume | Operating energy | Optical concentration factor |
|---|---|---|
| **(1 μm)³** (a) | | |
| laser | ~160 fJ | ~5 (b) |
| modulator | ~5 fJ | ~1 (c) |
| **(300 nm)³** = 0.027μm³ | | |
| laser | ~4300 aJ | ~200 |
| modulator | ~135 aJ | ~40 |
| **(100 nm)³** = 10⁻³μm³ | | |
| laser | ~160 aJ | ~5×10³ |
| modulator | ~5 aJ | ~10³ |
| **(10 nm)³** = 10⁻⁶μm³ (e.g., a quantum dot) | | |
| laser | ~160 zJ (d) | ~5×10⁶ |
| modulator | ~5 zJ (e) | ~10⁶ |

(a) E.g., an active (gain) region 50nm thick, 200nm wide, and 100 μm long, as in some hypothetical quantum well edge-emitting laser, or 300 nm thick, 350 nm wide and 10 μm long as in some hypothetical short modulator.

(b) If the gain material entirely filled the mode cross-section, a gain ~ 100 cm⁻¹ would allow a laser of ~ 100 μm length to work with only a very weak resonator. For the 50×200 nm hypothetical gain cross-section of note (a) above, in a hypothetical mode cross-section of 300×350 nm², so about ×10 larger than the gain cross-section, because of this mode overlap of only 1/10 with the gain material, we would only obtain a gain of about 10% in one pass, so we would need cavity mirrors of ~ 90% (power) reflectivity $R$ to reach threshold, which would correspond to a concentration factor $\gamma \sim 5$ (see below).

(c) No resonator is required for a 10 μm long QCSE modulator, as in [78], because the absorption in a single pass is large enough.

(d) This energy is equal to 1 eV and corresponds to one electron-hole pair in the quantum dot. 1 zJ ≡ 10⁻²¹ J.

(e) This energy corresponds approximately to a charge of one electron on one face of the dot and a charge of one hole (or one less electron) on the opposite face, with a corresponding voltage between the faces of ~ 100 mV.

Such laser and modulator devices are in wide practical use today in telecommunications and other applications, and they represent realistic examples of efficient device approaches with well-understood physics and technology. Both already exploit quantum-confinement benefits through the use of quantum well structures. QCSE modulators typically use III-V quantum wells, but they can also use germanium quantum wells on silicon substrates [41], [67], [68], [69], [70], [71], [72], [73], [74], [75], [76], [77], [78], [79], [80], [81], [82], [83], with performance comparable to or better than their III-V counterparts [82].

We estimate energies for different device active volumes in Table II. For laser energies, we presume that the device volume has to have an injected carrier (pair) density of 10¹⁸ cm⁻³ so that it has enough gain to lase and that the resulting gain is ~ 100 cm⁻¹. These are typical orders of magnitude for operating semiconductor lasers[25]. In calculating operating energies, we presume we require 1 eV of energy to inject or create each carrier pair. With such an assumed required carrier density to get the laser gain medium to be sufficiently above threshold, then the energy required to operate the laser is proportional to the volume of the device. Of course, to make a small device work, we may also need to concentrate the optical field, as in a resonator, and we discuss this point in Section III C below.

For modulator energies, we presume that the modulator requires an electric field $\mathfrak{E}$ of ~ 10⁵ V/cm to operate; this is a typical value of operating field for strong QCSE absorption edge shifts in a modulator. For a given operating field, there is therefore a corresponding electrostatic energy density, and this contribution to the operating energy is therefore proportional to the volume. For a semiconductor relative dielectric constant $\varepsilon_r$ ~ 12, we obtain the modulator energies shown[26] in Table II, presuming an energy equivalent to $(1/2)CV^2 \equiv \int_{volume}(1/2)\varepsilon_r\varepsilon_o\mathfrak{E}^2 dv$ where the integral is over the device volume.

## C.  Optical concentration factor

To make the optoelectronic devices work, especially for the smaller active volumes of material, we may need to increase the energy density in the electromagnetic field by some optical "concentration factor" so that there is enough interaction with the active material – e.g., for a laser operating above threshold, a modulator with enough contrast ratio, a light-emitting diode (LED) with strong enough spontaneous emission into a given mode, or a detector with enough absorption. That concentration might involve a resonator, a sub-wavelength waveguide (e.g., using metals), a structure with reduced group velocity, or some other approach (see Fig. 5).

There are many ways to define such concentration; several terms like cavity quality factor $Q$, cavity finesse $\mathfrak{F}$, and Purcell enhancement factor $F_P$, are well known from analysis of resonators. Partly because we want to include more than just resonator approaches, instead we use a simple and general "optical concentration factor" $\gamma$. Appendix B gives the relation between these various terms[27].

---

[25] Note that such a carrier density also corresponds to one carrier (pair) in a quantum dot of 10nm³ volume, which also makes physical sense for the approximate threshold for population inversion in a quantum dot.

[26] Electroabsorption modulators like QCSE devices can also have dissipation from the photocurrent that can be generated from absorbed photons. We have not included that here, though it has been analyzed elsewhere [41]. If the

devices are designed to operate with low drive voltages ~ 1V or less [41], then this dissipation is essentially just part of the optical loss in using optical modulators, and is best counted there rather than here. High bias voltages would, however, lead to magnification of that dissipation.

[27] Briefly, a cavity of finesse $\mathfrak{F}$ increases the concentration factor by $\mathfrak{F}/\pi$ (Eq. (19), and in a resonator structure, $F_P$ and $\gamma$ are essentially the same concept,

We define our optical concentration factor $\gamma$ as follows. We presume we have some material of refractive index $n$. The wavelength inside the material is $\lambda_n = \lambda / n$ where $\lambda$ is the free-space wavelength. First, we consider a "reference structure" that is a square dielectric waveguide of cross-sectional dimensions $\lambda_n$ in each direction, as sketched in Fig. 5 (a). We presume we are propagating unit optical power through this guide, and for simplicity we presume the power is all confined within this square cross-section. Essentially, this is like a dielectric waveguide near to the minimum practical size. There are, however, no mirrors or resonator structures in this reference structure. As a result of our unit power propagating through this structure, there is some average electromagnetic energy density $U_1$ inside the material[28].

Other structures might have some other average energy density $U_S$ when we are propagating unit power through them. Then, we define our optical concentration factor as

$$\gamma = \frac{U_S}{U_1} \qquad (4)$$

With this definition, our reference structure has $\gamma = 1$. Hypothetically, our reference device of any given kind (i.e., a detector, modulator or emitter) is one that runs using such a reference structure[29], with a length $L$ sufficient to give enough absorption or absorption change, refractive index change, stimulated emission gain, or other emission for a functioning device.

Devices such as photodetectors, lasers, LEDs, and modulators using changes in absorption coefficient or refractive index are all quantum-mechanically based on transition rates proportional to the electromagnetic energy density, as in (single-photon) emission or absorption processes (or the corresponding virtual transition rates in the case of changes in linear refractive index [84]). As a result, if we want to retain the same overall effect of the material on the light, reducing the active material volume by some factor $\beta$ requires we compensate by increasing the electromagnetic energy density with some optical concentration factor $\gamma = \beta$ to keep the device functioning. So, if we want to use a smaller volume of active material for the device, we need to increase $\gamma$ proportionally.

Note any approach that increases the electromagnetic energy concentration while reducing the device active volume by the same factor will reduce the operating energy for such devices. That increased electromagnetic energy density can be from resonators, from slower group velocity (which necessarily requires energy storage somewhere[30]), or reduced waveguide cross-sections. Generally, reducing the group velocity will reduce the operating energy as long as at least some, and ideally all, of the corresponding increase in energy density is in the active medium itself.

Nanometallic or plasmonic concentration in subwavelength waveguides could reduce the operating energy for devices like emitters or modulators, both by reducing the cross-sectional area in which the propagating light is confined (see, e.g., [85], [86]), and possibly also by leading to slower group velocity (see, e.g., [87], [88]). If the use of metals leads to greater loss, though, we may be losing overall in device performance, so such metallic structures need a careful analysis to be sure of their benefits. Dielectric waveguide structures can also reduce group velocity in devices (see, e.g., discussion in [49]).

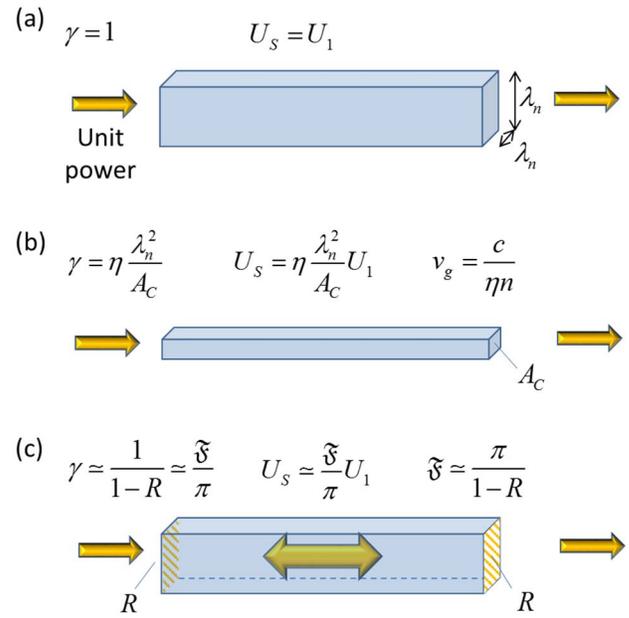

Fig. 5. Illustration of optical concentration factors $\gamma$ and electromagnetic energy densities $U_S$ for various example structures using dielectric materials of refractive index $n$. For free-space wavelength $\lambda$, the wavelength inside the device is $\lambda_n = \lambda/n$. $c$ is the velocity of light in free space. (a) Hypothetical "reference" device structure, with a dielectric guide of size $\lambda_n$ in both directions that confines the propagating light within it. By definition for this "reference" structure, $\gamma = 1$ and the electromagnetic energy density $U_S = U_1$. (We presume for simplicity that the light has phase and group velocities of $c/n$ in such a guide.) (b) A waveguide with the light confined in some smaller cross-section $A_C$, such as by metal walls. The light might also be propagating with some group velocity $v_g$ slowed down by some factor $1/\eta$ compared to the phase velocity $c/n$, i.e., $v_g = c/\eta n$, giving $\gamma = \eta A_C/\lambda_n^2$. (c) A high-finesse resonator structure with mirrors of intensity reflectivity $R$ and a corresponding finesse $\widetilde{\mathfrak{F}} \sim \pi/(1-R)$.

Fig. 5 illustrates the optical concentration factors corresponding to various simple situations. Fig. 5 (a) shows the

---

with $F_P \equiv 0.477\gamma$. Note that $Q$ is the finesse $\widetilde{\mathfrak{F}}$ multiplied by the cavity length in half-wavelengths.

[28] In waveguide structures, there may be low actual energy density at the walls and a higher density in the middle – e.g., perhaps twice as high as the average energy density – and in resonator structures there may be standing wave patterns in which the peak energy density is up to twice as high as the average. Though we could incorporate such effects more precisely in our definition here, for our order-of-magnitude arguments, we simply ignore such effects on the scale of factors of two, and work with the overall average energy densities. We also presume the phase velocity and group velocity in such a reference structure are both just $c/n$, where $c$ is the velocity of light in free space, so we are

neglecting minor possible effects on these from this waveguide structure with a wavelength-scale cross-section.

[29] Formally, a conventional laser cannot run with such a structure because there is no resonator, but we can equivalently presume a hypothetical device that is about one "gain" length long, i.e., a gain of a factor of $e$.

[30] After a pulse enters a structure, its energy has to be stored inside the structure somewhere until it exits the structure again. Unless we have some "side" resonator or other energy store, the energy will be stored as electromagnetic energy inside the material. If the light energy is propagating at a group velocity $v_g = v_p/\eta$, where $v_p = c/n$ is the usual phase velocity, so it has been slowed down by a factor $1/\eta$, then the energy density must have increased by a factor $\eta$ so the total power propagating remains the same.

reference structure. Fig. 5 (b) shows a structure with some hypothetical waveguide of some much smaller cross-sectional area $A_C$; such structures are possible with metals, for example (though in practice such approaches can lead to substantial loss if the guide is too long – see, e.g., [85], [86]). In such small waveguide structures, or in structures with slow-light propagation, the group velocity $v_g$ might also be reduced by some factor $\eta$, e.g., giving $v_g = c / \eta n$, which necessarily leads to an increase in energy density[31] of a factor $\eta$. Fig. 5 (c) shows a resonator with cavity mirrors of some at least moderately large (power) reflectivity $R$. That resonator leads to an increase of optical energy density by a factor $\gamma = 1 / (1 - R) \approx \mathfrak{F} / \pi$ (see Eqs. (19), (20), and (21) in Appendix B.)

For example, for an absorption modulator, for whatever is the absorption coefficient change $\Delta \alpha$ we can make to run the device, in our reference device the length $L$ needs to be such that $\Delta \alpha L \sim 1$ or larger to give a strong modulation. If we make the device shorter than this by some factor $\beta$ (i.e., a total length $L / \beta$) so as to reduce the active volume, then we could make some change to the optics, such as a resonator, to increase the effective optical energy intensity $\gamma = \beta$ to retain approximately the same overall device performance with this smaller active volume. Similarly, in a refractive modulator using our "reference" structure, we would need to make an optical path length change $\sim \lambda / 2$ in the length $L$ to make some useful device. If we reduce the device length to $L / \beta$, then we will need increased optical energy density $\gamma = \beta$ for the device still to work. (We will discuss use of resonators with absorptive and refractive modulation effects in greater detail in Section IV D below.)

In the case of a laser, the gain per pass has to be sufficient to overcome the loss through the mirrors. If we reduce the length to $L / \beta$, then we have increase the optical energy density in the cavity by $\gamma = \beta$ for the laser still to work. In some resonator structure, this is equivalent to reducing the leakage through the mirrors by a factor $\gamma$; for example, increasing the mirror reflectivity from 95% to 99% corresponds to increasing $\gamma$ (and finesse $\mathfrak{F}$) by a factor of 5.

Note, though, in these scaling arguments, that as long as we keep the same electromagnetic energy density interacting with the same total volume of active material, as far as operating energy is concerned, it does not matter what specific length $L$ of device design we have used to achieve this. In the device design, we could completely fill the cross-section of the device with the active material, or instead we could just fill some central slice or layer, but keep the total volume of active material the same by correspondingly increasing the length $L$.

Equivalently, the "fill factor" – the average fraction of the cross-section of the waveguide filled by the active material – does not matter for the energy as long as we are still using the same total volume of active material interacting with the same electromagnetic energy density. If we have a smaller "fill factor", we might, however, choose to increase the optical concentration factor rather than increase the device length.

We should note, too, that for resonators, if we make them longer for the same finesse (and hence the same concentration factor), then the $Q$ factor will rise in proportion, which leads to tighter requirements on resonator tuning. So, if we are using resonators, short structures in which the material fills the mode are preferable to longer ones in which the material only fills a small fraction of the mode. For the same reason of limiting the required $Q$, it is preferable to have strong microscopic optoelectronic effects that can give large absolute values of gain or of changes in absorption or refractive index since those can result in shorter devices and hence lower $Q$ structures for the same optical concentration factor.

The required optical concentration factors for the smaller volumes in Table II are based on a simple scaling from the (1 μm)³ case, in proportion as the volume of active material goes down. The order-of-magnitude energy numbers here for the (1μm)³ active volume are comparable to those of actual demonstrated devices. The 160 fJ for the laser with (1 μm)³ active volume should be comparable to the energy to turn on an efficient conventional edge-emitting laser. 56fJ/bit has been reported for surface emitting lasers [89] with a 3.5 μm diameter aperture[32]. Presuming an active region thickness $\sim 0.1$ μm or less, this number is also in reasonable agreement with the estimated 160 fJ in our approximate analysis[33].

Research demonstrations using photonic crystal resonators and/or quantum dot materials (e.g., [90], [91], [92]) can also be compared[34] with the projections in Table II.

For the modulator with 1 μm³ active volume, the 5 fJ is comparable with the operating energy (including the bias field[35]) for a compact QCSE modulator [41], [78].

One caution for using small volumes of active material is that crystal cavity. [92] shows a single quantum dot lasing in a cavity with a reported concentration factor $\sim 30,000$. Such dots may be somewhat larger in volume than our hypothetical 10 nm cube, by a factor of, e.g., 3 or so [93], so our simple scaling would suggest at least $\sim 10^6$ required optical concentration, a factor of 30 higher than used by [92]. An important difference here, though, is that the experiments in [92] are conducted at a temperature of $\sim 10$K, not at room temperature, and we could expect much greater gain per injected carrier pair as a result, so less optical concentration may be required. There may also be some additional benefit from the greater quantum confinement in the quantum dot as compared to the quantum well gain media presumed in our scaling.

---

[31] Here we presume the resulting increased energy density is all in the active material, though that may not always be the case in such guides; some energy might be stored in the metal guiding layers, for example.

[32] Such VCSEL technology is actively researched for optical interconnect applications [129], with impressive system demonstrations with total energies per bit in the range of a few picojoules [18].

[33] Such surface-emitting structures may also have higher concentration factors than we have suggested in Table II as being approximately the minimum required since they operate with very high reflectivity mirrors.

[34] For lower energy lasers, researchers have exploited photonic crystal cavity structures that allow particularly small active volumes and high $Q$ factors, allowing strong optical concentration. For example, [90] shows 13 fJ/bit operation in a laser with a 0.18μm³ active volume, a number in rough agreement with our scaling here for such a volume. [91] has shown a low-threshold electrically-pumped nanocavity laser using layers of quantum dots in a photonic

[35] Actual energy per bit can be lower because it is not necessary to swing over the entire bias voltage to run QCSE devices [41]. Sub-femtojoule per bit can be deduced in that case for this modulator.

in practice we may need high $Q$ cavities to exploit them. For modulators in particular that is problematic because we need to match the narrow resonance with some operating wavelength, to a precision $\sim 1/Q$. That poses fabrication and operational problems (e.g., temperature drift, feedback stabilization), especially for $Q$ values of 1000 or more. Even for lasers, if they are to be matched to specific operating wavelengths in some WDM system, we would have similar problems. Modulator devices with $Q < 100$ might be usable without such tuning problems, however. See Appendix B for a more detailed discussion.

Note in these scaling arguments that the operating energies of quantum well electroabsorption modulator devices are lower than those of lasers; generally, lower operating energy densities are required in these modulator cases. We could, for example, propose a quantum-well electroabsorption modulator of total volume $\sim (300\text{nm})^3$, which might correspond to some waveguide resonator with a cross-section of $200\times300$ nm and a length of 450 nm (about 1 wavelength in a typical semiconductor in a device operating at a free-space wavelength of 1.5 μm). Only a moderate optical concentration factor of $\sim 40$ would be required to run such a device, which could mean a relatively low-finesse resonator that therefore did not have to be fabricated to extremely high precision. According to the scaling in Table II, such a device would have an operating energy of $\sim 135$ aJ.

### D. Conclusions for scaling to attojoule optoelectronic devices

The key conclusion of this scaling argument is fundamentally optimistic for attojoule optoelectronics: even if we only consider known mechanisms already widely exploited technologically, sub-femtojoule optoelectronic output devices are physically quite possible. Whether the extreme case of the $(10 \text{ nm})^3$ active volume is practical is very much a speculative question, and that case here is included largely for comparison purposes. However, we can be cautiously optimistic that devices in the $(300 \text{ nm})^3$ active volume range are quite possible, and perhaps even the $(100 \text{ nm})^3$ active volume range are viable without drastic technological efforts.

The challenges are that we will have to make the devices small, into the range of 100's of nm or smaller, and they will have to be very well integrated with their associated electronics if we are to obtain the full energy benefits. We will also have to consider seriously and critically any required approaches to concentrating optical fields, such as the use of resonators or conceivably other approaches such as nanometallics (e.g., plasmonics) or slow light, with any associated loss being a major issue; furthermore, the issues of fabrication precision and operational stabilization for resonators with $Q > 30$ need to be carefully examined for any proposed approach.

## IV. OPTOELECTRONIC OUTPUT DEVICE APPROACHES

In our argument so far, we considered only two example approaches to optical output devices; as we think about potential low energy optoelectronics, we should look at the broad range of available approaches to optical output devices generally. Here, we briefly summarize and compare various of the options and their properties and requirements. See also [94] for another discussion of potential low-energy optoelectronics.

Fig. 6 shows various device configurations with conventional waveguides, ring or disk resonators, and "surface-normal" structures in which the light comes in and/or out perpendicular to the surface, either with or without a resonant cavity. There can be many variations in such structures, and photonic crystal or nanocavity structures are also possible.

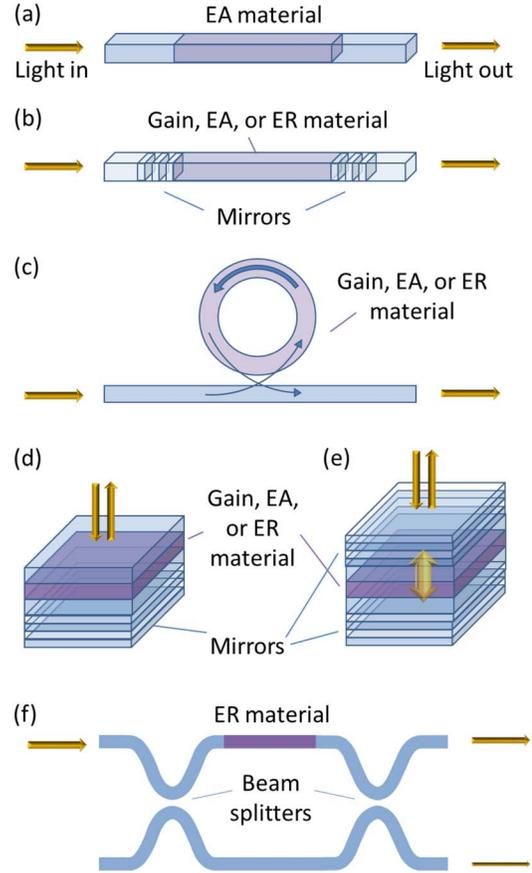

Fig. 6. Various configurations for emitter and modulator devices. (a) A waveguide containing active electroabsorptive (EA) material for a "single-pass" optical modulator. (b) A resonator structure for a laser or a cavity-enhanced electroabsorptive or electrorefractive (ER) device. (c) A disk (or ring) resonator with active gain, electroabsorption or electrorefraction material side-coupled to a passive waveguide. (d) A reflection modulator for use with electroabsorption or electrorefraction material for a "surface-normal" device (the top surface may also be anti-reflection coated). (e) A vertical cavity structure for a surface-emitting laser or a resonant cavity modulator. The bottom mirror may be designed for near 100% reflection so light only leaves from the top. (f) A classic Mach-Zehnder waveguide interferometer structure for an electrorefractive modulator. Two beamsplitters, nominally with split ratio 50:50, split an input beam along two arms. Changing the phase shift in one arm compared to the other changes the division of power between the two output ports on the right, allowing amplitude (or power) modulation from a simple phase shift.

### A. Qualitative comparison of light-emitters and modulators

Before comparing specific device mechanisms, we can make some general comparisons between the two approaches of light-emitters and modulators, as summarized in Table III. In general, the choice between these two approaches is not simple because it involves benefits and problems that emerge when we consider the larger system in which we are using the devices; as a result, a simple comparison on one parameter or on one strength or weakness is not generally sufficient to make a choice.

TABLE III
COMPARISON OF LIGHT-EMITTERS AND MODULATORS AS OUTPUT DEVICES

**Light emitters - Pro**
- No additional optics to get the light to the output device
- Only need to turn the lasers on for the channels in use

**Light emitters - Con**
- Difficulty of monolithic integration with electronics
- Difficulty of wavelength control for individual emitters, limiting the use of any dispersive optics (e.g., diffractive optics) and wavelength division multiplexing
- May require optical isolation
- May require polarization and mode control
- Relaxation oscillation limit to frequency response, increasing power densities at high speeds
- May have timing issues from turn-on delay [95]
- All power dissipation is on-chip
- Issues with temperature variation because of
  - different shifts of bandgap and resonator wavelengths
  - decrease of laser gain with increasing temperature

**Modulators - Pro**
- Centralized wavelength, mode, and polarization control, and optical isolation, all at the laser power source
- Can be driven by optical pulses for precise signal timing, including whole arrays of modulators synchronously
- Only the modulator drive power is on-chip
- Many approaches tolerant to high-temperature operation
- Can be compatible with wavelength division multiplexing, even for untuned or low Q modulators

**Modulators - Con**
- Separate light source required
- Needs optics to split and deliver the power to the many modulators
- All illuminated modulators consume at least the optical drive power even if not driving any signals

Some features that might be viewed as weaknesses can also be strengths. For example, modulators obviously require an external light source and optics to distribute that to the devices, but that also means that we only need to perform the optical isolation and stabilize the polarization, mode form, and wavelength of that one source; we may also be able to exploit the optics to distribute a synchronized set of readout optical pulses, derived from pulsing the one source, to all of the modulators [63], [64]. We will return to such points when we discuss systems in Sections VII, VIII and IX.

### B. Efficiency

#### 1) Device power efficiency

Any optical output device that is to allow low energy optoelectronics must both operate at low total energy, and be very efficient in delivering the necessary modulated optical output power; if it is not, we have to increase its optical output power, and hence its overall power dissipation, so that we can deliver sufficient energy to the photodetector at the other end of the interconnect [41]. As we will see later when we discuss receivers, simply increasing the sensitivity of the receiver to make up for low emitter efficiency or high background loss or absorption in a modulator itself also leads to greater power dissipation. Hence, we have to try to

- avoid light emitters where substantial efficiency compromises have to be made to allow integration
- avoid low-efficiency emitters, even if they have low operating energies
- avoid modulators with significant background loss

#### 2) Single spatial mode operation

A second point about efficiency is that the emitted power must be in such a form that it can be efficiently delivered to the photodetector at the other end of the link. For reasons that will become clearer once we discuss optics and receivers below, this argues strongly that all light emitters in the system emit into a single spatial mode, whether they are the optical output devices themselves or are the optical power source for modulators; indeed, on one argument, only the optical power in the most strongly coupled optical mode from the output device to the photodetector is useful, and the rest of the power is wasted (see Section VII A below).

For modulators, since they will likely be powered by some external optical power source laser, it is relatively easy to make such a laser operate in a single spatial mode; as long as the intervening optics is of reasonably high quality, then modulators will anyway be operating on single-mode beams.

For light-emitters as output devices, we should make sure that they emit with high efficiency into one spatial mode. For laser output devices, we will have to take some care to make sure the spatial mode is controlled, most likely to be in the lowest spatial mode.

If we want to use light-emitting diodes (LEDs) as the optical output devices, we need to construct them in such a way that they emit predominantly into one spatial mode. Typical LEDs are not constructed this way, though as we consider the possibility of making very small LEDs, it becomes more feasible to consider such single-mode devices.

### C. Light emitters as output devices

#### 1) Lasers

Most lasers in use today in information processing and communication are semiconductor lasers. These have the advantages of small size, which in turn is because of the very high gain per unit length possible in semiconductors. They can operate at high speed in direct modulation, though with some limits from the relaxation oscillation frequency (see, e.g., [96] for a recent discussion) that tends to require higher power dissipation for faster modulation rates.

As we have argued above, if such lasers are to have sufficiently low operating energies, then we may need to change from conventional edge-emitting or surface emitting cavities towards nanoresonator structures [97], as in research examples like [91] and [92], or other structures with greater optical concentration. Whether lasers with metallic confinement are viable depends on loss; though demonstrated examples may be small [97], [98], [99], their efficiency or operating energy may be limited as a result (see, e.g., [85] for an analysis of metallic loss in small semiconductor structures in nanometallic waveguides). Small size is of little use here if it results in larger overall operating energy.

Whether we can exploit gain media other than semiconductors is an open question; if their gain is lower, then it may be difficult to get the required performance. Higher-dimensional quantum confinement, as in semiconductor quantum wires and quantum dots, can offer somewhat better gain because of the more concentrated densities of states and possibly improved electron-hole overlap, though these advantages may be somewhat offset by the lower "filling factor" in the use of such structures. Possibly an ideal material would be some relatively dense collection of uniform quantum dots (see, e.g., [100] for a recent example of improving fabrication approaches); size and shape uniformity is, however, important if all the gain is to be concentrated at one operating wavelength so that the device remains efficient.

### 2) Light-emitting diodes

As mentioned above, normally LEDs would be ruled out because of their typical optical inefficiency from emitting into large numbers of spatial modes. One of the major opportunities for light emitters, however, is that LEDs intrinsically become more interesting as we make them small. One reason is that a small LED cannot avoid emitting into only a small number of modes; indeed, an LED with a subwavelength volume can only really emit into one spatial mode (or two, including polarization), which is the mode that is essentially in all directions at once.

A second reason for small LEDs is that the use of strong optical concentration as discussed above will lead to Purcell enhancement of the spontaneous rate emission into the modes with strong optical concentration; indeed, as discussed in Appendix B, the Purcell enhancement factor $F_P$ is essentially[36] the optical confinement factor $\gamma$ we mentioned above in the discussion of optical concentration.

Such Purcell enhancement can also avoid speed limitations that otherwise apply to LEDs because it correspondingly reduces the spontaneous emission lifetime that governs the dynamics of LED modulation. See, e.g., [101] for an example of a nanocavity LED exploiting Purcell enhancement for single mode operation at low energy. Another interesting recent example [102] uses a nanoantenna to enhance spontaneous emission, and [103] uses nanometallic guides. LEDs have the additional advantage that, unlike lasers, they are not "threshold" devices – no particular level of drive is required to get them to work.

Of course, any serious proposition for the use of LEDs would need to show substantial levels of efficiency in the generation of light as well as emission into predominantly a single spatial mode, but LEDs become a serious candidate for low-energy light emitters as we move to smaller sizes and energies.

### D. Modulators

Modulators come in two basic types: ones that operate by changes in the optical absorption of a material (electroabsorption), and ones that use changes in optical path length or refractive index (electrorefraction).

Both kinds of devices can provide amplitude modulation. In devices without resonators, an electroabsorption device in which the increase in the absorption coefficient corresponds to ~ 1 or more absorption lengths in the device length will give useful modulation; in the case of electrorefraction, simple two-beam interference, as in a Mach-Zehnder interferometer [104] (see Fig. 6 (f)), for example, allows amplitude modulation by changing from constructive to destructive interference by inducing ~ 1 half wave of relative path length difference between the two arms[37].

Electrorefraction devices have the advantage that they can be used to switch a light beam from one path to another, as in Mach-Zehnder or directional coupler devices, for example. Generally, electroabsorption devices cannot efficiently switch beams between different paths, for the relatively obvious reason that in one state they are absorbing the beam power.

### 1) Materials criteria for optically efficient modulators

For modulators, we obviously must care about the ability to make some change in absorption coefficient or in refractive index; but, we also care about any overall loss. For example, a particularly important criterion for using a modulator in a system is the absolute difference $\Delta T$ in the transmission of the modulator in its two states [41]; indeed, for some optical input power $P$ to the modulator, the useful optical signal power that leaves the modulator is $P\Delta T$. So, if $\Delta T$ becomes smaller, we will have to increase the power $P$ in proportion. Hence, background loss becomes very important in a modulator.

In Appendix C, we give an extended discussion of the consequences for modulator materials of this requirement of high $\Delta T$. Here we can briefly summarize the key results.

- For electroabsorptive materials, presume we have a material with a background absorption coefficient (i.e., the absorption coefficient in the "transmitting" state) of $\alpha_{trans}$ and a larger "absorbing" state value of $\alpha_{abs} = \rho\alpha_{trans}$, so that the ratio of the "off" to "on" absorption coefficients is $\rho$. To avoid a rapidly increasing system loss penalty, in practice we require

$$\rho = \frac{\alpha_{abs}}{\alpha_{trans}} \geq 2 \qquad (5)$$

- For electrorefractive materials with a background optical absorption coefficient of $\alpha$, so that we get enough path length change without absorbing too much power, for the available change $\Delta n$ in refractive index, we require

---

[36] Formally, $F_P \cong 0.477\gamma$ in resonator structures, as discussed in Appendix B.

[37] Other electrorefractive approaches without resonators, such as devices that might deflect a beam out of the way by changing the optical path in one half of the beam, or devices in which we cause a beam to "leak" out of a waveguide by making a mode unguided as a result of an index change, tend to have similar requirements on effective required path length change.

$$\frac{\Delta n}{\alpha} \geq \frac{\lambda}{2} \qquad (6)$$

- These materials criteria remain essentially the same in devices with resonators. Use of resonators does not help us avoid these materials criteria.

The criteria (5) and (6) can be quite difficult to meet, and various otherwise promising mechanisms cannot achieve them.

### 2) Microscopic mechanisms for optical modulation

There is a broad range of mechanisms that have been proposed and investigated for modulating light in response to electrical drive. We are not aware of a broad comparative review of these in the literature. Because of the breadth of this topic and the level of discussion of physical mechanisms required, we give this detailed treatment in Appendix A, and summarize some key conclusions here as they relate to energies.

#### a)    Electroabsorption mechanisms

The strongest modulation mechanism overall is likely the electroabsorption from the QCSE [65], [66], which is seen in quantum well layered semiconductor structures and other quantum-confined structures; we have already given estimates of the required energies in Table II. It is a mechanism that results directly from the electric field applied to the structure. It is seen in direct gap semiconductor materials and near the direct gap of indirect-gap materials like germanium.

A related electroabsorption mechanism, commonly called the Franz-Keldysh effect (FKE), is seen in bulk materials near to their direct bandgap; it is somewhat weaker and shows less abrupt changes in absorption[38], but is still a viable strong mechanism.

The other main category of mechanisms for changing absorption in semiconductor structures involve band-filling – that is, filling up the "bottom" of a band (usually the conduction band) with carriers (usually electrons) so as substantially to eliminate the possibility of any absorption into those states, thereby removing substantial absorption from some region of the spectrum for photon energies near to the semiconductor bandgap energy.

The resulting magnitude of the changes in absorption from band-filling are similar to, or, under strong excitation, larger than those of the QCSE and FKE; the carrier densities required for operation are similar to those required to turn on lasers, so the operating energies of those devices would be similar to the laser energies in Table II. This category of mechanisms has various other names, including Pauli blocking, Burstein-Moss shift and phase-space filling, and there are some subtleties to the physics, including the influence of excitonic effects, that are not conveyed by these names.

None of these relatively strong electroabsorption mechanisms appear either to be available or usable in silicon itself, however, because they are only seen at or near direct band gaps[39].

#### b)    Electrorefraction mechanisms

Any change in optical absorption spectrum results in a change in the refractive index spectrum through the Kramers-Kronig relations. Hence, there are relatively strong electrorefraction mechanisms associated with the QCSE and band-filling electroabsorption mechanisms, and these can make functioning devices that are competitive with other electrorefractive approaches. One difficulty with such mechanisms is that, in practice, to satisfy the condition (6), the operating photon energy has to be moved to significantly below the band gap energy (i.e., to longer wavelengths) to get away from strong background absorption near the band gap energy. The usable strength of the refractive effect is therefore weaker because the refractive effects fall off as we move away from the region where the absorption is being changed.

Hence such purely electrorefractive devices using these mechanisms have to be longer (e.g., 10's to 100's of microns instead of a few microns) and can therefore have $\sim \times 10 - 100$ higher operating energies than their purely electroabsorptive counterparts. The combination of electroabsorptive and electrorefractive effects can lead to an attractive low energy modulation mechanism in resonant devices, however, somewhat enhancing performance compared to purely electroabsorptive devices (see, e.g., [79]). Again, this class of bandgap resonant electrorefraction mechanisms is not practically available in silicon.

A mechanism that does exist in silicon, and has therefore been widely investigated and used very successfully in devices (see, e.g., [105]), is the "free carrier plasma" (FCP) refractive index change associated with changes in carrier (electron and/or hole) densities [106]. This mechanism is not resonant with any bandgap energy[40]. It is, however, relatively weak, being a further $\sim \times 10$ weaker than the index changes per unit carrier density in the bandgap-resonant "band filling" mechanisms.

Overall, this FCP electrorefractive mechanism in silicon is $\sim \times 1000$ weaker for making a device than the best electroabsorption mechanism (QCSE), as is borne out in device performance; a simple Mach-Zehnder FCP modulator without any optical concentration will require a few pJ/bit [104], whereas a short QCSE electroabsorption modulator with no resonator requires a few fJ/bit or less [41], [78]. As a result, for low-energy devices, the FCP requires very high optical concentration to operate, as in high-Q ring [107], [108] or disk [105] resonator structures, with all of the problems, such as tuning, associated with that.

The final main electrorefractive mechanism of interest is the Pockels effect – a linear change of refractive index with electric field. This mechanism is seen in materials like lithium niobate

---

(which is widely used in telecommunications modulators), in III-V semiconductors, and in electro-optic polymers, with all of these mechanisms being strong enough to demonstrate viable devices. It is not, however, seen in bulk silicon because of silicon's crystal symmetry properties.

The energy required for Pockels effect does not have the same scaling as the other mechanisms discussed; in fact, in the absence of background losses such as waveguide propagation loss, there would be no actual minimum energy – doubling the modulator length would actually halve the energy required[41]. As a practical matter, for reasonable lengths of devices the energies required to operate Pockels-effect devices are not likely to be lower than hypothetical similar devices using other good electrorefractive mechanisms. With very good device engineering, however, including optical concentration from nanometallic waveguides and slow group velocity, devices with ~ 25 fJ/bit have been demonstrated [87], [88], [109] using electro-optic polymers in a device ~ 10 μm long.

### c)    Use of two-dimensional materials

Two-dimensional (2D) materials like graphene or $MoS_2$ have emerged in recent years as intriguing new opportunities for optoelectronic devices. We compare the resulting mechanisms to others in Appendix A; the comparison to quantum well structures is particularly useful because 2D materials and quantum wells share much basic physics.

The simplest way to state the conclusions of this comparison is to say that, broadly, the useful strengths of mechanisms like band-filling, in terms of the energies required, are essentially the same in 2D materials and quantum wells, though 2D materials may offer the possibility large total changes in absorption in small overall volumes, which could help in avoiding high-Q structures. But, electroabsorption mechanisms like QCSE, if they exist at all in given 2D materials, are practically weaker there. For the QCSE, the 2D materials are actually too thin; the ~ 10 nm thickness of quantum wells is close to some kind of optimum.

Therefore, 2D materials may offer many interesting opportunities, such as the ease of applying them to diverse substrates, but they do not currently appear to offer large energy advantages for optoelectronic devices, and are in practice missing a key strong mechanism (the QCSE).

### d)    Conclusions on energies for modulator mechanisms

With the exception of the particularly strong QCSE or the somewhat weaker FKE effects, other electroabsorptive mechanisms will require operating energy densities and optical concentration factors comparable to the those for lasers in Table II. The corresponding electrorefractive mechanisms are generally effectively weaker for device operation than their electroabsorptive counterparts (e.g., by ~ ×10 − 100 ), so would require either longer lengths (and larger energies) or higher optical concentration factors. The widely-used FCP effect in silicon is about another factor of 10 weaker from the point of view of operating energy, so needs particularly long devices or high optical concentration factors. Pockels-effect devices can work well, though they do not appear in practice to offer effects for devices that are stronger than the other electrorefractive effects considered. 2D materials may be interesting for many reasons, but they do not currently appear to offer substantially lower device energies compared to quantum well structures.

Overall, modulator mechanisms can offer operating energies ranging from somewhat worse than laser energies to much better, including the lowest energy microscopic mechanisms for output devices.

## V.    PHOTODETECTORS AND RECEIVER CIRCUITS

If we think about qualities of a good photodetector, considered as a device on its own, we might look for good efficiency, in terms of photocurrent or photovoltaic power generation for every incident photon, and very low intrinsic noise; both of these attributes would obviously contribute to the ultimate sensitivity possible in some optical receiver. For long distance communications, such ultimate sensitivity is very important. The size and capacitance of the photodetector would be secondary attributes; a good receiver design can give very good sensitivity even with large photodetector capacitance (see, e.g., [19], [110]).

As we think about short distance interconnects, however, the requirements change substantially. Specifically, we need to minimize the *total* energy to communicate a bit. That energy must include the energy of all circuits, including the output driver circuit and, especially, the receiver circuit. Receiver circuits can dissipate substantial energies, in some cases possibly even being the largest single contributor to the power consumption overall in a link [17].

When we optimize for minimum *total* energy per bit, the required criteria for the photodetector change substantially. One key and surprising conclusion is that we will likely *not* run the interconnect in a noise-limited fashion [111]. This is a very different approach compared to that in long-distance or even medium distance communications. We should remember, however, that short electrical wire interconnects also do not run anywhere close to a noise limit, so this is a common aspect of short distance connections.

Indeed, one goal in the design of short optical interconnects could be to make them appear as close to the behavior of an electrical short wire interconnect as possible; there is no overhead on such a connection for low-noise amplification, line coding, CDR or SERDES – we simply put the signal on one end of the line and it appears at the other. That simplicity is essential for minimizing energy dissipation in short connections; use of low-energy optoelectronics might enable us to extend that simplicity and low overall energy to much longer connections.

### A.    Receiver circuit energies

The issue of increased power dissipation for high-sensitivity

---

[41] Doubling the length and therefore halving the required refractive index change would double the active volume. Since the change in refractive index in the Pockels effect is proportional to the applied electrostatic field $\mathfrak{E}$, halving the

required refractive index change would halve the required field. But, the electrostatic energy density is proportional to $\mathfrak{E}^2$, so it would reduce by a factor of 4, hence halving the required electrostatic energy overall.

receivers is well understood from classic receiver design analysis. [110] shows[42] that for a field-effect transistor (FET) front-end amplifier circuit, the minimum overall noise from thermal (Johnson) noise is obtained when the total of the photodetector capacitance and any stray and/or wiring capacitance at the input is equal to the physical input capacitance of the FET. This means that such a receiver designed for optimum sensitivity with respect to thermal noise will have an FET size that grows with the size of that photodetector and wiring capacitance, with a corresponding increase in the static current in the FET channel when it is biased as an AC amplifier.

So, even if we consider a noise-limited approach, to reduce power dissipation overall, it can be useful to reduce the total input capacitance connected to the transistor, including the detector capacitance. (See also [19] for a recent analysis of noise in optical interconnect receiver circuits[43].)

To understand the energies involved in receiver circuits, consider, for example, a recent low-energy photodiode and receiver design [112]. The photodiode has $\sim$ 8fF or less capacitance and the hybrid (solder-bump) packaging technique adds about another 25 fF for a total capacitance of $\sim$ 30fF. This example gives a receiver circuit operating at 170 fJ/bit at 25 Gb/s with -14.9 dBm noise-limited sensitivity. Such a receiver circuit energy per bit is impressively low; other circuits (see [112] for comparisons) can dissipate as much as several pJ/bit.

In the work of [112], including the input coupling loss of $\sim$ 6 dB, the effective responsivity of the photodetector is 0.2 A/W. -14.9 dBm is equivalent to a power of 32.3 $\mu$W, so at 25 Gb/s the photodetector is receiving an optical energy of $\sim$ 1.3 fJ/bit, which will be generating $\sim$ 260 aC/bit of charge in the photodetector. In a capacitance of $\sim$ 30fF that charge will give a voltage swing of $\sim$ 8.6 mV. So the effective voltage gain of this amplifier system, including the front end amplifier and the sense-amplifier circuits, is $\sim$ 50 − 100 to get a final logic level output swing that is a substantial fraction of a volt. But, the energy cost of this sensitivity and noise-limited operation is $\sim$ 170 fJ/bit when working with this $\sim$ 30fF input capacitance.

### B. Low-capacitance front ends and receiverless operation

Now suppose that we were able to make a small photodetector (as discussed above in Section III A), integrated very close to the input of a CMOS gate, with a total capacitance of the photodetector, the connecting wiring and the transistor input of, say, $\sim$ 300 aF. Then that same 260 aC of optically-created charge in the receiver of [112] would itself generate a logic-level swing $\sim$ 0.8V [6] to drive the CMOS gate. That would completely eliminate the 170 fJ/bit of receiver circuit energy, allowing the receiving system to operate at $\sim$ 1 fJ/bit total energy. Such an extreme system with no voltage amplifier, and relying on a full logic voltage swing from the photodetector itself, can be called a "receiverless" system [45], [46].

This receiverless approach can be a good starting point for considering designs and energy savings from low photodetector capacitance. The resulting electrical input circuits can be extremely simple, being just CMOS gates, for example.

### C. Near-receiverless operation

approach[44]

Such a "receiverless" approach may not represent the very lowest possible total energy per bit for such links with low detector capacitance; it may be that we can take what we can call a "near-receiverless". In such an approach, conceptually we start with a receiverless design, and then add some receiver gain, but only insofar as we are reducing the total energy per bit of the system. The energy required to give the additional receiver gain must be lower than the energy saved as a result of needing less optical source power.

It might seem obvious that adding more receiver gain would always reduce the total energy per bit because it would allow lower transmitted power. But, adding gain stages does increase receiver power dissipation as well. And, if we increase receiver gain so much that we start to approach a noise-limited design, the receiver power dissipation can rise substantially [111]. We discuss this point in more detail in Appendix D. There is therefore a balance between receiver gain and power dissipation on the one hand and transmitter power dissipation on the other. For long links with high loss and/or high bit rates, then such noise-limited receivers typically are required for functioning links, but once we consider short links and more limited bit rates, optimizing for minimum total energy per bit can lead to quite different conclusions, especially if we have low photodetector capacitance.

One conclusion from our analyses in Appendices D and E and previous discussions [111] is that possibly about one gain stage might be advantageous in such a near-receiverless design for low-loss optical links with low photodetector capacitance, and this gain stage design would still lead to a noise-limited one; this optimum design would still lead to voltage swings that the receiver input that are much larger than any effective noise voltage. The conclusion that only about one such simple gain stage would be required is why we can call this approach "near-receiverless". With the example numbers we consider here, that receiver amplifier circuit could consume up to a few fJ/bit of energy and still lead to overall energy reductions.

### D. Low-capacitance photodetectors

To understand the possibilities for operating with low-capacitance photodetectors, we can examine some orders of magnitude for capacitance, as shown[45] in Table IV.

Historically, photodetectors in telecommunication systems had relatively large capacitances such as $\sim$ 1 pF; the detector and the receiver circuit might be made in different technologies with different materials, and a simple wire bond between the two (with a capacitance that could easily also be $\sim$ 1 pF) allowed

---

[42] See Eq. 4.65 of [110] and associated text.
[43] See, for example, the terms proportional to the photodetector capacitance and inversely proportional to the square root of the transimpedance amplifier power dissipation in determining the minimum possible received optical power

in Eqs. (12) of [19]; low total input capacitance and high amplifier dissipation improve sensitivity in such a noise-limited receiver.
[44] This term "near-receiverless" is one that we are introducing here.
[45] Many of the capacitance numbers here are as discussed Section IIIA above when we were considering electrostatic energies of output devices.

simple manufacture. Receiver power dissipation was also a relatively unimportant issue in such systems.

We see, however, from Table IV that, if we could make detectors with size scales $\sim 1{\times}1{\times}1\,\text{nm}^3$ to $100{\times}100{\times}100\,\text{nm}^3$, the detector capacitance can be comparable to or lower than the input capacitance of the small transistor to which it would be connected. Fortunately, if we use a direct absorption mechanism in a semiconductor, we can obtain strong absorption typically in one to a few microns of length, so even without any optical concentration to increase the absorption per unit length, relatively compact and effective photodetectors are possible (e.g., [113], [114]). Such direct absorption mechanisms are available at commonly used telecommunications wavelengths in III-V semiconductors (e.g., InGaAs) and across the direct gap of germanium.

### TABLE IV
### CAPACITANCE ($C$) OF SMALL STRUCTURES

| Structure | $C$ | References and notes |
|---|---|---|
| $100{\times}100\,\mu\text{m}$ square conventional photodetector | $\sim 1\text{pF}$ | (a) |
| $5{\times}5\,\mu\text{m}$ CMOS photodetector | $4\text{fF}$ | [46]; (b) |
| Wire capacitance, per $\mu\text{m}$ | $\sim 200\text{aF}$ | [6] |
| FinFET input capacitance | $\sim 20 - 200\,\text{aF}$ | [35]; (c) |
| $1{\times}1{\times}1\,\mu\text{m}^3$ cube of semiconductor | $\sim 100\text{aF}$ | (d) |
| $100{\times}100{\times}100\,\text{nm}^3$ cube of semiconductor | $\sim 10\text{aF}$ | (d) |
| $10{\times}10{\times}10\,\text{nm}^3$ cube of semiconductor | $\sim 1\text{aF}$ | (d) |

(a) Assuming a 1$\mu$m thick depletion region and a semiconductor with dielectric constant $\sim 12$.

(b) This is a lateral p-i-n silicon detector, operated at ~425nm wavelength where silicon has strong optical absorption.

(c) The ~20aF capacitance is simulated [35] for a single-fin FinFET, at fin widths of $\geq$8nm. The larger number of 200aF is to account for the possible use of FinFETs with more fins, as is common in circuits, and some parasitic capacitances.

(d) Assumes only the plane-parallel capacitance between two opposing faces, neglecting any fringing capacitance, and assuming a typical semiconductor dielectric constant of $\sim 12$.

We would, however, have to keep the connection to the transistor short – e.g., $< 1\mu\text{m}$ – if wiring capacitance is not to dominate. That means that we need monolithic or at least very intimate integration of the photodetectors with the electronics to which they are connected.

A good example of a detector that could be monolithically integrated with silicon for receiverless operation is a germanium waveguide detector on silicon, with a $1.3{\times}4\,\mu\text{m}^2$ footprint, $\sim 1\,\mu$m height, and ~1.2 fF capacitance [113] Such a device has no optical concentration, so the prospects for reduced capacitance in a smaller device with some concentration, such as a low-$Q$ resonator, are promising.

Avalanche gain in the detector itself is another approach to reducing the required optical input energy. Such detectors have been demonstrated in germanium structures on or with silicon [115], [116], with gains of up to 12 [115], for example. See also work with III-V nanoneedle structures [117], [118], [119], including examples on silicon [117], [119].

Nanometallic resonator photodetector structures have been demonstrated, allowing high responsivity structures in germanium [120]. This work showed up to ~1 A/W responsivity in a lateral resonant cavity structure 975nm wide and ~ 300 nm thick. In such structures, $Q \sim 100$. Such structures can also exploit photoconductive gain, an alternative approach to avalanche gain for useful current gain from the detector itself.

Another approach for concentration with metals uses nanometallic (or plasmonic) dipole antennas to concentrate into a $\sim 100{\times}100{\times}100\,\text{nm}^3$ detector volume [121]. Nanometallics can also enhance other photodetector structures [118], [119], including those with avalanche gain. Mie and other resonances in dielectric structures such as nanowires [122], Fano resonance modification of those [123], and nanocavities [124] are other possible approaches for moderate $Q$ resonances for photodetection.

Note, incidentally, that nanometallic or plasmonic concentration into such small detector volumes is one of the cases where such use of metals can make overall sense despite the loss problems with such use of metals. Suppose metallic optical concentration allows a detector volume that is smaller by a factor of 10, which might reduce the capacitance by a factor of 10 as a result. Even if that metallic concentration is only 30% efficient because of metallic losses, then we may still be winning by a factor of 3.3 in reducing the energy of the system. See, for example, [85] for an analysis of a photodetector in a structure with metallic concentration, including metallic losses. Note also that the use of metals, with their very large effective dielectric constants, is likely the only way to concentrate light into deeply subwavelength volumes.

In general, we can conclude that the concept of very low capacitance photodetectors with reasonable efficiency is quite viable, especially with some moderate amount of optical concentration from resonators or nanometallic (plasmonic) structures. A key point, however, is that such photodetectors must be integrated very close to the electronics. The required closeness of integration here (e.g., $< 1\,\mu$m given the $\sim 200$ aF/$\mu$m wiring capacitance) may mandate a monolithic integration approach if we are to get the major benefits possible here. If we take this approach of small photodetectors and tight integration, though, we can avoid the dissipation of noise-limited receivers (see Appendix E for a discussion of noise in these cases).

### VI. COMPARISON OF LONG, MEDIUM AND SHORT DISTANCE SYSTEMS

Table V summarizes some of the key attributes and technologies for the use of optics in sending information at different length scales. Optics has been overwhelmingly successful in long-distance telecommunications; arguably nearly the all the information we send over nearly all the distance we send it travels over optical fiber. The modern internet would be impossible without the dramatic increase in information transmission optical fiber technology has enabled. A key requirement for long distances is that we get the maximum information over a given fiber over the longest possible span.

Optics is increasingly used at medium distances, such as those between racks inside data centers and large information processing machines. Here one key driver for the use of optics is that otherwise we run out of space for wiring the connections – connection and bandwidth density become important.

At shorter distances, such as inside racks and down towards the edges of chips themselves, optics is not yet a dominant technology, but increasing the density of connections and reducing energy per bit communicated become major system requirements.

No such simple table can be comprehensive, of course, and there are many technologies not mentioned in Table V that underlie the entire table, such as semiconductor electronics and optoelectronics. The details of such a table are also open to debate.

A main point, though, is that the requirements on the technology change substantially as we move to shorter distances, especially for the shortest distances. This is important because much of the investment and technological development has obviously been for the longer distances, but we cannot simply take the same approaches at the shortest distances. We need to view components and systems very differently at short distances, and there are substantially different challenges and opportunities.

We should not doubt that there are significant interconnection problems currently constraining systems at medium and short distances. For example, the "byte per FLOP" problem in supercomputers is well known [17]; it is very desirable in computer architectures to be able to access a byte of information from memory for each floating-point operation (FLOP), but modern machines fall well below this goal. This problem has proved quite intractable so far by electrical approaches; such machines are unable to transfer enough information between the memory and the processors – they operate as if they are in a permanent and severe information "traffic jam". At the present time, there appears to be no physical solution other than optics for major improvements in the information density for such relatively short interconnects.

In the following Sections VII and VIII, we will look at some of the key different requirements and opportunities for short interconnects. Specifically, we consider optics for dense, short interconnects, and issues and opportunities related to clocking, timing, and time-multiplexing. We need to minimize the total energy per bit communicated while also enabling very high densities of interconnections; these two requirements tend to work with, not against, one another, though they lead to approaches quite different from current medium and, especially, long interconnects.

TABLE V
COMPARISON OF LONG, MEDIUM AND SHORT DISTANCE
OPTICAL COMMUNICATION

**Long-distance telecommunications (> 1 km)**
- Key benefits of optics
  - Very large data rates over very long distances
- Key requirements
  - Maximum capacity (b/s) over the longest span
  - Maximum capacity per fiber
- Key technologies
  - Single-mode fibers for low dispersion communication
  - Optical amplifiers for maximum distance
  - WDM for maximum capacity
  - Low-noise receivers for maximum distance
  - Coding and error correction for maximum distance
  - Advanced modulation formats for maximum capacity
- Emerging possibilities
  - SDM for higher fiber capacity

**Medium-distance data links (~ 10 m – ~ 1 km)**
- Key benefits of optics
  - High density of connections
  - Enable flat networks within data centers [14]
  - Reduce overall power dissipation in data centers
- Key requirements
  - High density, low cost, connections between racks
- Key technologies
  - Dense integrated optics and optoelectronics
  - Array optics (e.g., linear arrays of fibers)
  - Line coding to avoid AC coupling issues
- Emerging possibilities
  - SDM for higher density connections

**Short-distance interconnects (< 10 m)**
- Key benefits of optics
  - Very low energy per bit communicated
  - Very high density of connections
  - Signal integrity
    - Signal timing, voltage isolation, low pulse distortion
- Key requirements
  - Very low energy optoelectronics
  - Minimize energy per bit overall, including dissipation in any electronic circuits
  - Integration for very low energy, very high density, very low cost per connection
  - Tolerant to component and operating condition (e.g., temperature) variations
- Key technologies
  - Silicon-compatible integration
  - Very low capacitance photodetectors and integration
  - Very dense, array optics
- Emerging possibilities
  - Free-space and/or SDM for very high densities, allowing moderate clock rates that minimize energy per bit
  - Large synchronous zones to eliminate retiming power

WDM – wavelength-division multiplexing

SDM – space-division multiplexing (as in multiple modes or cores per fiber)

## VII.  OPTICS FOR SHORT-DISTANCE INTERCONNECT SYSTEMS

A key guiding principle for short distance interconnects is that we must optimize the entire interconnect for minimum total energy. That principle leads to some consequences and novel opportunities for optics.

- First, the need to minimize energy overall, and hence minimize optical loss, pushes us to use what we could call "mode-matched" and/or diffraction-limited optics.

- Second, optics also offers the opportunity at short distances to work with very large numbers of channels, which, obviously, can improve interconnect density.

- Third, and less obviously, we can trade off numbers of channels to reduce energy by eliminating electronic link circuitry.

We will discuss the first two of these here, and we will return to the third point in Section VIII when we consider clocking and time-multiplexing.

### A.  "Mode-matched" and diffraction-limited optics

Long-distance communication uses single-mode fiber in part because it avoids the problems that arise from light in many different spatial modes propagating at different speeds, which would lead to pulse dispersion. At medium distances, such pulse dispersion is less important, and multimode fibers can be used; multimode fibers are more tolerant of alignment precision, allowing lower cost systems, and they can also be designed to minimize dispersion.

In such multimode systems, it makes little difference in which mode or modes the signal propagates; a large detector can collect the power in all the spatial modes. With an appropriate receiver amplifier, there is no sensitivity penalty for using such a large detector, and we can let the light scatter into the many modes of a multimode fiber or waveguide.

At short distances, however, to reduce or eliminate receiver energy dissipation, we want to work with the smallest possible photodetector to reduce capacitance. In the receiverless limit, the operating energy is proportional to the photodetector capacitance until that capacitance becomes comparable to the capacitance of any wiring that connects the photodetector to the transistor, and/or to the transistor input capacitance itself. Given our discussion of capacitances above in Table IV, the size scale at which the photodetector capacitance will be comparable to transistor input capacitance in a well-integrated system is at a wavelength scale or smaller.

Suppose we design a photodetector so that it is "minimum-sized" – that is, it has small an area as possible to collect essentially all the light in at least one form of input beam. In conventional optics, it will then have some size of the order of a square half-wavelength in area, as sketched in Fig. 7 (a), to absorb the light as efficiently as possible from one specific tightly focused spatial mode or "spot". A key point, though, is that it will not then efficiently absorb much light at all from any other spatial mode (in the same polarization) [125].

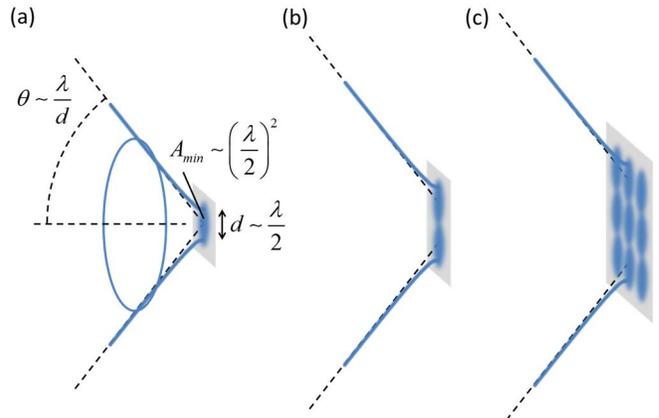

Fig. 7. Sketch of (a) a single beam focused with an appropriately large convergence angle $\theta$ towards an approximately minimum sized spot, of area $A_{min} \sim (\lambda/2)^2$, (b) two beams focused to two spots, and requiring a total detector area $\sim 2\ A_{min}$, and (c) $N$ (= 9 here) beams focused to $N$ spots on a total area $\sim NA_{min}$.

To absorb a second spatial mode efficiently, we would have to at least double the detector area, as sketched in Fig. 7 (b). That doubling is relatively obvious if we think in terms of "spots" that should not overlap; we might think there is some other set of propagating beam shapes that could avoid this problem, but in fact that cannot be done[46] [126] (see Appendix F).

Quite generally, then, for plane absorbing surfaces on photodetectors, their area has to grow proportionately with the number $N$ of modes they are to detect, which means their capacitance also grows by a factor $N$.

In a receiverless system dominated by photodetector capacitance, if we increase the detector area by $N$ to collect the power from $N$ spatial modes, this corresponding growth in capacitance means the voltage swing generated by the optical input energy is therefore reduced by $N$, so we have to increase the total transmitted optical energy by $N$ to restore the voltage swing. So, making a detector that is efficient for detecting a signal in any of $N$ modes can lead to an increase in required system energy per bit by a factor of $N$ in a receiverless system.

We might imagine that we could make some piece of optics ahead of the photodiode that would somehow recombine the incoherent power from multiple different spatial modes into one; that, however, would violate the Second Law of Thermodynamics[47] [125], as well as some basic optics [125].

Hence, in a receiverless system, if we create the light in multiple different modes or if we let it scatter into many different modes, effectively we can make little or no use of that power in different modes. Essentially, we cannot usefully get back any power that we launch incoherently into other modes[48].

---

[46] The approximate "counting non-overlapping spots" heuristic approach is backed up by a much more general and rigorous theory of coupled channels between surfaces and volumes [126]; that approach is based on a sum rule of coupling strengths for the optimum orthogonal channels (or "communications modes"), and can be regarded as a generalized theory of diffraction [126].

[47] If we could do that, we could combine the power from two cool black bodies to heat up a warmer one, for example.

[48] For optical concentration into a minimum-sized photodetector, the best we can do in general in a multimode system where we do not know the relative coherence between the power in the different modes is to concentrate the power from the most powerful mode into the minimum-sized photodetector; all power

In general, we can only use the power in the most strongly coupled mode; power in other modes is wasted. So, we want to run with optics that creates and retains the power of a given signal in one spatial mode. In free-space optics, this requirement equivalently means we want to run with diffraction-limited optics since otherwise we are leaking power (by aberrations) into other spatial modes.

This conclusion also has implications for the use of LEDs. If we allow the LEDs to emit into multiple spatial modes, then only the power in the most powerful mode is useful to us; the rest is wasted in a receiverless system. So, LEDs become interesting in receiverless systems if and only if they are essentially emitting into only a single spatial mode. That is by no means impossible, however, and the smaller we make LEDs, the easier it becomes to move towards such a situation. LEDs with significant Purcell enhancement in a specific mode become quite attractive options (see, e.g., [96] for a recent discussion).

### B. Beam couplers

One other consequence of the need to work with such "mode-matched" optics is that any optics that is to couple from one form of light beam to another, such as a grating coupler to couple from free-space to waveguide optics, has to be mode-matched; that is, if we are only coupling *into* a single mode, as in a single-mode waveguide or a minimum-sized detector, then we can only couple *from* a single mode, that is, from a specific beam shape and alignment, and the resulting coupler has to couple exactly only these two modes to one another [125]. It is not sufficient that a coupler has no absorption losses, for example. Any coupler that is to be efficient must be matched specifically to the modes it is coupling. The alignment tolerance of an efficient coupler is fixed by the sizes of the beams being coupled; that tolerance is not something we can design to any better than that [125].

Beam couplers have received considerable attention (see, e.g., discussion in [49]), based especially on approaches like grating couplers and inverse tapers. It is an interesting question whether nanophotonics could enable other approaches. Novel mode converters based on arbitrary and computational approaches in compact nanophotonic structures [128], [129], [130], [131], [132] have been designed. Extending this approach could be a promising direction for improving coupler efficiency yet further.

There are also novel possibilities for self-aligning couplers that could adjust themselves after fabrication [133], [134], [135] and compensate for aberrations, imperfections, misalignment, and even some mixing from scattering between modes.

### C. Large numbers of channels

Optics has at least two ways[49] in which we can substantially increase the number of available channels: (i) wavelength-division multiplexing (WDM); and (ii) space-division multiplexing (SDM). In WDM, we exploit the very high carrier frequency of light (e.g., 200 THz at 1.5 μm wavelength); we can put many channels of different carrier frequencies on one spatial mode, but still close enough in frequency that their propagation behavior is essentially the same or similar, as in the use of ~ 50 channels on 100 GHz spacing in the telecommunications C-band. In SDM, we might try to exploit some moderate number of different orthogonal spatial modes in a single-core or multiple-core fiber [12] or in a free-space link between buildings, or a very large number of modes, such as 1000's to 10,000's of channels in optical imaging links between chips [2], [136].

Incidentally, the issue of the number of available spatial channels in an optical system, either in free space or in fibers, and the optimum choice of the optical modes for communications in optical systems is one over which there has been some confusion recently; for example, orbital angular momentum modes are sometimes discussed as if they represent an additional set of degrees of freedom for SDM communication, beyond conventional spatial or polarization degrees of freedom, which is not the case. These points are discussed in Appendix F. A simple formula [126] for the diffraction limit to the number of separable channels between two parallel surfaces of areas $A_T$ and $A_R$, separated by a distance $L$ and operating at a wavelength $\lambda$, is (for a given polarization)

$$N_C = \frac{A_T A_R}{L^2 \lambda^2} \tag{7}$$

which is derived as Eq. (35) in Appendix F.

#### 1) Wavelength-division multiplexing in short distance interconnects

There are two basic approaches to WDM for short-distance interconnects: we can use passive optics to split different wavelengths to photodetectors and to combine signals on different wavelengths from modulators that do not themselves need to be tuned or resonant (see, e.g., [137], [138], [139]); or we can use resonator modulators and/or photodetectors that themselves extract the WDM channels by tuning to specific wavelengths (see, e.g., systems using sets of microring or microdisk resonators, each tuned to a chosen different wavelength [14], [17], [20], [105]). See also [140] for a critical analysis of WDM approaches for dense interconnections.

To use either approach for short distance interconnects, we may need to use micro- and/or nano-photonic approaches; the wavelength separator must be very compact if we are to achieve the large number of interconnect channels we would need off a chip. For passive splitters, conventional approaches like arrayed waveguide gratings may be too large to allow one for every spatial channel at short distances, though compact devices have been demonstrated [141]. Echelle gratings are another relatively compact passive micro-optical approach [139]. Solving this problem could be a promising direction for nanophotonics; there are several novel possibilities here, including superprism wavelength splitters [142], waveguide in other modes is useless [125]. Even if all the scattering is coherent, undoing arbitrary coherent cross-coupling into other modes, though now understood to be possible in principle [127], would be hard to apply to complex scattering.

---

[49] We can also use different polarizations, but that only gives a doubling of the number of channels.

nanophotonic wavelength splitters [143], [131], [144], as well as conventional approaches exploiting nanophotonic fabrication (see, e.g., [140]). Such systems could have the additional advantage of being able to interface directly with medium- or long-distance WDM systems.

Whether we can use dense WDM techniques (e.g., with many 10's of different wavelengths) for large-scale short distance interconnects is an open question (see, e.g., [140]); we re-encounter the issue of fabricating or adjusting large numbers of systems with high precision that we found above when considering high-$Q$ resonators. Note that 100 GHz in 200 THz is 1 part in 2000, and any system that pulls out one such channel needs at least that precision to operate. Possibly we can adjust systems in real time to allow such tuning precision, at some cost in complexity and power (see, e.g., analysis by [105]).

### 2) Dense waveguides

One obvious form of SDM is to use multiple separate waveguides. Technologies like silicon photonics can operate with wavelengths that might be as small as $\sim 200 \times 300 \, \mathrm{nm}^2$; such an approach allows quite dense waveguide circuits. In a planar structure, we can therefore have dense waveguide arrays, possibly up to many thousands per centimeter of overall width. If we do use such small waveguides, there are some other considerations, such as loss and crosstalk, and the issue of just what waveguide size to use in various applications is a matter of debate [145].

Such a waveguide technology can be used within either a set of waveguides on a chip, or possibly on some "interposer" secondary waveguide structure onto which multiple chips are attached (see, e.g., [58], [146], [147]). Just what density of connections we could make to some such interposer structure is an open question; couplers between chip waveguides and waveguides on some interposer might require sizes larger than the waveguides because of diffraction. If we tried some simple butt-coupling approach of face-to-face coupling of guides, we would require alignment tolerances between guides on different chips on a scale much smaller than the guide cross-section; that could be challenging with small guides at micron or sub-micron sizes. So, whether it is practically possible to have 1000's of waveguided connections off a chip to such an interposer is still arguably quite speculative.

Plasmonic or nanometallic guides can operate with even smaller cross-sections, such as $\sim 80 \, \mathrm{nm}$ (see, e.g., [86]); at such sizes much smaller than dielectric guides, their losses are, however, relatively high, such as a loss-limited propagation distance $\sim 10 \, \mu\mathrm{m}$ (see, e.g., [85], [86]). Such nanometallic or plasmonic waveguides and related "antenna" concentrator structures (see, e.g., [121]) could be very useful at distance scales of microns or shorter; they represent the only way to guide light controllably at few-micron or sub-micron scales, and the only way to concentrate light directly into sub-wavelength structures. For longer distances, however, to reduce loss, they would have to be made with larger cross-sections, and then it is no longer clear that they offer advantages compared to the dielectric guides we could then make at similar cross-sectional sizes (see, e.g., [148] for a critical discussion).

### 3) Free-space and space-division multiplexed optics in short distance interconnects

The core idea of free-space optics, and more generally of SDM optics in which beams may overlap as they propagate, is that with one optical system, we can handle multiple beams of light or spatial modes at once. We start out with signals in separate "spots" or single-mode guides at the transmitter end. In the middle of the optical system, the resulting beams may all be in modes that overlap, but the optical system will separate these out to similar spots or single-mode guides at the receiver end, giving multiple separate channels for communication, as in Fig. 7.

Of course, this idea is routine in classical optics – imaging optics with a simple lens does exactly this function. Such imaging optics can form the basis for free-space optics for interconnection with many 1000's or 10,000's of beams [136], and we will return to this point below.

#### a) Few-mode SDM systems

For small numbers of modes, e.g., from a few modes up to possibly 10's of modes, it may also be possible to run separate spatial channels through a single optical fiber. That possibility is relatively straightforward if the fiber has multiple separate cores with negligible optical coupling between the cores. More intriguing is the possibility of operating with overlapping modes in fibers. That possibility requires some way to transform in and out of the overlapping fiber modes to connect to separate spots or waveguide at the ends of the system, which is an interesting area for novel optics [10], [11], [12], [149], [150]. Recently, it has been understood, at least in principle, how to solve such separation problems even in the general case of arbitrary overlapping but orthogonal beams [133], [134], [135].

A subtler issue is that, with overlapping modes or even loosely coupled cores in one fiber, there will in general be scattering between the modes. That scattering is not in general predictable; it can result from imperfections and it can change in time because of, for example, temperature fluctuations or mechanical bending or vibrations.

Scattering in and out of different modes can additionally lead to variations in group delay, which can impact the use of SDM in long connections [151]. In short connections, such group delay variation might not be as much of a problem, but we would still need to undo the scattering to separate the overlapping information channels again. Use of electronic techniques to undo the effects of the coupling, such as MIMO[50] algorithms in digital signal processing (DSP) circuits [152], can handle both group delay variation and separation of cross-coupled channels. Those MIMO algorithms and processing might make sense at longer distances. The power consumption of such circuits could rule out such approaches in short

---

[50] MIMO – multiple-input, multiple-output – approaches come originally from wireless communications technology, where many transmitting and receiving antennas may be used at once. Signal processing techniques can

separate out the channels from the signals from the multiple antennas, including undoing the effects of the delay variation from signals propagating along different paths in a scattering environment.

interconnects, however.

It is possible in principle to undo such scattering [127] using purely optical self-configuring techniques running with low power feedback loops [133], [134], [135], and such a scheme has recently been demonstrated based on these architectures and algorithms [153]. For such schemes to be practical for short distances, we would, however, require optical phase shifters that can run at very low power; possibly micromechanical approaches could achieve such low-power phase shifting [154], [155], though this remains speculative. Such schemes also might take up significant chip area, which could limit their use somewhat.

*b)      Systems with very large numbers of modes or beams*

If we consider free-space optical systems, we can consider very large numbers of modes or beams. With reasonable design we can suppress most undesired scattering between such modes (for example, any good imaging system, like a camera lens, will have very little scattering between different image pixels). Such systems can routinely support millions of modes, pixels or resolution elements, even in quite compact, millimeter-scale optical systems like cell-phone cameras.

Free-space optical approaches have been researched in various functioning systems and technologies. For example, a six-stage digital system with more than 65,000 light beams, using imaging interconnects between stages, has been successfully demonstrated [136], as have various other free-space optical systems and approaches [156], [157], [158], [159], [160], [161], [162].

Generating arrays of 1000's of light beams with low loss from one source is straightforward using Dammann-grating spot array generators[51] [163]. Other diffractive optics in planar structures [161] can offer more complex interconnection patterns, and further approaches are available for specific regular interconnection networks [162], [163]. Various other micro-optical techniques are also possible, including lenslet arrays. (See, e.g., [164] for an extensive discussion of such free-space optics, including various micro- and nano-optical approaches and technologies.)

Though such approaches have been successfully researched, they have not yet been exploited to any great degree in short interconnects, in part because we have not yet needed the densities of connections they can provide. The time when we may need such densities may be approaching, however, and there are other benefits, including reduction of energy for clocking and timing that we will discuss below.

We might think there would be problems with letting the light leave the waveguides and propagate through free space, but, as stated, we routinely do this in imaging systems without major difficulties. Furthermore, though we use the term "free-space", we do not necessarily mean we are propagating through air; instead we could use bulk glass or plastic, so we can readily avoid problems such as dust or turbulence.

We might think it would be difficult to align so many beams.

In fact, though, approaches like Dammann grating spot array generators [163] easily and efficiently generate customizable and very regular arrays, with a geometric precision guaranteed by lithography. In using such arrays, we only need to ensure alignment of a few parameters; once we have set those, the entire array is aligned. As with any optical alignment even of one beam, we need to set the overall position in three spatial dimensions and in two angles, and we need to focus the beam; but, then to align the entire array we only need one additional angle (rotation about the beam array axis) and one additional factor, which is the overall physical size scale of the array of spots. Such spot array generators are diffractive elements, and as such will have some wavelength dependence, but, as discussed in Appendix F, these do not appear to be major limitations.

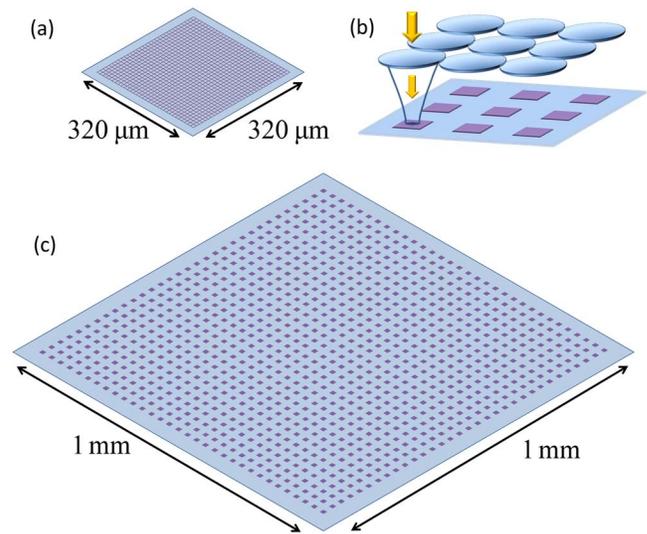

Fig. 8. Illustration of 32×32 arrays of 10×10μm² areas for optical spots on a chip or other substrate. (a) Directly side-by-side, taking up 320×320μm² area. (b) An optional array of lenslets, e.g., on 31.25μm centers, shown above the array of spot areas. Such lenslets can take an array of larger side-by-side spots and focus them onto the small spot areas on the chip. (c) A 32×32 array, spaced apart on 31.25μm centers, possibly using lenslets as in (b), taking up 1×1mm² area.

A simple calculation can show the orders of magnitude possible with such "free-space" optics (See Fig. 8). Suppose, for example, we allocate a surface area of ~ 10×10μm² for each optical "spot" on the surface of the chip. (Such an area corresponds approximately to the size of the optical spot in a single mode fiber, and could correspond to the area of a grating coupler or some other structure for converting between free-space and waveguide propagation.) If we arranged such areas side by side, a 32×32 array of such spot areas, giving 1024 spots or channels, such channels would only occupy a chip area of ~ 320×320μm² altogether. Even at an on-chip clock rate ~ 2 GHz on each such channel, the bandwidth density here would be 20 Tb/mm² (2000 Tb/cm²). Also shown in Fig. 8 is the possible use of arrays of lenslets to concentrate from larger

---

[51] Such an approach uses one lens to collimate a beam from a source like a fiber output or a laser, a diffractive optical element, which is a lithographically fabricated plane structure, that generates beams at multiple different angles

from the collimated beam, and then a second lens to turn the multiple different angles into spots on the output plane. See [163].

spots onto spot areas space apart, e.g., on 31.25μm centers, leading to an expanded total area of ∼ 1×1 mm² .

Even if we expanded to 62.5 μm center-to-center spacing, the total area required would only be ∼ 2×2 mm² for these 1024 channels. Such a spacing would allow significant room between the spot areas or corresponding output couplers for waveguides to route optical signals and/or optical power beams, a concept we will discuss in Section IX below. Such a 2x2 mm² cross-section system could carry thousands of channels over distances of many centimeters with only simple lenses. See Appendix F for a calculation of the numbers of channels available in free space systems.

These areas are much less than the overall surface area of a chip, which can be up to a few square centimeters. Even running at only an on-chip clock rate of ∼ 2 GHz, and even allowing 2 physical "beams" or channels for each data channel (as in "dual rail" operation – see below in Section VIIIB) such a system with 1024 beams in ∼ 2×2 mm² area corresponds to 1 Tb/s of data and a bandwidth density of 25Tb/cm².

Suppose even that we wanted to couple directly into a 32×32 array of conventional optical fibers, which would therefore imply 125 μm center spacing; such an array butt-coupled or imaged at unity magnification onto the surface of a chip would only require a 4×4 mm² chip surface area.

Hence, operating with ∼ 1000s of channels of free space connections in an out of the surface of a chip corresponds to relatively straightforward optics that also need not use a large fraction of the chip surface area. Such a number does not approach any limits in optical design, required area, alignment or wavelength precision. Significantly larger numbers of channels, e.g., up to 10's of thousands, might be possible if desired. Such optics need not occupy a large fraction of the chip area, leaving considerable room for other functions, such as heat-sinking or electrical connections.

Communication of 10,000's of channels over distances of many meters is also straightforward with a single free-space optical system that could be similar to two telephoto lenses "staring" at each other (see Appendix F).

## VIII.   CLOCKING, DATA RETIMING, AND TIME-MULTIPLEXING

### A.   Timing problems and resulting power dissipation

There is an important aspect of dissipation in interconnect systems that so far we have overlooked – the energy required for clocking, data retiming, and time-multiplexing in interconnect links. One reason we have not considered these aspects so far is that they are mostly not problems in transmitting and receiving devices themselves[52]; rather they arise as problems from electronic circuits.

In conventional digital logic systems, not only do we need well-defined logic levels for "1" and "0" in terms of some signal amplitude like voltage; we also need logic signals to fit into well-defined time slots. Obviously, if some 2-input AND gate is to give meaningful outputs, the two inputs must be representing valid logic levels at the same time, and we must only look at the gate output at a time when both inputs are valid. In typical logic systems, we do this by applying a "clock" in the system to define the valid time windows, and we may also use additional circuitry like latches to "freeze" signals so they are valid in the desired time slots. The distribution of the required clock signal itself can be regarded as in interconnect problem, and that distribution can also take up a significant fraction of the chip power (see, e.g., [45], [165]).

If we think of relatively "long" interconnects, which here could be as short as across a chip or our "short distance" interconnects between chips, boards or cabinets, then two further issues arise:

(1) the interconnects themselves can have significant delay, the delay is likely not an integer number of clock cycles, and that delay is also somewhat unpredictable in electrical wiring[53];

(2) the clock frequencies in use at the two ends of the interconnect may not even be the same.

In data links, the first of these two problems can be handled by circuitry that performs just clock phase recovery, effectively from the data itself; the second problem can be addressed by recovering both the clock phase and the clock frequency from the data. Both clock phase and clock frequency recovery can require significant circuitry, including delay- and/or phase-locked loops and data buffering for retiming; such circuitry obviously dissipates power. Collectively, these issues of recovering the clock phase and/or frequency and of retiming the data are referred to as "clock and data recovery" (CDR).

Typically, on links we have also wanted to get the maximum amount of data on a given physical channel; so we may time-multiplex the data from the lower frequency of the circuit's basic logic operations to some higher frequency, and similarly time-demultiplex it at the receiving end. Hence, we have the additional power consumption of the time-multiplexing and demultiplexing circuitry (otherwise known as serialization and deserialization or "SERDES" circuitry). That circuitry necessarily has to run at some significant multiple of the logic circuit frequency, which typically will mean it is consuming more energy per bit operation than a logic gate itself does. Furthermore, with such time-multiplexing, the problems of clock recovery become worse; now we must recover a higher frequency clock, which will also mean we need an even better timing precision in the recovery of the clock window.

For example, [24] shows that the electronic circuit functions of line coding (for receiver AC coupling), CDR, and SERDES can together consume ∼ 20 mW for a 10 Gb/s channel, so ∼2 pJ/bit. Of this energy, more than half (so, ∼ 1 pJ/bit) is consumed by the SERDES circuitry. All this energy is in addition to any energy to run the optical signal receiver and transmitter circuits and devices. 12 - 14 % of the power is in the CDR, so > 100 fJ/bit for just that portion, in addition to the SERDES dissipation.

---

[52] Turn-on delay in lasers can contribute to timing variability, however [95].

[53] The rise times of signals on electrical lines depend on the line resistance, but the temperature coefficient of the resistance of, e.g., copper is such that the

rise time is not reliably predictable, and hence the effective signal delay is not predictable in practice on long electrical lines [22], at least not to within some small fraction of a clock cycle.

The reason for such energies in SERDES and CDR is clear from our earlier discussion of energies to run logic gates (see Table I). Because running just one gate to perform just one logic operation requires several femtojoules at a minimum (and possibly considerably more), every time we "touch" a bit in some operation or perform some other logical operation, we dissipate at least such femtojoule energies. Each bit is "touched" multiple times in a time-multiplexed link; for example, SERDES circuits typically require clocked latching and time (de)multiplexing of each bit at transmit and receive, as well as other logic operations such as byte realignment. As a result, no such time-multiplexed link can approach the energies of a simple local interconnect in an electronic circuit.

So, in some hypothetical future link using low-energy optoelectronics, in which we may have eliminated the receiver circuit power dissipation by our receiverless or near-receiverless approaches, unless we somehow also reduce SERDES and CDR powers by orders of magnitude, we cannot take much advantage of the benefits of the new optoelectronics approaches.

Fortunately, however, there are ways in which optics can eliminate both CDR and SERDES and their associated power dissipation. These approaches are somewhat radical from the perspectives of interconnect systems as we currently know them, but because of the growing importance of these issues, we need to consider these optical approaches seriously.

### B.  Optical approaches to eliminating line coding, CDR and SERDES

Optics has three major advantages that are not yet greatly exploited in short interconnects:

(1) optical delay is very predictable, allowing possibly larger synchronous systems, such as an entire rack or set of racks [22];

(2) optics can support short pulses over moderate lengths, allowing very precise clocking from the fast rise times of the optical pulses [45], [62];

(3) in systems with modulators, we can read out the modulators using optical pulses, automatically retiming the data as it is read out [63], [64].

We also need to consider two other aspects of optical receivers – namely,

- AC coupling, which typically leads to the requirement of line coding circuitry, and

- gain control.

We would also like to use optics to avoid both of these additional circuit issues.

#### 1)  Optical delay variability and precision

Optical fibers have a change of refractive index with temperature of $\sim 10^{-5}$ per K (or per degree Celsius). With a temperature range of 100 K (or Celsius degrees) for the system, we could, for example, have optical fiber connections as long as $\sim 3m - 10m$ for only $\sim 10$ ps $- 30$ ps timing uncertainty from thermally-induced propagation delay variation in the fiber. Delays of this magnitude are likely small compared to the clock period of typical logic circuits; clock frequencies of $\sim 2$ GHz would have total clock periods of $\sim 500$ ps.

#### 2)  Short pulse propagation in fibers

From the usual relation between frequency bandwidth and pulse time duration, as in Fourier transforms, a pulse of full width at half maximum (FWHM) $\Delta\tau$ has a minimum frequency FWHM bandwidth $\Delta f$ given by an "uncertainty principle" relation [84], which, for a Gaussian pulse shape as an example, takes the form

$$\Delta f \, \Delta \tau = 0.44 \qquad (8)$$

Mode-locked lasers can generate pulses of quality comparable to such minimum uncertainty principle limits[54], for example, and likely a well-designed low-chirp modulator can also generate such high-quality pulses from a continuous-wave beam.

For example, a $\sim 10$ps pulse has a bandwidth $\Delta f \approx 44$ GHz . Near 1.55μm wavelength, this is equivalent to a wavelength spread $\approx 0.35$ nm . Typical long-distance telecommunications fiber is designed[55] to have a dispersion $\sim 10 - 20$ ps/nm-km [10]. So such a 10 ps pulse would have a spread of $< 7$ ps in one kilometer length[56]. Hence over lengths even up to 100's of meters, such pulse dispersion may present no problems, and for distances of meters or 10's of meters, it is essentially completely negligible. Even pulses $\sim 1$ ps duration would show only moderate spreading over 10 m.

We also know that we can use such short pulses to deliver very precise clocking to electronic systems[57], with sub-picosecond precision demonstrated [62]. So optics, then, is a very good way to deliver precise and accurate clocking to electronic systems, even up to overall size scales $\sim 10$ m. Once caveat is that we would not in general be able to inject the clock signal optically for all the points on a chip that need to be clocked; on a chip there is a very large number of such points, and we would not have enough optical power in practice to clock all such points directly. We could, however, eliminate some of the upper layers in the clock distribution tree on a chip with optics [45]. The main benefit of optical clocking for large systems, though, may be in its ability to run that entire large system synchronously, avoiding the CDR and SERDES power on the longer links (e.g., off chip) as discussed above.

#### 3)  Data retiming by pulsed optical readout of modulators

One additional benefit of using short-pulse optics with a modulator-based approach is that we can read the data out of modulators and retime the data as a result, with no additional power dissipation required for that retiming [63], [64].

---

[54] Such pulses are known as "time-bandwidth limited".

[55] It is also possible to design fibers with lower or even near-zero dispersion [10]; finite dispersion in long-distance fibers can also be a deliberate system choice to avoid various problems in fiber transmission.

[56] This calculation is a simplistic linear addition of the calculated pulse dispersion spread to the pulse width, and may therefore be an over-estimate.

More correctly for Gaussian-like pulses, we might add in quadrature (the square root of the sum of the squares of the pulse widths or spreading).

[57] Note that in general, not only can optical clocking deliver very precise timing in terms of predictability and length of the optical pulses; effectively, the optical pulse also leads to a much faster rising voltage edge than can be generated by conventional electrical means on chip, with further improvements in the resulting circuit performance [64], [166].

The idea here is shown in Fig. 9. The data from the electronic logic circuit drives a modulator or array of modulators. Then we read out the modulator(s) with a short pulse, or an array of optical short pulses, as might be generated using a Dammann grating from one short pulse source. As long as the optical pulse readout comes at some time that the electrical data is valid, all the data read out now acquires the timing of the readout pulse.

Hence we can remove the timing skew (different fixed delay on different logic paths) by simple choice of optical path lengths, and we can largely eliminate the jitter (statistically varying delay from noise or power supply fluctuations, for example), retiming the data precisely to the optical clock. Effectively, the optical pulse readout is also removing the need for a set of data registers and their clocking.

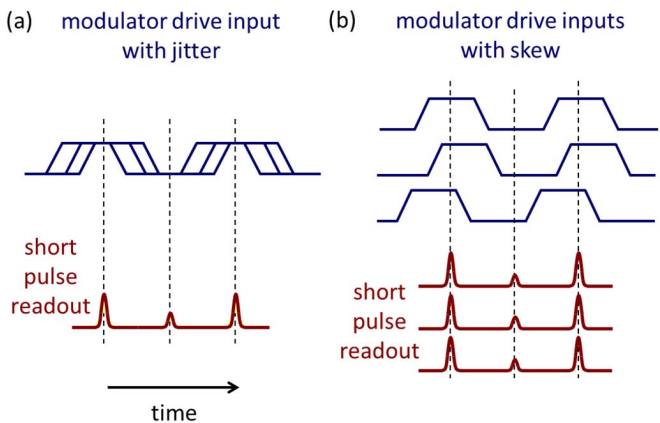

(a) modulator drive input with jitter

(b) modulator drive inputs with skew

short pulse readout

short pulse readout

time

Fig. 9. If optical modulators are read out with synchronized optical pulse trains as inputs, then we can remove (a) jitter (random variations in signal timing) and (b) skew (different timings on different signal channels) that are present in the original electrical drive inputs to the optical modulators, leaving retimed and synchronized signals on the optical pulses after the modulators. (After [64]).

Note that, as long as we design the optical system so that readout pulses arrive within the clock window, we need no electronic circuitry at all to achieve this retiming. We also need make no change to the optical modulators as long as they are capable of handling the optical bandwidth of the pulses (which devices like electroabsorption modulators can certainly do); specifically, we do not need to change the way we drive them electronically or speed them up in any way. There need be no change in the modulator design or increase in its size or decrease in packing density to exploit this approach. We also do not need fast receiver amplifiers or electronic sampling circuitry. Simple "receiverless" operation (see Fig. 10) or integrating receiver front-ends need no modification to work with such pulsed input, and indeed can then perform better than they otherwise do [64], [166]. Of course, large numbers of spatial channels are required if we eliminate time multiplexing, but as we have discussed in Section VII, free-space optics can offer such numbers.

### 4) Optically modulo-synchronous volumes

We could extend the use of the precision of timing available in optics and optical fibers to what we could call optical modulo-synchronous volumes (an idea and a terminology that we are introducing here[58]). By modulo-synchronous we mean that all propagation delays are either one clock cycle or integer numbers of clock cycles, to some accuracy of a small fraction of a clock cycle, so they have the same or similar time delay, modulo a clock period.

The idea of such modulo-synchronous volumes is that we would completely remove the need for clock phase and frequency recovery on all interconnects, from a chip-sized length scale up to a scale of possibly many cabinets in size (e.g., $\sim 1$ cm up to $\sim 10$ m). All signals on such interconnects would be delivered within a known fixed part of the clock cycle window throughout this modulo-synchronous volume, with effectively integer numbers of clock cycles of delays. Within what we could call a module, that delay be within one clock cycle, or whatever substantial fraction of that we normally consider for reliable combinational logic operations. Between modules and racks, the additional delay would be integer numbers of clock cycles. The overall optically modulo-synchronous volume could be of order $\sim 10$ m in size, eliminating all clock recovery within a rack or even a set of racks.

The optical requirements for such a modulo-synchronous system are quite modest, especially if we decide to run the system at the moderate, few-GHz clock rates of modern electronics (clock rates that are chosen so as to minimize and control power dissipation)[59]. The simple action of cutting optical fiber cables to specific lengths within $\sim$ a few mm is then sufficient to allow modulo-synchronous operation over $\sim 10$m size scales, with the additional propagation delays precise to timescales $\sim 10$ ps. Pulses from modulated conventional or mode-locked lasers provide suitable optical sources. We give some specific example calculations for such systems in Appendix G.

### 5) Avoiding AC coupling and gain control problems

When we make a receiver circuit, especially one with a small-signal amplifier at its front end, we have to consider the referencing of the voltage "midpoint" or signal "zero" of the amplifier input and the corresponding equilibrium voltage output from the photodetector (i.e., the average voltage output through some long string of 1's and 0's); in general, these voltages will not be the same. This DC offset could cause significant problems, especially if it is comparable to or larger than the input sensitivity of the receiver amplifier.

A typical solution to such a problem is to AC-couple the amplifier input to make it insensitive to such static DC offsets, for example by putting a capacitor between the photodetector output and the amplifier input. That AC coupling leads to

---

another problem, however; if the data corresponds to a very long string of 1's or a very long string of 0's, the capacitor will essentially block such sequences[60]. As a result, we may add "line coding", in which the actual data signal is "coded"[61] before transmission into a different one that avoids such strings, and then "decoded" at the receiver end. One example is "8b/10b" coding [24]. That line coding adds circuit complexity and power dissipation. In the example we quote above [24], the power dissipation associated with that line coding was ~ 15 – 20 % of the energy per bit, so ~ 300 – 400 fJ/bit. This is a significant energy, so it is important to try to eliminate it also.

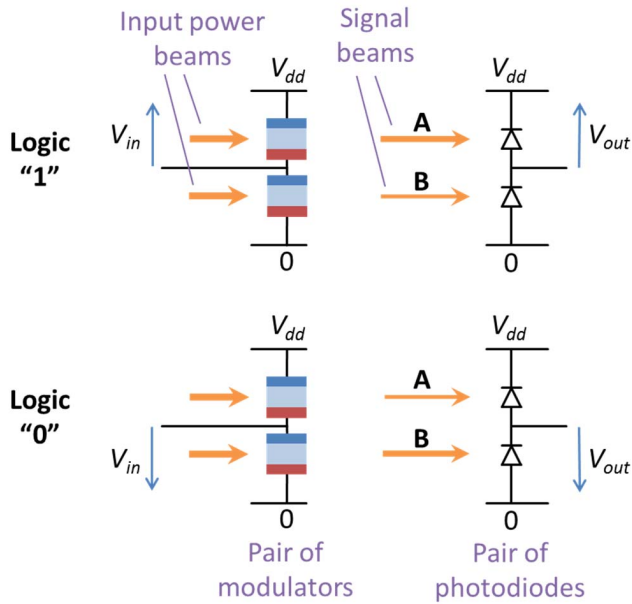

Fig. 10. Dual-rail signaling. Equal input power beams are modulated by a pair of modulators, electrically stacked and driven at their center point by a voltage $V_{in}$. The resulting pair of modulated beams are transmitted to a pair of electrically stacked detectors, where they lead to an output voltage at their center point of $V_{out}$.

If we have large numbers of optical channels available, as in some free-space system, for example, there is one interesting optical option to avoid these problems – namely, "dual-rail" operation, in which we use a pair of beams $A$ and $B$ to represent one signal. Here, a logic 1 is represented by beam $A$ being bright and beam $B$ being dark (or less bright), and a logic 0 corresponds to the opposite. Then if we use a pair of photodetectors in a "stacked" configuration at the receiver, we can avoid AC coupling and all of the associated coding and decoding power (see Fig. 10). Such dual-rail optically approaches were successfully employed in large digital optical system demonstrations [136], [156], [157].

This approach has several additional benefits:

- it avoids any requirement of high on/off contrast in a light beam, because it is a differential approach that only works with the difference between the powers, not the absolute values;
- it avoids any need for gain control when used in a receiverless or near receiverless mode; even with arbitrarily large over-drive of the actual optical inputs, the output voltage at the center point will either saturate at the supply rails (for photoconductors) or at voltages no more than the diode forward voltage past those supply rails (as in so-called "diode-clamped" receivers [167], [168])
- it can operate as an analog data latch when using photodiodes; if we receive optical pulses into such a detector pair driving a high-impedance input like a CMOS FET gate, then, in the "dark" between the arrival of the pulses, there is essentially no path for the charge to leak off the photodetectors – at least one of the diodes is always in reverse bias in such a scheme – so the logic state voltage is remembered until being reset by the arrival of the next pair of data pulses.

## IX. AN EXAMPLE PHYSICAL ARCHITECTURE FOR ATTOJOULE OPTOELECTRONICS

To illustrate how these various optical techniques could be used in a large system to reduce power dissipation, we sketch a physical architecture here. This example exploits the various approaches outlined above to eliminate the energies of receiver amplifiers, line coding, CDR, SERDES, and a significant portion of clock distribution power generally, while allowing large interconnect bandwidth densities. This is not meant to represent some optimum architecture, or to exclude other approaches; instead, it is just an example to show potential viability and performance. If we could generate the necessary low-energy optoelectronics integrated with their electronic circuits, this example is otherwise one that reasonably could be engineered; the other optics required are well within the capabilities of current engineering if we chose to pursue them.

### A. System interconnect energies

We presume first that we are going to run the entire system at 2GHz clock rate[62], consistent with power-efficient silicon chips, and we presume the optical modulo-synchronous approach for the system. We drive all longer interconnects using optical pulses through modulators. Such interconnects would predominantly be off-chip, but could include some of the longer on-chip interconnects also in the optically hybrid waveguide/free-space architecture we will discuss. Note that these interconnects could be as long as ~ 10m within the modulo-synchronous approach.

Hypothetically, we would operate receiverless or near-receiverless photodetector pairs integrated very close to minimum-sized transistors, using a dual-rail approach. We presume the total capacitance of the photodetector pair, the input transistors, and any wiring connecting them is ~ 100 aF.

---

[60] Such sequences also cause problems with clock recovery because there are no "transitions" to use to estimate the clock cycle time.

[61] This "line coding" is quite different from coding we might use for error correction to counteract the effects of noise, and would be in addition to that. In general, in short interconnects, we would try to avoid running near any noise

limits anyway, and would certainly want to avoid the yet further complexity and power dissipation of error correction on every link.

[62] Note that the major electrical interconnect connections to memory chips, such as the DDR4 specification, are specified to run just at such low GHz rates, simply running large numbers of lines to achieve large aggregate data rates.

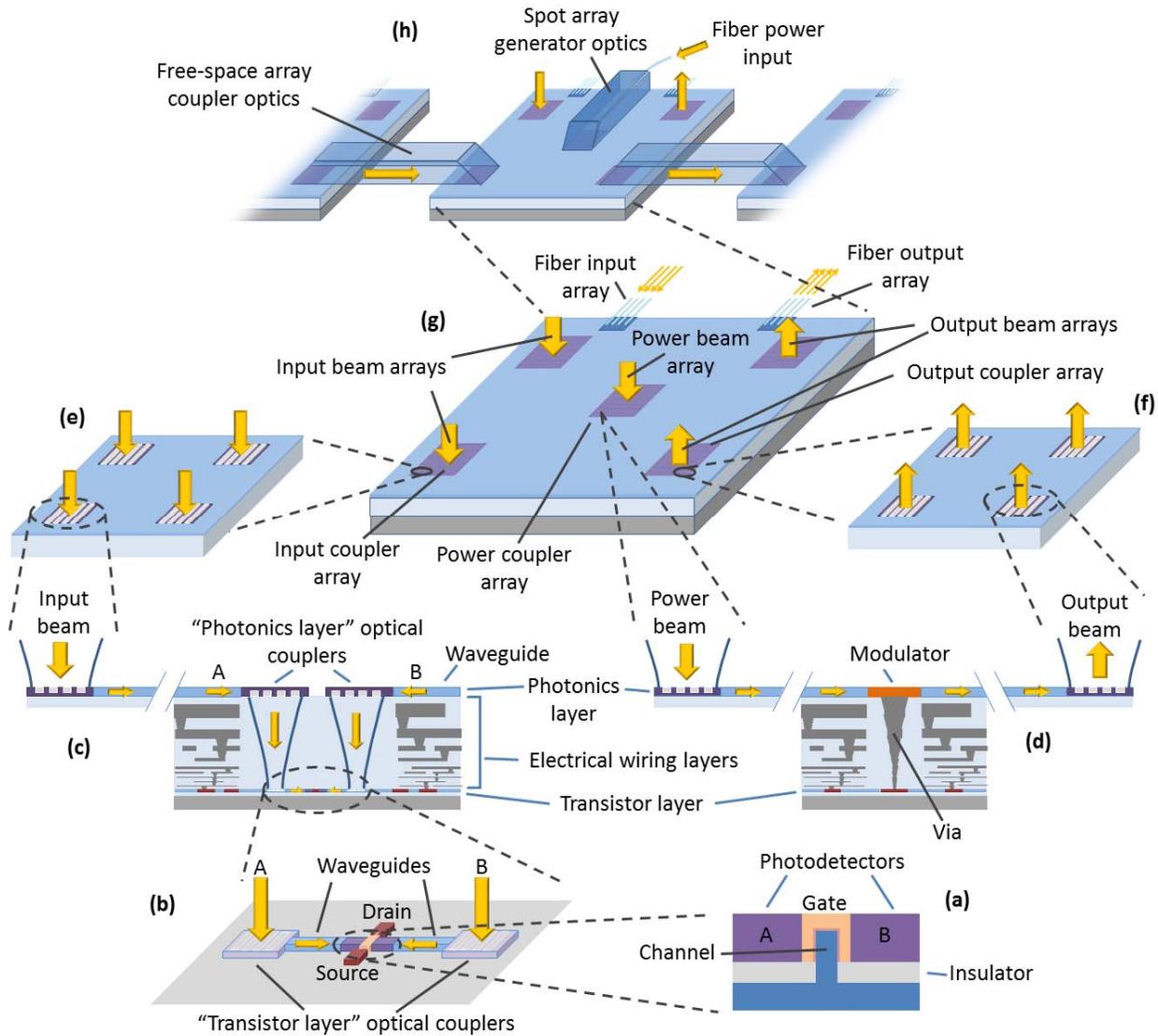

Fig. 11. Sketch of an optical platform for dense, low-energy interconnects, shown at multiple different length scales, from the transistors up to free-space arrays off a larger chip. (The figure is not to scale, especially for the size of the transistors, which would be relatively much smaller than depicted here.) (a) A pair of photodetectors is integrated beside the gate of the corresponding transistor input (here shown in the form or a FinFET structure). (b) A dual-rail optical beam pair *A* and *B* are connected though "transistor-layer" couplers and short (e.g., ~ 1μm) waveguides to the photodetectors. (c) A photonics layer (e.g., as in silicon photonics) sits on top of the electrical wiring layers of the chip. Here it contains couplers to couple the input beam pair A and B though waveguides in the photonics layer, to a pair of "photonics layer" optical couplers that here focus the light through transparent regions in the electrical wiring layer onto the "transistor layer" optical couplers. (d) Elsewhere on the chip, electrical "via" connections through the electrical wiring layer connect from output transistors to modulators that are in waveguides in the photonic layer. Optically, power is fed into the modulator waveguide from a power light beam through an input coupler, and an output coupler couples the resulting modulated power to an output beam. (e) and (f) show portions of input and output couplers and beam arrays. (g) shows a larger picture of the photonic layer on top of the entire chip. Here we envisage various 2 dimensional coupler arrays: input and array coupler and beam arrays; a power array coupler and beam array; and linear arrays of fiber inputs and outputs. (h) shows spot array generator optics fed by input power from some central optical power source through a fiber, and how multiple chips might be connected laterally and vertically using free-space connections. Such connections could include array coupler optics laterally between adjacent chips, as well as other array connections possibly vertically in and out from other modules or boards.

This is a moderately aggressive target, but not unreasonable as a stretch goal given our arguments above.

By use of some moderate optical confinement (e.g., 10) in the photodetectors and some photodetector material (e.g., germanium) operated at its direct gap, we presume these photodetectors have close to unit quantum efficiency (1 electron of current for each incident photon). Hence a received energy of ~ 100 aJ would be sufficient to swing a logic level at the transistor input, even without additional gain.

We now further presume the minimum required optical input energy per bit could be reduced to ~ 30 aJ with some avalanche, photoconductive or transistor amplifier gain, without substantial energy cost compared to the optical transmitter energies; so, we are taking a "near-receiverless" approach at the input.

As in Fig. 11, we presume that we have one or more free-space array units, each of, say, 1024 optical spatial channels, coming on or off each chip. We could configure these as 512 logical channels in a dual-rail approach. At 2 GHz clock rate, that would correspond to ~ 1 Tb/s data rate on or off the chip in such a unit. We also presume that the total optical system loss from the power source laser to the photodetector, including loss from finite modulator contrast, is 19dB (a factor of 80)[63] (see, e.g., [19]). Then the required optical energy per bit would be 30aJ×80 = 2.4 fJ. The total optical power for one such unit of 512 channels would therefore be 2.4fJ x 1Tb/s=2.4mW.

If we assume the laser source driving this system has a "wall-plug" efficiency of 30% (which is an aggressive target), then the total power to run the laser is 8 mW (or 8 fJ per bit). If we use optical modulators that themselves operate with energy < 1fJ/bit, which is already possible with quantum well modulators [41], [78], and we assume a similar electrical circuit energy per bit to drive the modulators, then we end up with a total system energy per bit < 10 fJ/bit, or < 10 mW to drive 1 Tb/s of interconnect.

Note that this hypothetical interconnect can drive connections over the entire modulo-synchronous volume at the same energy per bit. Hence the potential here is to reduce interconnect energies by ∼ 2 orders of magnitude or more compared to current approaches. The energy per bit here is so low that it would be energetically favorable to use it for longer on-chip interconnects, which may otherwise take 100's of fJ/bit (see Table I).

### B. Optical platform concept

Here, we sketch an optical platform approach that could provide the necessary bandwidths, connections and energies. See Fig. 11. This is very much in the spirit of a "straw man" proposal, i.e., one that is intended to generate discussion, rational criticism and comparison, and stimulate improved or alternative proposals. We will describe this progressively from the smallest, "transistor" level up to the largest meter-scale level and beyond.

In this example, we presume first that we have integrated photodetectors right on top of the transistors, in what we call the "transistor" layer (Fig. 11 (a) and (b)). We use dual-rail signaling, so these photodetectors are electrically "stacked" as in Fig. 10 (though they are physically side-by-side in the integration here), and we presume the center point of this stack directly drives the gate or gates of a CMOS stage (here depicted like a FinFET with a single "fin").

Since the detectors will require to be driven by two beams A and B that may need optical spacing of ∼ a micron or more just to separate them, the light from these beams is optically routed from couplers in the "transistor" layer through waveguides in the "transistor" layer to the two detectors, thereby avoiding the capacitance of electrical wiring (at ∼ 200 aF/μm) that would be required if we spaced the detectors themselves by microns.

Possibly the waveguides here are nanometallic or plasmonic, or some combined metal-dielectric guide. Possibly there is some optical resonance in the overall detector structure (e.g., Fabry-Perot, Mie or other shape resonance). Possibly the detectors use germanium or III-V materials integrated with an underlying silicon electronics platform.

Overall, the choice of integrating the detectors right with the transistors is made so as to eliminate high electrical receiver power dissipation. The goal is to achieve total capacitance at the input, including photodetectors, parasitics and transistor input capacitance ∼ 100 aF per detector while achieving reasonably efficient optical coupling into the detectors. The precise design of this integrated transistor/ photodetector/ waveguide/ coupler structure to achieve efficient coupling to the detectors with low parasitic capacitance is an interesting and substantial research challenge for nanotechnology and nanophotonics.

In this proposal, above the electrical wiring layers on the chip we add a photonics layer (see Fig. 11 (c), and (d)), such as a silicon photonics layer. This layer contains waveguides, optical couplers to the photodetectors, optical couplers to external beams (in free space or other guided wave structures like fibers), and optical output devices (modulators or lasers). Possibly this optical layer is hybrid-attached after separate fabrication on another temporary substrate.

This approach of putting an optical layer on top of the electrical wiring layers may allow some separation of the electrical and optical fabrications requirements. For example, optical waveguides work with lower loss with a relatively thick dielectric layer (e.g., microns) underneath the waveguides, but electronic processes typically do not use such thick layers. Additionally, it may be somewhat easier to manufacture sophisticated integrated photonic structures, such as those requiring advanced materials like quantum wells for modulator or laser structures, if we separate them from the electronic fabrication itself.

Functionally, putting this optical layer on top means we do not have to route optical waveguides in between wires inside the electrical wiring layers themselves; we only need to allow occasional transparent regions vertically in the wiring layers to pass light beams through to the detectors and/or local couplers and waveguides on the "transistor" layer. The use of this separate layer on the top also ensures that the entire area is available for optical waveguides, couplers, and output devices.

One disadvantage of putting the optical output devices in the photonic layer is that we will necessarily have some capacitance to connect to them. For a ∼ 5 μm vertical "via" through the electrical wiring layers, we should expect a capacitance of ∼ 1 fF. That may be tolerable for an output device that itself might have 1 fJ of operating energy anyway, though it would be undesirable for the photodetectors, which is why we have put them here on the transistor layer in this example; moving the photodetectors up to the photonics layer might make integration easier, at some cost (∼1 fF) in the input capacitance and required optical energies, though it would avoid the additional loss of optical coupling down to the transistor layer. For the output devices, the energy to charge and discharge this "via" capacitance is also not "magnified" substantially by the system, being essentially just an additive energy. (In contrast, increasing the required optical energy at the photodetector is likely essentially to scale up the entire energy of the system proportionately.) We are presuming here that the electrically-driven devices in the photonic layer are otherwise attached without additional substantial capacitance, however.

---

[63] Possibly we could presume less loss than this. This number, though, is one estimated for real systems in current research demonstrations [19].

One concept as shown in Fig 11 (e), (f) and (g) is that we would group input and output couplers in arrays. We could drive an entire chip optically with a single pulsed optical power source, for example delivered through a fiber from the central laser, as shown in Fig. 11 (h), and distributed to 1000's of power input couplers using Dammann grating spot array generator optics [163], here presumed miniaturized to a millimeter scale. This optical power would then provide the input power to modulators through waveguides (and could also be used for clocking inputs to the chip).

The modulators would be driven electrically as in Fig. 11 (d), and the optical output from those modulators would be fed through waveguides to an array of free-space output signal couplers. From modulator arrays on other chips we would have free-space arrays of input signal beams into the chip, which would eventually be coupled to the photodetectors as in Fig. 11 (a), (b) and (c).

Signal arrays could be fed from chip to chip using free-space optics, such as the "free-space channel couplers" in Fig. 11 (h), which could be plastic or glass channels (or possibly even mostly empty space), with appropriate mirrors. Possibly the optics would use lenslet arrays, as in Fig. 8, directly above the optical input and output couplers. The optics might use only the lenslet arrays, mirrors, and an imaging lens, or could also include additional imaging optics. The lenslet array can also likely be aligned very precisely using some planar alignment technique to micron or even sub-micron accuracy to the couplers, reducing positional alignment tolerances in the rest of the free-space optics. (See Appendix F for a discussion of the optical design of such free-space coupler optics.)

The free space connections do not need to go through solid channels, nor do they need to be only between adjacent chips. (Remember, too, that light beams in free space can pass right through one another, so crossing arrays of light beams pose no problem.) Also shown in Fig. 11 (e) are arrays of inputs and outputs for other free-space array connections, possibly to adjacent boards, for example. A silicon photonics platform is of course also capable of making fiber connections on and off the edge, which could be useful for making particularly long connections or connecting to external networks; we have sketched those also in Fig. 11 (g) and (h).

Note in an optical system like this that the optical loss in propagating from one device to another is essentially determined by coupler losses, not propagation loss. There is essentially no loss on propagating either through "free-space" or through optical fibers over the distances up to 10 m considered here.

There could be many reasons why we might consider fiber connections over the longer distances of meters inside system, and indeed we have presumed some fiber-based distribution of optical power here. But we should note also that free-space connections of thousands of channels can work over longer distances even up to 10's of meters if we choose to engineer them.

Actual free-space connections over meters pose no basic problems for optics (see Appendix F for calculations of numbers of channels as limited by diffraction in such longer connections). Conventional imaging optics routinely handle millions of resolution elements. We might need some autofocus and autoalignment approaches, but since those would be done on entire optical systems, the amortized cost of those per channel would be relatively small. Note again that consumer cameras routinely operate with many millions of pixels, and with both autofocus and image stabilization performed optomechanically in the optical system.

Note, incidentally, that with our system here hypothetically requiring 2.4 mW of optical power for every 1 Tb/s of interconnect, it is quite conceivable to run the interconnects for an entire large system from one centralized laser source. A 1 W source, such as a single semiconductor laser amplified by an erbium fiber amplifier, would provide enough power for over 200 chips, and support a total interconnect bandwidth of over 200 Tb/s – a bandwidth that, incidentally, is comparable to the entire long-distance internet bandwidth.

## X.  CONCLUSIONS

### A.  Using optics to reduce the energy for handling information

In this paper, we have argued that energy consumption and dissipation are the dominant limit on our ability to continue to scale information processing and communications; if we do not reduce the energy per bit processed and/or communicated, we will not be able to continue the exponential growth in the amount of information we consume.

We have next argued that most of that energy is in the communication of information, especially over the distances within an information processing or switching machine. We have seen that it is difficult to reduce that energy if we stay with purely electrical approaches.

Progressively, we have then argued that, because of the different physics of optical communications compared to that of electrical wires, optics can reduce that communications energy. This potential reduction comes in two forms.

First, we can avoid the charging and discharging of lines that leads to the majority of the dissipation in electrical connections at short distances; we propose to do this by substituting optical interconnects, which have no such dissipation, for essentially all off-chip interconnects (and possibly some connections on chips). The technical challenge then becomes one of reducing the energies required to run the optoelectronic devices themselves. That challenge leads us to the need for attojoule optoelectronics, both in photodetection and in optical output devices like lasers, LEDs and modulators.

If we can eliminate most of the detector capacitance, down to levels ~ 100 aF or smaller, with such attojoule optoelectronic approaches, and integrate them directly on electronics, then we can largely also eliminate the substantial power dissipation of receiver amplifier circuits; we would then move to operational modalities that we call "receiverless" (no electronic receiver amplifier) or "near-receiverless" (only simple low-energy receiver amplifiers). (The receiver energy this eliminates is currently of the order of 100's of fJ/bit or higher.)

Second, we can go on to propose the use of other features of optics, especially its abilities (i) to deliver very precise and predictable timing in volumes up to ~ 10 m in size and (ii) to offer very numbers of channels, especially in free-space connections. As a result, we can eliminate other high-dissipation electronic circuits normally associated with interconnect and data links – circuits that currently can dissipate picojoules or more per bit; specifically, we argue we can eliminate line coding, CDR and SERDES circuits entirely.

The net result of these eliminations of line charging and of most or all of the circuitry commonly associated with longer links is that we can propose that we could make essentially all links within a system look like short on-chip interconnects, up to and beyond entire cabinets of electronics, both functionally and in their energy use.

A stretch goal for such an approach is a total energy of ~ 10 fJ/bit communicated, and we have sketched a "straw-man" system that arguably could work towards such a goal. Note that such a goal, if achieved, would correspond to 10mW of total dissipation for each Tb/s of communication inside an entire system up to ~ 10 m in size. That energy per bit is therefore 2 to 3 orders of magnitude lower than current approaches at length scales from chip-to-chip interconnections or longer connections. Such an energy is even less than that of current electrical interconnects across a chip itself.

In the proposed "straw-man" approach, the optics can also operate at very high interconnect bandwidth densities. Particularly if we make the transition to free-space optics for some of the connections, we may be able to break the interconnect "byte-per-flop" limits that severely constrain architectures today.

With this example approach, we can see that we are substantially addressing all four goals originally set out in Section I for our attojoule optoelectronics interconnects.

### B.  Key research directions

This proposal is certainly speculative, but it is meant to be one that is physically realistic and could reasonably be engineered. It does not require the discovery of any new physical mechanisms beyond those we already understand and in materials we currently use. Indeed, part of our analysis shows that existing known mechanisms used in current devices and applications offer energies at least as low or lower than more exotic recent proposals such as 2D materials (see Appendix A). That is not to say we should not continue to explore novel material approaches, especially if they are somehow more convenient in operation or integration, for example, but we do not apparently fundamentally require either them or any other more fundamental breakthrough to meet the kinds of targets discussed here.

Note, incidentally, that we have focused here exclusively on the use of optics and electronics to reduce energy by solving problems of interconnects. We have not proposed optical or optoelectronic approaches to logic itself. We have addressed this point elsewhere [34], showing the challenges in such logic for any mainstream use[64]; arguably, the case for more optics in interconnects is much stronger.

There are, however, various areas of technological research that will be very important if we are to work towards realizing the goals we set out for interconnects.

### 1)  Nanoscale integration of photodetectors and electronics

Perhaps the most important direction and opportunity in nano technology required here is the intimate integration of photodetectors right beside or even on top of transistors (see, e.g., [169], [170]). Such an approach would seek to minimize capacitance, towards the range of 10's of attofarads, while also combining good optical coupling into the detector, possibly including nanoresonator structures and/or nanometallic or plasmonic elements.

We note that, once we reach "receiverless" or "near-receiverless" operation, the overall operating energy of the system can scale down, largely in proportion, as this input capacitance is reduced and the optical coupling efficiency is increased. For maximum benefit, this photodetector integration should be directly within the fabrication of the logic technology or "transistor" layer; moving it to higher layers of the fabrication adds the capacitance of the resulting longer electrical connections between the photodetector and the transistor.

### 2)  Low-loss mode coupling

The overall operating energy improves, essentially in proportion, as we reduce optical loss in the system. Most optical loss is in the couplers between one device or optical layer and another, not in the actual propagation of light within guides or free space. Optical coupling devices themselves generally are not lossy in the sense of having optical absorption. Rather, the losses could all be viewed as mode mismatch. Not all the incident light in its input mode (e.g., a free-space beam) is coupling into the output light in its output mode (e.g., a single-mode guide); the shape of the actual output beam does not match the shape of the desired output mode. (In this sense, all uncoupled or scattered light is merely light left in some other, undesired mode.) Such precise mode-conversion has been a problem in optics for some time.

---

[64] One major challenge is that nearly all such optical proposals do not meet even the qualitative requirements for logic devices and systems [34]. With the techniques discussed here, we could, however, make new lower-energy versions of previous functionally successful devices [64], and we could even argue that we could now make such functionally viable optoelectronic devices operate with possibly only hundreds of attojoules. But, at that point we would merely just be competitive with the transistor for logic operations. We would also have to create other technologies such as dense local optical wiring. Now, we could conceive of some solutions there, such as nanometallic concentration and waveguides. But, we would need very large numbers and very high yields

for all aspects such a technology if we were to supplant CMOS logic; this does not therefore seem a particularly promising direction with substantial and unequivocal benefit. We are not arguing against research here on truly novel ideas; the promise, too, of some fully quantum operations for some possible quantum computing systems certainly remains a worthwhile long term goal for fundamental research. But, we have argued here that we have a clear and convincing case now for advancing and exploiting low-energy optoelectronics to solve the problems of interconnects for all longer wires. Those problems have existed for some time, with no apparent path to better solutions other than a change to such optics.

Recently, however, there have been substantial advances in techniques to allow arbitrary design of optical nanostructures [128], [129], [130], [131], [132]; such design, together with nano-scale fabrication techniques could allow a new generation of low-loss couplers, in part because such nano-fabrication could allow the incorporation of the full design complexity needed to match precisely from one mode shape to another. Additionally, there are approaches to self-aligning couplers that could adjust themselves after fabrication [133], [134], [135].

Such low-loss coupling – from large beams to small beams, from free space to waveguides, from one guide to another – is both a critical requirement and a major opportunity for these emerging design opportunities. Since there are likely many such mode conversion interfaces in the whole optical path, the research target here for a coupler is to move from loss of a few decibels to loss of a few percent.

### 3) Free-space micro-array optics and systems

Free-space array optics would allow very high densities of connections in and out of chips and modules, solving the bandwidth bottleneck, and enable us to save energy by eliminating much of the electronic circuitry of current links. Compact, dense, self-aligning, free-space systems are now quite feasible, and a broad range of micro-optical technology exists. Following on previous successful laboratory demonstrations of free-space digital systems, the research goal now would be to generate technology for arrays of 1000's to 10,000's of beams (i) with millimeter cross-sections and centimeter lengths for on-board or on-module connections, possibly in rigid and manufacturable structures and (ii) in self-aligning free-space array optics for board-to-board or even cabinet-to-cabinet connections.

### 4) Extending integrated optics technologies

We need to be able to make large numbers of optical devices, such as waveguides and beam couplers, ideally integrated with active optical devices, such as photodetectors, modulators, lasers and LEDs. An integration platform like silicon photonics [47], [48], [49], [50], [51], [52], [53], [54], [55], [56], [57], [58], [59], [60] gives a good basis, allowing large numbers of optical components and complex optical circuits.

A key research direction will involve augmenting such a platform with monolithic or heterogeneous integration of other materials or structures so we can reach energy and performance targets especially for output devices. Such additions could include

- III-V materials
- quantum well or other quantum-confined structures [171], [172] in III-Vs or germanium [41], [78]
- integration of materials other than silicon, either in monolithic form, including novel nanoscale integration approaches that can avoid problems with lattice mismatch [117], [119], [124], [173], [174], [175], or using heterogeneous integration of III-V device structures [50], [171], [172], [176], [177] or other materials such as organics [87], [109]
- technology for electrically connecting such optoelectronics onto (or into) electronics with negligible additional capacitance, such as some direct-bonding technique right on top of the chip wiring layer
- micro- and nano-mechanical technologies for tuning and adjustment of optical devices and circuits.

### 5) Low-energy output devices and their integration

Devices exploiting the relatively weak optical modulation mechanisms available in silicon have been engineered to a remarkable degree and their feasibility and challenges for systems have been deeply analyzed (see, e.g., [105]). Other microscopic mechanisms are much stronger, as we have discussed. For example, a hypothetical QCSE electroabsorption modulator using germanium [41], [78] or III-V quantum wells with a $(300 nm)^3$ active volume could be an attractive approach. There are also many promising directions such as nanoneedle and nanocavity growth on silicon for lasers [54], [174] and LEDs (as well as photodetectors) [117], [119], [124], [174], [175] that could address integration issues.

The research goal here should be to exploit the stronger microscopic physics of such effects to achieve a sub-femtojoule device working over the entire C-band while eliminating the need for any post-fabrication trimming or active temperature stabilization. Any new device approach here would, however, have to have some credible path by which an integrated system could be made, with very large numbers of devices at high yields.

## C. Final conclusions

We have taken a broad view here of the motivations and technological opportunities, from environmental limits on information processing and computing through to fundamental optics and quantum mechanical mechanisms, for using optics and optoelectronics to reduce energy in handling information. As we said earlier, this article cannot be a deep review of any topic; its main goal is rather to clarify research directions, questions, and opportunities.

We have considered novel and even radical approaches to complete systems; having such a complete system proposal is important because it enforces an intellectual honesty on our optimistic conclusions of real benefit – we cannot just push show-stopping difficulties "under the rug" in the hope that someone else will deal with them. Though we have proposed an entire platform example here, from the transistor level up to long fiber connections, it is just that – an "existence proof" example. There may be many other valid approaches.

Though we have identified many technological challenges that would need to be addressed to realize the full benefits envisaged here, a solution to any one of these challenges, such as better integration, lower energy devices, or lower loss coupling, will be useful on its own. Complete success in all aspects at once is not necessary for useful progress.

Overall, our conclusion here is strongly optimistic: optics offers real opportunities for substantial reduction in energy and improvements in performance in systems that handle information, and these opportunities should stimulate many exciting and worthwhile research and technology directions in optics and optoelectronics. Indeed, without optics, we may have no other solutions to eliminating much of the energy we use to handle information.

In this Appendix, we will give a more detailed discussion and comparison of the energy requirements of various mechanisms currently understood for making optical modulators that could operate at GHz or higher rates.

Such mechanisms fall broadly into two categories: those that work by electrically-induced changes in optical absorption (i.e., electroabsorption), and those that work by electrically-induced changes in refractive index (i.e., electrorefraction).

### D. Electroabsorption mechanisms and approaches

There are two main categories of electroabsorption mechanism: (1) absorption changes as a direct result of electric field in the material, and (2) absorption changes resulting from electrical control of carrier (i.e., electron and/or hole) density in the material.

#### 1) Electric field mechanisms

A set of related mechanisms are found for electroabsorption with photon energies near the direct band-gap energy of a semiconductor, usually exploited for photon energies in the region just below the that nominal band-gap energy (so at wavelengths longer than the bandgap wavelength). These are (i) the Franz-Keldysh effect (FKE) [178], [179], [180], [181], [182], [183], (ii) exciton broadening (bulk excitonic electroabsorption) [66], [181], [182], and (iii) the quantum-confined Stark effect (QCSE) [65], [66].

The FKE and exciton broadening are seen in bulk semiconductors. Exciton broadening is also seen in quantum well layered structures for applied electric fields parallel to the layers [66], and the QCSE is observed for applied electric fields perpendicular to the quantum well layers [65], [66]. The QCSE is also present in quantum wires and quantum dots when the field is applied along one of the confinement directions [183]. Fig. 12 compares experimental data for germanium[65] bulk FKE and for germanium quantum-well QCSE.

Incidentally, it is not necessary that the lowest bandgap energy in a semiconductor is the direct gap in order to see such electroabsorption mechanism. These electroabsorption effects can be seen at the direct gap even in materials that are themselves indirect. A good example here is germanium, which shows all these electroabsorptive effects at its direct gap energy in appropriate structures [41], [67], [68], [69], [70], [71], [72], [73], [74], [75], [76], [77], [78], [79], [80], [81], [82], [83], [185], as in Fig. 12.

Optical absorption across the direct gap in semiconductors is described in the simplest model as being between plane-wave "Bloch" states for electrons in the valence and the conduction band; this is a "non-excitonic" model. Though it is simple, and does describe some features, it is both qualitatively incorrect – it does not actually predict the spectral shape of the absorption – and quantitatively quite inaccurate – it substantially underestimates the strength of that absorption.

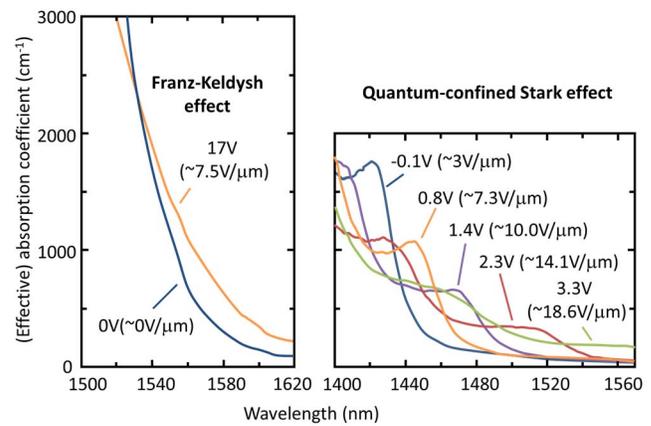

Fig. 12. Comparison of FKE (data after [184]) and QCSE (data after [82]) electroabsorption. The FKE data is taken in a SiGe diode structure with a ~2µm thick depletion region with ~ 0.6% fractional Si content (so technically a $Si_{0.006}Ge_{0.994}$ alloy) so as to shift the absorption edge slightly to shorter wavelengths from that of pure Ge. The QCSE data is taken in a Ge/SiGe heterostructure diode with a ~220nm thick depletion region containing 5 Ge quantum wells, each ~ 14nm thick, with 18nm $Si_{0.19}Ge_{0.81}$ barriers between them. The effective absorption coefficient for the QCSE is calculated using the total thickness of the wells plus barriers for the effective optical thickness of the structure, so an effective absorption coefficient of 314cm$^{-1}$ in the figure is equivalent to ~0.1% probability of a photon being absorbed as it tries to pass through one quantum well from one side to the other. For the QCSE diode, fields are calculated from voltages by adding on a built-in field equivalent to 0.8V across the 220nm-thick depletion region; this built-in field, which would correspond to that in a homostructure diode with a ~0.8 eV bandgap energy (like the direct bandgap of Ge), is only an estimate because this is a heterostructure diode that contains contact regions with direct bandgaps larger than this, but at the same time there are lower, indirect gaps present from the Ge materials and possibly in the contact regions also.

What is missing from this "plane wave" approach is that the actual final state is that of an electron-hole pair; because of their Coulomb (electrostatic) attraction, they are much more likely to be in the same place than is estimated based on the "plane wave" approach. In this electron-hole pair model, the probability that we will absorb a photon to create a pair in a given state is proportional to the probability that the electron and hole will be found in the same unit cell in the resulting state [186], [187]. There are both bound states ("excitons") of these electron-hole pairs that appear just below the bandgap energy [186] as strong absorption lines, and also so-called "Sommerfeld" enhancement of the absorption above the bandgap energy (see, e.g., [188] for expressions for both aspects for 2D and 3D cases).

In many bulk semiconductors at room temperature, the exciton absorption peaks associated with the bound states are already so broadened by lifetime effects (such as ionization by optical phonons [189]) that they are often not clearly resolved at room temperature; the excitonic effects are, however, still strongly affecting the shape and strength of the optical absorption spectrum. When we quantum-confine electrons and holes in semiconductors at sizes comparable to or smaller than the size of the lowest-energy ("1s") exciton – so ~ 10 nm, for example – in one or more dimensions, we increase the probability of finding the electron and hole in the same place[66].

---

[65] Technically, this data is for a $Si_{0.006}Ge_{0.994}$ alloy.

[66] Somewhat surprisingly, confining in that one direction also leads to the exciton being smaller in the other two directions (see, e.g., the analysis by [188] for the 2D case), which further enhances excitonic effects.

As a result, excitonic effects are enhanced in such quantum-confined structures, often allowing the associated peaks to be clearly resolved at room temperature even when they are barely resolved in the equivalent bulk material [187], [189], [190]. Hence enhanced excitonic effects in optical absorption are a particularly important consequence and benefit of quantum confinement in nanostructures.

### a)   Franz-Keldysh and exciton broadening electroabsorption

If we neglect the excitonic effects for the moment and consider the effects of electric fields on the absorption near the direct bandgap energy in bulk semiconductors, then we calculate the FKE [178], [179], [180], [183], which leads to a "tail" on the absorption that extends into the bandgap region.

The electroabsorption very near to the direct bandgap energy in bulk semiconductor materials can be dominated by another effect – what we are calling exciton broadening electroabsorption; this is the lifetime broadening of the excitonic absorption lines resulting from the field-ionization of the bound excitonic states in the electric field [181], [182]. Nonetheless, the qualitative effect is similar once we are significantly below the energy of the main exciton absorption peak, with the appearance of an electrically-controllable absorption tail that extends smoothly below the bandgap energy to longer wavelengths.

Though the exciton broadening electroabsorption is quite sensitive to field in the region very near to the (main) exciton absorption peak, it is likely not usable there because it does not have enough absorption coefficient contrast, as required in criterion (5) above, so this general category of electroabsorption effects in bulk semiconductors near the bandgap energy is only usable at energies moderately below the bandgap energy where the "zero-field" background absorption is small (and where the mechanism is typically described and modelled as being the FKE even if there may be some excitonic broadening effects also present). This mechanism is exploited successfully for optical modulators (see, e.g., [185] for a recent example).

### b)   Quantum-confined Stark effect

In a quantum well structure, such as a ~ 10 nm thick layer of a narrower bandgap semiconductor sandwiched between layers of wider bandgap semiconductors, the allowed states in the quantum-confinement direction (the direction perpendicular to the layers) become quantized. If we neglect excitonic effects for the moment, the absorption between the resulting valence and conduction sub-bands would lead to an absorption spectrum that is a set of "steps" [187]. Excitonic effects are also strong in such quantum wells, however, and as a result we can clearly see strong excitonic peaks associated with each such step, even at room temperature.

If we apply an electric field in the direction perpendicular to the layers, we can shift the energies of the confined states inside the well, leading to energy shifts of the sub-bands, which in turn would lead to shifts in the "steps". In particular, the lowest "step" would move to lower photon energies. Fig. 13 shows the quantum mechanics behind the majority of the shift in the absorption edge in an example calculation. With applied field, the lowest electron and highest hole confined states (which are the edges of the sub-bands) in the well move towards one another in energy, thereby moving the lowest absorption step to lower photon energies. We see in this typical quantum well that the separation between these states changes from 819meV ($\cong$ 1514nm wavelength) at zero field to 799meV ($\cong$ 1552nm wavelength) for $10^5$V/cm (10V/μm) applied perpendicular to this 12nm thick layer, which would shift the absorption edge to lower energies by 20meV ($\cong$ 38nm change in wavelength).

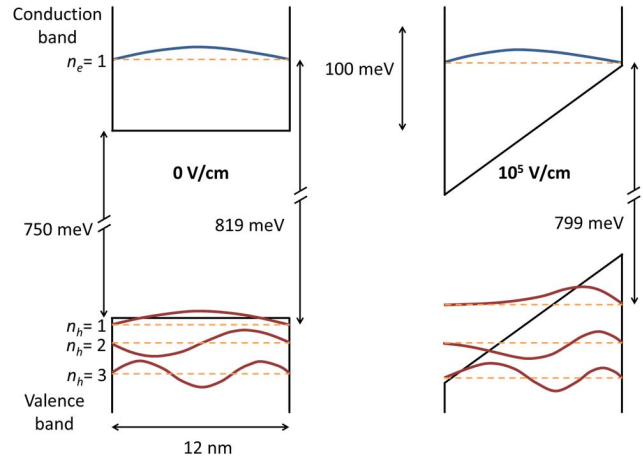

Fig. 13. Calculations of the electron (conduction band) and (heavy) hole (valence band) energies and wavefunctions for the edges of the sub-bands, numbered $n_e$ ($n_h$) for the conduction (valence) sub-bands, in a 12nm-thick In$_{0.47}$Ga$_{0.53}$As quantum well (the composition that lattice matches to InP). These calculations use the simplifying approximation of infinitely high potential barriers on either side, using the analytic model for "tilted" potential wells [84], at 0V/cm and at $10^5$V/cm (10V/μm). The bandgap energy of the unconfined material is taken as 750 meV, with effective masses of $0.041m_o$ (electron) and $0.46m_o$ (heavy hole), where $m_o$ is the free electron mass.

We see also that the wavefunctions for the electron (i.e., in the conduction band) and the hole (i.e., in the valence band) are distorted by the applied field. For these closest electron and hole levels, this distortion reduces the "overlap" integral between the wavefunctions, which lead to some loss in the corresponding "height" of the absorption step with field. This kind of behavior is clear in the QCSE spectra of Fig. 12 for the longest wavelength "step" in the absorption. In our discussion of the QCSE so far, we have neglected[67] excitonic effects. A key additional point, however, is that, unlike the behavior with bulk materials, the excitons are not rapidly field-ionized even for strong fields applied perpendicular to the layers; that is because the walls of the quantum well hold the exciton together. Hence, we see clear shifting of the absorption steps while retaining strong and relatively sharp excitonic peaks, which is the mechanism known as the quantum-confined Stark effect

---

[67] Formally, if we neglect the excitonic effects in the QCSE model, then the resulting behavior is essentially the quantum-confined version of the (non-excitonic) FKE mechanism. We can show that the electroabsorption spectrum

of the (non-excitonic) quantum well electroabsorption would tend towards the (non-excitonic) FKE spectrum as we increased the width of the layer [194].

(QCSE) [65], [66][68]. These excitonic peaks are visible, for example, in the QCSE data of Fig. 12, where they are seen as the slight peaks in the various spectra (e.g., near 1420 nm in the QCSE spectrum at -0.1V).

This mechanism can equivalently be regarded as a giant Stark shift of the exciton, and is formally equivalent to the electric field shift of the ground state of a hydrogen atom if we were able to confine it between two "walls" less than ~ 1Å (0.1nm) and apply an electric field of ~ 10 – 100 V/Å.

From a practical point of view, the QCSE offers an electroabsorption in which we can shift a relatively abrupt and strong (e.g., 100's to 1000's of cm$^{-1}$) absorption by large amounts (e.g., even as much as ~ 100 meV). The required electric fields are in the range of $10^4 - 10^5$ V/cm $(1 - 10$ V/$\mu$m), which can be applied using reverse-biased diode structures, for example.

Comparing the QCSE to the FKE in similar materials, as in Fig. 12, we see first that both effects are capable of producing absorption coefficient changes ~ 100cm$^{-1}$ to nearly ~ 1000cm$^{-1}$ for photon energies in the region just below the bandgap energy (wavelengths longer than the bandgap wavelength). With the QCSE it is easier to get large contrasts in the absorption coefficient between the "on" and "off" states, which is an important criterion for devices. The abruptness of the QCSE absorption edge means that, unlike the case of the FKE, the device can be tuned by biasing so that the absorption edge is shifted close to the operating wavelength, and then the device can be operated by applying a small additional bias to shift the absorption edge just past the operating wavelength.

This level of electroabsorption can be exploited in waveguide structure that, for the case of the QCSE, can be similar to those used for semiconductor lasers; indeed, QCSE modulators are widely used today in optical telecommunications, where they are often integrated with semiconductor lasers.

The QCSE electroabsorption effects are also large enough to give strong modulation of light in micron-thick structures, allowing modulators that can operate directly on light propagating perpendicular to the surface either in (e.g., [75], [79]) or without resonators, or enabling particularly compact low-energy waveguide modulators. Indeed, a short (10 μm long) waveguide QCSE modulator, without any resonant cavity, has already shown sub-femtojoule operation [41], [78]. Such devices can be run with the ~ 1V drive swing readily available from CMOS electronics [41], [78], [79], and test structures show the potential for total voltage ~ 1V [82].

The QCSE may represent the strongest and most energy-efficient high-speed optical modulation mechanism available. Physics experiments confirm its operation to picosecond time scales [69], and the speed limit is likely sub-picosecond [191].

QCSE modulators can exploit other forms of quantum well structures, such as coupled wells [192], [193], which can offer some improved electroabsorption in specific cases [193]. Whether such coupled wells offer substantial benefits can depend on the abruptness of the optical absorption edge since their strongest effects correspond to "clearing out" a region of absorption as the coupling is turned on and off with field; if the edge is not abrupt, then the "cleared out" region may not have sufficient absorption contrast.

Such electric-field electroabsorption devices can have some temperature dependence because, like lasers, the bandgap energy does move with temperature, generally by ~ 0.3 – 0.5 meV/K. In the case of QCSE modulators, this may be less of a problem because the modulator can be voltage-tuned to compensate for temperature variations, and if we operate at high field, the absorption change may be sufficiently broad in wavelength range that no temperature compensation is necessary. For example, the QCSE electroabsorption in germanium quantum wells on silicon can be voltage tuned to work with good absorption coefficient contrast over ~ 125 nm of wavelength range [82], which corresponds to ~ 150 K temperature range given a measured 0.46 meV/K (0.84 nm/K in wavelength shift) [68]. Even with somewhat lower applied fields, as in [79], allowing ~ 500cm$^{-1}$ absorption change at high absorption contrast over ~60 nm wavelength range would be sufficient for a 70 K operating temperature range.

Because such modulators can run well even when hot (e.g., 100 °C [69]), such modulators can also be temperature tuned by heating, which is generally easier to achieve and more energy-efficient than cooling.

### 2)   Carrier density mechanisms

#### a)      Free-carrier plasma

For photon energies far below the bandgap in direct gap materials or for indirect materials with populations only in the "indirect" valleys, there are absorptive and refractive effects associated with "free" carrier densities $N_e$ and $N_h$ (conventionally given in units of "per cubic centimeter" – cm$^{-3}$) in the conduction and valence bands respectively. For silicon, the absorption coefficient for such "free-carrier plasmas" at an example (free-space) wavelength of 1.55 μm is given approximately by [106]

$$\alpha_{fc} \simeq 8.5 \times 10^{-18} N_e + 6.0 \times 10^{-18} N_h \qquad (9)$$

So, a carrier density of $10^{18}$ cm$^{-3}$ leads to absorption coefficients of ~ 10 cm$^{-1}$.

Such models are approximately justifiable from a Drude free-carrier plasma approach, though the situation with holes is more complicated, in part because of absorption between different valence bands.

For a typical III-V material, free-carrier absorption associated with holes is thought to dominate, for example in the operation of lasers [195], and the hole absorption numbers are comparable to those in silicon (e.g., 13 cm$^{-1}$ at $10^{18}$ cm$^{-1}$ in InGaAsP near its bandgap wavelength).

Such absorption coefficients are essentially too small to be attractive for compact electroabsorption modulators, but are large enough to be a nuisance in giving background absorption in high-Q structures. The change in absorption with carrier density can also influence the behavior of refractive modulators based on high-Q structures.

---

[68] There is a small additional shift of the binding energy of the exciton itself, though this is relatively small compared to the shifts of the electron and hole "single-particle" levels [66], [68].

### b)  Band-filling mechanisms

As we add electrons or holes to a semiconductor, in one simple (non-excitonic) view, we can start to fill up the bands, and that band filling blocks the possibility of further absorption into the states that are already occupied, as sketched in Fig. 14. In direct bandgap materials, such as many III-V semiconductors, the electron effective mass can be quite small, and hence the density of available electron states per unit energy is also quite small. As a result, with moderately large densities of carriers, such as $\sim 10^{18}$ cm$^{-3}$, a "pool" of electrons collects in the bottom of the conduction band, effectively blocking absorption over a substantial spectral range.

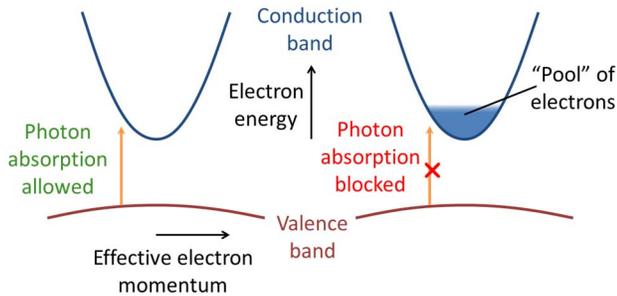

Fig. 14. On the left we see a simple picture of valence and conduction bands in a direct-gap semiconductor. On the left, the valence band is full of electrons, and the conduction band is empty. A photon of energy just above the bandgap energy can be absorbed to take an electron from the valence band to the conduction band. If, however, we add a large number of electrons to the conduction band, they will fill the lowest states in a kind of "pool" of electrons that collects at the bottom of the conduction band. Now the absorption of the photon is blocked because the final state for the electron is already occupied.

The detailed physics of such mechanisms is somewhat more complicated than this non-excitonic description suggests. The presence of free carriers also effectively screens the interaction between the electron and hole in excitons, so there is an additional benefit from the disappearance of the excitonic peak from such screening. There is also some bandgap renormalization – a shrinkage of the bandgap with increasing carrier density – that partly counteracts the band filling. See, e.g., [196] for a discussion of such mechanisms. A general term to cover the resulting changes in absorption spectrum is "phase-space absorption quenching" [197], though the more informal and less accurate "band-filling" is more common. Band filling is also sometimes called Pauli blocking or Burstein-Moss shift.

A good example of band filling is given by a quantum well in a field effect transistor structure [196], [197]. Relatively complete quenching of the absorption with $\alpha_{abs} / \alpha_{trans} \geq 2$ is possible over at least 40 meV in photon energy range near 0.8 eV at room temperature [196] with sheet carrier concentrations just under $10^{12}$ cm$^{-2}$ in a 10 nm thick quantum well (an effective volume density therefore just under $10^{18}$ cm$^{-3}$). Such quenching corresponds to $\sim 1\%$ change in the transmission of light through a single quantum well [196], [197]. If we presume it takes $\sim 1$ eV of energy for each added electron in the structure (consistent with bias voltages $\sim 1$ V), then the energy density to run such a device is comparable to that of the light emitter device in Table I (e.g., 160 fJ/(μm)$^3$. The absorption changes here are somewhat larger than in the QCSE, so the devices could be somewhat smaller even than the QCSE devices. Hence, devices made using this mechanism would lie somewhere between the light-emitter and modulator numbers in Table I.

Optical modulators based on band-filling in graphene have been proposed – see, e.g., [198], [199], [200], [201]. Based on current understanding, however, the required operating energies for these would be considerably higher than those for quantum wells, for example. We will discuss the comparison of 2D materials and quantum wells below.

### E.  Electrorefraction mechanisms and approaches

There are several different mechanisms for changing the refractive index of a material under some kind of electrical control. We will consider two basic categories: (i) electric-field mechanisms that work as a result of some microscopic polarization of the electron wavefunctions; and (ii) band-filling mechanisms that work as a result of the change of carrier (electron and/or hole) density in the material. There are other ways of changing refractive index, such as heating (which is quite a useful mechanism in silicon photonics for tuning and slow switching), molecular reorientation (as in liquid crystals), and change of physical state (as in phase change materials like GST), but we will not consider these further here, mostly because they will not generally be fast enough for modulating interconnect or communications signals.

All these refractive mechanisms can be understood through the Kramers-Kronig relations (see, e.g., [202] for a classical discussion and [84] for the relation to quantum mechanical approaches) as resulting from changes in the optical absorption spectrum. Indeed, these relations show that any change in absorption at any wavelength will in general lead to changes in refractive index at all other wavelengths (and *vice versa*).

Classic electrorefraction mechanisms such as the Pockels effect and the Kerr effect are not usually described in terms of absorption changes, in part because these mechanisms are typically employed in a spectral region far from the wavelengths where any absorption changes are taking place (the absorption changes may be at very short wavelengths). These mechanisms, as a result, are in practice generally not resonant and vary little with wavelength.

One mechanism associated with free carriers is a result of the plasmon absorption peak that results from free carrier densities; in semiconductors at normal carrier densities, that plasmon absorption is at long, far infrared wavelengths, and a direct calculation using a Drude model for the plasma behavior can be useful, at least for electrons.

Other mechanisms like refractive index changes from band filling and from electroabsorption near to the bandgap energy are generally best understood and calculated working directly from the known changes in absorption spectrum near the band gap energy and using the Kramers-Kronig relations explicitly.

When we are working at photon energies or wavelengths close to where the major absorption changes are occurring, such Kramers-Kronig calculations will typically show changes in the real and imaginary parts of the dielectric constant that are of comparable magnitude. (The real part is responsible for refractive effects and the imaginary part for absorptive effects). See, for example, the dielectric constant or susceptibility near a typical atomic absorption line to understand this point [84]. But the absorption itself in such regions is usually too large to make much direct use of such large refractive changes because of our criterion (6). Hence we typically need to move to a spectral region where the absorption and/or induced absorption are lower, which means we also get lower refractive index changes. As a result, for a given such resonant mechanism, refractive devices tend to have to be longer, and hence have lower energy efficiency, than the corresponding absorptive devices.

### 1) Electric field mechanisms – Pockels effect

The Pockels effect is a linear change of refractive index with applied electric field, and is an example of a second-order nonlinear optical effect (sometimes described in terms of a coefficient $\chi^{(2)}$). Since the sign of the refractive index change would obviously therefore be reversed if we reversed the direction of the electric field, any material that shows a Pockels effect must look different in two opposite directions. A classic Pockels effect material like lithium niobate, which has a strong Pockels effect, has such a property, and lithium niobate modulators are extensively used in telecommunications. III-V materials like GaAs have potentially usable Pockels effect for electric fields in certain directions. Silicon, however, because it does not have the right symmetry properties, does not show a Pockels effect.

If we strongly strain silicon, such as by depositing layers of a material like SiN on it under appropriate conditions, it then acquires the necessary asymmetry. Such strained silicon [203], [204] can show refractive index changes up to $\Delta n \sim 3.5 \times 10^{-5}$ with effective applied electric fields $\sim 5 \times 10^3$ V/cm . The corresponding effective electrooptic coefficient $r_{33} \simeq 2.2$ pm/V can be comparable to that of III-V materials, though it is about an order of magnitude smaller than that in lithium niobate, which has $r_{33} \simeq 33$ pm/V [204]. One other current approach is to try to hybridize lithium niobate on silicon [205], [206] for such electrorefractive modulators.

Organic materials can have larger electro-optic coefficients of $r_{33} \simeq 170$ pm/V [207], and they can be successfully exploited to demonstrate relatively low energies in optical modulators [87], [88], [109]; for these demonstrations, using a plasmonic waveguide with a 90 nm gap (and electrode spacing) and exploiting additional field concentration effects from the slow group velocity in the guide, this work shows a 10 μm long device with ±3V drive, on an estimated capacitance of 2.8 fF, for an energy of ∼ 25 fJ/bit.

One interesting point about Pockels effect devices is that, in principle, there is no specific minimum energy required to run them, even without resonators. To understand this, suppose we decide to double the length of some Pockels effect device; in that case, we can get the same path length change with half the electric field. But that means we only need ¼ as much electrostatic energy density, and hence half the energy overall. Equivalently, we may have doubled the capacitance $C$ by doubling the length, but we have halved the voltage $V$, hence halving the resulting $(1/2)CV^2$ operating energy. In some waveguide design, there is no specific limit to how low the energy can go if we can make the waveguide arbitrarily long. In practice, however, Pockels effects are sufficiently weak that the length of the waveguide is set by other practical considerations, such as waveguide loss or other practical limits on length, and devices like that of [88] may well represent the limits of low-energy operation for known materials in devices without resonators.

It is possible to make asymmetric quantum well structures (e.g., [193]), which would technically give Pockels effects in refractive index, though it is possibly simpler just to regard those as variants of the QCSE electrorefraction.

### 2) Electric field mechanisms – Kerr and QCSE

The Kerr effect is a quadratic variation of refractive index with electric field, and is technically a third-order nonlinear optical effect (sometimes described in terms of a coefficient $\chi^{(3)}$). No particular material symmetry is required for the Kerr effect, and it will exist in principle in essentially any material. Because it is third-order, however, at least for non-resonant mechanisms, it is generally weak, and therefore not of great interest for low-energy modulators in conventional materials.

The electroabsorption mechanisms discussed above all have electrorefraction associated with them, and that electrorefraction can be calculated quite effectively based on the Kramers-Kronig relations, usually from empirical absorption spectra. If the change in the absorption coefficient spectrum when we apply the field is $\Delta \alpha(\omega)$, then in practice, we can deduce the change $\Delta n(\omega)$ in refractive index at some (angular) frequency $\omega = 2\pi f$ (where $f$ is the conventional frequency in cycles per second) using the integral [189]

$$\Delta n(\omega) = P \int_0^\infty \frac{\Delta \alpha(\omega')}{\omega'^2 - \omega^2} d\omega' \qquad (10)$$

The "$P$" here means to take the principal value, which means technically we have to avoid the singularity at $\omega = \omega'$. The integrand just on the two sides of the singularity will actually cancel out so there is no actual divergence in the resulting integral[69].

Writing $\Delta \omega = \omega' - \omega$, we can rewrite the resonant denominator as $\omega'^2 - \omega^2 = (\omega' + \omega)\Delta \omega$. Hence a change in absorption at one frequency $\omega'$ gives rise to a change in refractive index that falls off approximately as $\Delta \omega$ as we move away in frequency.

For the resulting refractive index changes induced by the QCSE, see, for example, the calculations in [79], [208]). In the vicinity of the exciton resonance itself, the resulting index changes are quite large, in the range of $\Delta n \sim 0.01$ to $0.04$ [79], [208]. In that region, however, it is difficult to satisfy the criterion (6) for refractive devices because the absorption is too high. It is worth noting, however, that a "hybrid" resonator modulator using both electroabsorptive effects combined with simultaneous electrorefractive shifts of the cavity resonance can be quite effective in this region, with the electrorefractive effects significantly improving the performance of the modulator [79].

---

[69] One practical way to handle this numerically is to add a small positive quantity $\delta$ to the denominator when performing the integral in Eq. (10), decreasing the value of $\delta$ until it makes no further significant difference to the result in some wavelength range of interest.

If we want to make a more purely electrorefractive modulator, we need to move to photon energies somewhat below the bandgap energy where the background absorption is smaller [79], [208]. This is quite a viable strategy for electrorefractive devices based on the QCSE, which then are quite competitive with, say, lithium niobate approaches [209], [210]. [209] shows a switching device operating with a 675 μm long active region and 2.5 V drive swing, in a device operating with photon energies significantly below the bandgap energy, and made to satisfy the additional design constraint of polarization-insensitive operation.

The data and calculations of [79] suggest a non-resonator electrorefractive device with germanium quantum wells might be possible at 1.55 μm wavelength, with a background absorption of $\sim 30 \, cm^{-1}$ [77] and an index change $\Delta n \sim 2.3 \times 10^{-3}$ (satisfying condition (6)) at an operating field $\sim 10^5$ V/cm and a length $\sim 330$ μm. (Note, incidentally, that the indirect absorption tail in germanium [77] is generally not strong enough to preclude such refractive modulators.) In a hypothetical $200 \, nm \times 300 \, nm$ waveguide, the operating energy would be $\sim 100$ fJ with a drive voltage swing of $\sim 2$ V. Since there is no optical field concentration in such a hypothetical device, we can see that the basic energy requirements of such QCSE electrorefractive mechanisms are comparable to the lowest-energy demonstrated Pockels-effect devices [87], [88], [109], which have significant optical field concentration from nanometallic waveguides and group velocity effects.

### 3) Carrier density mechanisms

#### a) Free-carrier plasma

The refractive index change in silicon from the presence of free carriers at an example (free-space) wavelength of 1.55 μm is given by [106]

$$\Delta n_{fc} \simeq -\left[ 8.8 \times 10^{-22} N_e + 8.5 \times 10^{-18} \left( N_h \right)^{0.8} \right] \quad (11)$$

For a representative carrier concentration of $10^{18} \, cm^{-3}$ electrons or holes, which corresponds to moderately strong doping or carrier injection, we would have changes of refractive index of $\sim -8.8 \times 10^{-4}$ for electrons and $\sim -2.1 \times 10^{-3}$ for holes.

Devices based on this mechanism have been extensively researched (see, e.g., the reviews of silicon optical modulators [211] and of silicon photonics generally [49]). Simple Mach-Zehnder devices based on this approach tend to have energies in the low picojoule range [104]. In addition to the use of high-Q resonators, such as rings, other approaches, such as photonic crystal waveguides, can allow the devices to be shortened, reducing operating energies.

The lowest energies demonstrated may be in microdisk resonators [105], a version of the ring-resonator approach. With a Q-factor of $\sim 10,000$, $\sim 1$ fJ/bit can be obtained in a 4.8 μm diameter device. Some degree of tuning is possible without use of thermal mechanisms, and system-level choice of devices at run-time can avoid other thermal tuning at some cost of electronic energy dissipation. Overall, with additional feedback and control electronics, $\sim 10$ fJ/bit operating energy is projected for such a device used in conjunction with $\sim 10$ nm CMOS electronics; this energy is dominated by the monitor receiver energy.

#### b) Band-filling mechanisms

As we approach the bandgap energy, the changes in refractive index from band filling start to dominate over the simple free-carrier refractive effects discussed above. These band-filling effects can be much stronger.

At low carrier densities in direct gap III-V materials, neglecting excitonic effects, [212] gives

$$\Delta n \sim -1.7 \times 10^{-17} \frac{N_e}{\left( \hbar \omega_{AV} \right)^2 T} J\left( \frac{\hbar \omega - E_G}{k_B T} \right) \quad (12)$$

where the electron density $N_e$ is in $cm^{-3}$, $k_B$ is Boltzmann's constant, $E_G$ is the bandgap energy, $\hbar \omega$ is the photon energy (necessarily in eV in the denominator expression), $T$ is the temperature in kelvin, and $J\left( \varepsilon \right)$ is a resonant function that is $\sim 0.5$ at a photon energy an amount $\Delta E \sim k_B T$ below the bandgap energy, and falling off with photon approximately $\propto 1/\Delta E$ as the separation $\Delta E$ below the bandgap energy increases. So for a photon energy of $\sim 0.8$ eV (corresponding to $\sim 1.5$ μm wavelength), at room temperature this expression gives

$$\Delta n \sim 4 \times 10^{-20} N_e \quad (13)$$

at about $\Delta E \sim k_B T$ below the bandgap energy.

[213] estimate a modal index change of -0.0005 in increasing the carrier sheet concentration from $10^{12} \, cm^{-2}$ to $2 \times 10^{12} \, cm^{-2}$ in a 10 nm thick InGaAs quantum well with a $\Gamma$ factor of 0.03. Hence, the index change is equivalent to $\sim 0.0005/0.03 = 0.017$. Now, $10^{12} \, cm^{-2}$ is equivalent to a volume density of $10^{18} \, cm^{-3}$, so the index change here is

$$\Delta n = 1.7 \times 10^{-20} N_e \quad (14)$$

[187] also estimates $\Delta n \approx 4 \times 10^{-20} N_e$ in GaAs about $k_B T$ below the absorption edge at low densities in a quantum well.

Note here we are attributing this index change to the electron density; because of its low effective mass and the resulting low density of states for electrons, the filling of the conduction band is largely responsible for the band-filling effect that is behind this index change. Note also these estimates of the band-filling index change are $\sim \times 10$ larger than the silicon free-carrier index change in (11).

### 4) General conclusions on electrorefractive modulators

Electrorefractive effects can certainly be a viable choice for low energy modulators, though the devices will generally have somewhat higher energies than the best electroabsorption devices. For those devices based on the refractive consequences of changes in absorption spectra near the bandgap energy (so, band-filling and QCSE devices), for the same operating energy density (e.g., $10^{18}$ cm$^{-3}$ carrier density or $10^5$ V/cm operating field), the electrorefractive devices generally have to be longer (e.g., 100's of microns long without resonators) to work than the corresponding electroabsorptive device (e.g., microns long). The resulting operating energies for such electrorefractive devices are therefore going to be correspondingly larger than the electroabsorptive devices. So we might expect QCSE electrorefractive devices to more comparable to those of light emitters in Table I, and band-filling electrorefractive devices to have higher energy than the light-emitter numbers in Table I.

Devices based on the best conventional Pockels-effect materials, such as the organic materials of [88], could have operating energies comparable to those of the QCSE electrorefractive devices in comparable structures, but still significantly larger than the best electroabsorption modulators [41], [78].

For silicon devices based on free-carrier plasma effects, for the same energy densities, devices without resonators would need to be ~ 10 times longer than the band-filling or QCSE electrorefractive devices, with correspondingly larger energies of operation (hence the picojoule energies [104] of simple Mach-Zehnder modulators in silicon). Hence, low energy silicon modulators have to use large amounts of optical energy concentration to reach low operating energies (e.g., Q-factors of 1000's or higher).

In general, though electrorefractive devices remain an interesting option, it is harder to scale them down into deeply sub-femtojoule operating energies without substantial optical field concentration, and the silicon free-carrier mechanism may already be operating at close to the lowest possible energies in recent impressive demonstrations [105].

### F. Comparison of quantum wells and 2D materials

There has been considerable recent interest in 2D materials like graphene or MoS$_2$ for their potential in optics [214]. One often-cited attribute is that graphene, a material in the form of sheet that is only one atom thick, can have an absorbance (the probability that a photon would be absorbed in passing through the sheet)

$$A \simeq \pi \alpha_{fs} = \frac{e^2}{4\varepsilon_o \hbar c} \simeq 2.3\% \qquad (15)$$

which raises many interesting questions and possibilities for optical and optoelectronic devices.

This absorption is particularly broad band, and we can expect many novel possibilities for integration in which such a layered material can be conveniently integrated with other electronic and optical structures. Optical absorption modulators based on band filling have been proposed and demonstrated [198], [199], [200], [201]; [201] shows ~ 1 pJ operation in a 40 µm-long device integrated with silicon technology, for example, which is competitive with silicon Mach-Zehnder modulator [104] approaches, for example.

Our main interest here is in the possibilities of low energy devices. To understand the energy requirements, we can usefully compare these 2D materials with a quantum material structure, which is itself already in many ways a 2D material.

First, we note that an expression similar to (15) can also apply to quantum wells. If, for example, we take a simplified model of direct gap optical absorption in semiconductors (neglecting excitonic effects) [84], use a simple two-band k.p model for the semiconductor [84], and use the 2D density of states (as appropriate for a quantum well), then we derive the same expression as (15) for absorption just above the bandgap energy, with the only difference being that the result is divided by the refractive index of the material in which the quantum well is embedded. (That same factor would likely apply also to a graphene layer embedded in another material since it just comes from the electromagnetics of such a problem.)

A quantum well empty of carriers shows strong excitonic enhancement of absorption near the bandgap energy. Graphene does not show corresponding excitonic effects in the infrared or visible for two reasons: first, we are not operating near to a bandgap energy; and, second, the high carrier densities we need to use in devices when shifting the Fermi level to coincide appropriately with the operating photon energy would strongly screen any excitonic effects. In quantum wells, likely at least partially as a result of their excitonic enhancement, even with the reduction in absorption from the refractive index, experimentally, a single quantum well can show a measured > 2% absorption near its bandgap energy ([196] shows > 4% relative change in transmission in a double pass through a single quantum well as it is filled with sufficient carriers for band filling). Hence, a quantum well can have similar absorption as a single graphene sheet.

#### 1) Band-filling modulation energies

Both the quantum well and graphene show absorption modulation by band filling, but the sheet carrier density required in graphene is much higher. To make the graphene transparent at a photon energy of 0.8 eV, we would need to fill the conduction (or valence) band up to a Fermi energy $E_F$ of 0.4 eV. In graphene, using the standard expression $E_F = \hbar v_F \sqrt{\pi n_e}$ where $n_e$ is the sheet (i.e., per unit area) electron (or hole) density and the Fermi velocity $v_F = 10^6$ m/s [214], we would require $n_e \sim 1.2 \times 10^{13}$ cm$^{-2}$, which is significantly higher than the $< 10^{12}$ cm$^{-2}$ for the quantum well structure, as discussed above.

Possibly a fairer comparison is to ask by how much we would have to increase the sheet carrier density in the graphene for modulation, compared to some starting concentration. With sufficient carrier density to shift the transparency edge to ~ 0.8 eV photon energy, the edge of the absorption spectrum of graphene has a width of ~0.15 eV for a factor 2 change in absorption [214]. Moving $E_F$ from 0.3625 eV to 0.4375 eV to move the transparency edge by ~ 0.15 eV requires an additional sheet carrier density of ~ $4.4 \times 10^{12}$ cm$^{-2}$, which is still ~ 5 times the required sheet density in the quantum well case.

Graphene does have the significant qualitative feature that the precise operating wavelength can be set as necessary over a very wide range, just by changing the bias. Nonetheless, this mechanism in graphene does not offer lower energies than the quantum well approach, and its operating energies would lie somewhat above those shown for the light emitter in Table I.

### 2) Electroabsorption mechanism

It does not currently appear that 2D materials like graphene or $MoS_2$ offer useful electric-field-driven electroabsorptive effects for modulators. Graphene itself does not have a bandgap that would allow the excitonic and band-edge electroabsorption mechanisms, at least in the visible or near-infrared. Single layer $MoS_2$ does have a direct bandgap and strong excitonic effects [215]. [216] shows QCSE in $MoS_2$ with fields > MV/cm, corresponding to electron-hole pairs effectively confined within each layer of $MoS_2$. Though shifts of up to 16 meV are observed here, even with these large fields, these shifts are not apparently large compared to the exciton linewidth [217]; hence, they may not be particularly useful for optical modulators, because the background absorption is quite an important parameter as in criterion (5) above.

In effect, $MoS_2$ is arguably too thin for good QCSE. In quantum well materials there is effectively an optimum thickness for QCSE electroabsorption, which is typically $\sim 10$ nm. If the layer is thicker, the QCSE shifts are larger, but the absorption strength of the shifted absorption steps falls off too quickly with field (because of the separation of the electron and hole states to opposite sides of the well, decreasing the overlap of their wavefunctions that is necessary for optical absorption). If the layer is too thin, the quantum confinement energies become larger and the wavefunctions are too difficult to perturb, requiring larger fields. Also, often such thin layers have larger broadening of the absorption edge for any of a number of different reasons, effectively eliminating the necessary absorption coefficient contrast between "absorbing" and "non-absorbing" states as required by criterion (5).

The conclusion here is that, though there may be some viable and interesting prospects for modulators using 2D materials, and these may have some qualitative advantages, they currently do not appear to offer any basic energy advantage over structures like quantum wells, and may actually require larger operating energies. Possibly other such materials not yet investigated for optoelectronic device use may offer additional opportunities. We note, for example, that the related layered material $WS_2$ [218] does show very strong excitonic effects, with a particularly strong and clearly resolved peak.

## APPENDIX B – OPTICAL CONCENTRATION AND USE OF RESONATORS

### G.  Optical concentration factor

Here we briefly discuss the relation between our concept of an optical concentration factor $\gamma$ and various other terms used for concentrated electromagnetic fields. Formal definitions of finesse $\mathfrak{F}$, quality factor $Q$, and Purcell enhancement factor can be found in standard references, so here we will concentrate on an informal approach emphasizing the physical meanings.

For electromagnetic fields at the resonance frequency of a cavity, $Q$ can be thought of as

$$Q = 2\pi \times \left( \frac{\text{energy stored within the cavity}}{\text{energy lost during one cycle of oscillation}} \right) \quad (16)$$

and cavity finesse $\mathfrak{F}$ can loosely be considered either as

$$\mathfrak{F} = 2\pi \times \left( \frac{\text{energy stored within the cavity}}{\text{energy lost during one cavity round trip}} \right) \quad (17)$$

or equivalently as

$$\mathfrak{F} = 2\pi \times \left( \begin{array}{c} \text{number of cavity round trips a photon makes} \\ \text{before being lost} \end{array} \right) \quad (18)$$

at least for high-finesse cavities.

For both finesse $\mathfrak{F}$ and quality factor $Q$, the loss in question can be from absorption, scattering, escape through the mirrors, or any combination of these.

From our statement Eq. (18) above, instead of a photon just making just one pass through the material in the cavity, it will now make $\mathfrak{F} / \pi$ passes (note that one round trip corresponds to two passes through the material), so the average energy density in the cavity is magnified by this amount. If the optical concentration factor in some propagating mode was originally $\gamma_o$, then adding some cavity of finesse for that mode, then the new optical concentration factor is

$$\gamma = \frac{\mathfrak{F}}{\pi} \gamma_o \quad (19)$$

Consider a cavity of length $L$ in which the only loss mechanism is the transmission of light through mirrors, with (intensity) reflectivities $R$ at each of the two ends of the cavity. The probability that the photon leaves the cavity on hitting one of the mirrors is $1-R$, which will be a small number for high-reflectivity mirrors. So the probability that the photon is lost to the cavity in a round trip is approximately the sum of these small probabilities for the two mirrors, giving a probability of loss per round trip of $2(1-R)$, and therefore an average number of round trips before being lost of $1/[2(1-R)]$. So, we arrive at the expression for such a cavity

$$\mathfrak{F} \simeq \pi / (1-R) \quad (20)$$

and from Eqs. (19) and (20), the optical concentration factor is

$$\gamma \simeq \frac{1}{1-R} \quad (21)$$

The relation between $Q$ and $\mathfrak{F}$ for high-finesse cavities can be stated as

$$Q = \frac{2L}{\lambda_n} \mathfrak{F} \quad (22)$$

where $\lambda_n$ is the wavelength inside the material. For a refractive index $n$, and a free-space wavelength $\lambda$, $\lambda_n = \lambda / n$. So $Q$ is larger than $\mathfrak{F}$ by a factor that is the length of the cavity in half-wavelengths in the material. We can see this relation also from Eqs. (16) and (17). Light propagates one wavelength in the material (i.e., $\lambda_n$) in one cycle. It therefore requires $2L / \lambda_n$ cycles for a round trip; to get to $\mathfrak{F}$ from $Q$, we need to divide by $2L / \lambda_n$.

We see from Eq. (19), incidentally, that finesse $\mathfrak{F}$ rather than the quality factor $Q$ is a more direct measure of the increase of optical concentration factor resulting from the use of a cavity.

The Purcell enhancement factor $F_P$ is typically defined in terms of the ratio $Q / V_{\lambda n}$ where $V$ is the cavity volume expressed in in units of $\lambda_n^3$, in which case the definition is

$$F_P \equiv (3/4\pi^2) Q / V_{\lambda n} \qquad (23)$$

Substituting from Eq. (22)

$$F_P \equiv \frac{3}{4\pi^2} \frac{2L}{\lambda_n} \frac{\mathfrak{F}}{V_{\lambda n}} = \frac{3}{2\pi^2} \frac{\mathfrak{F}}{A_{\lambda n}} \qquad (24)$$

where $A_{\lambda n}$ is the cross-sectional area of the cavity in square wavelengths. A guide of cross-sectional area $A_{\lambda n}$ without a resonator would have a field concentration factor $\gamma_o = 1/A_{\lambda n}$. So, using Eqs. (19) and (24), we have, for some resonator structure,

$$F_P = \frac{3}{2\pi} \gamma \qquad (25)$$

Hence, for resonator structures, the concept of Purcell enhancement factor $F_P$ and our optical concentration factor $\gamma$ are essentially the same, differing only by a numerical factor $3/2\pi \simeq 0.477$. Equivalently, Purcell enhancement factor is effectively defined for a somewhat smaller cross-sectional area than our reference structure, e.g., a square cross-section of area $(3/2\pi)\lambda_n^2$, or a circle of radius $\left(\sqrt{3/2\pi^2}\right)\lambda_n$, for example, instead of the square $\lambda_n^2$ reference cross-section we use for $\gamma$.

One could argue that we should just use the Purcell factor rather than introducing our optical concentration factor; in response, we would argue that our factor is more directly intuitive and applies to a wider range of structures, not being restricted to resonators.

The term "local density of states" is sometimes used to cover broader cases that do not necessarily involve resonators, but it is arguably a deeply confusing and unfortunate terminology[70], especially for situations that do not involve resonators, so we avoid it. Essentially, the ratio of the local density of states to the density of states (modes) in free space would correspond loosely to our optical concentration factor $\gamma$, however.

### H.  Use of high-Q resonators

Though it is the finesse $\mathfrak{F}$, rather than the cavity $Q$, that determines the concentration factor, to make small devices work using resonators, in practice we typically need to increase the $Q$ factor, not just the finesse. With most microscopic mechanisms that we use for devices, we are limited in the absolute values we can have for processes such as absorption or absorption changes, gain, or refractive index change. For light emitters or modulators, beyond some level of excitation or drive, we will reach some limit on these changes; either the basic properties of the material itself or our practical inability to drive it more strongly (such as practical voltage limits) may prevent us from increasing the amount of emission or gain or of absorption or refractive index changes.

Hence, even if we fill the active cross-section of the waveguide or resonator with the active material, we will still need some product of length and concentration factor to get the device to work. For resonator approaches, that product is essentially the $Q$ of the resonator – $Q$ is finesse $\mathfrak{F}$ multiplied by the length of the cavity in half-wavelengths, as stated above. So $Q$ is often the quantity quoted in devices rather than finesse $\mathfrak{F}$. It is still correct, however, as implied by Table II, that we need specific levels of concentration factor $\gamma$ (and hence of finesse in cavity approaches) to make devices using specific active volumes. The energy numbers in Table II presume we are operating the microscopic mechanisms at some typical practical level of excitation or drive.

Note, though, that it is the cavity $Q$ that determines how precisely the resonator has to be tuned. The resonant frequency $f$ is proportional to the cavity length $L$, and the frequency width $\Delta f$ of the resonance is $\Delta f = f/Q$. So, to hit the resonant frequency within a resonance width, the length of the cavity has to be correct to within a precision $\Delta L$ given by

$$\frac{\Delta L}{L} = \frac{\Delta f}{f} = \frac{1}{Q} \qquad (26)$$

so, a fractional precision of $\sim 1/Q$. Hence, if we require high-$Q$ resonators, we have to deal with this tuning precision either in the original fabrication, in some post-fabrication trimming, or in some feedback adjustment in operation.

---

[70] In quantum mechanics, as in Fermi's Golden rule (see, e.g., [84]), the transition rate for a process like optical absorption or emission can be proportional to the square, $|d|^2$, of a matrix element between initial and final states, and to the density $\rho$ of available final states. One view of resonators is to say that they concentrate the optical density of states by some factor, and that concentration therefore enhances the transition rate; and this is a common view in discussing Purcell enhancement (introduced in Purcell's original description [219]). However, if we consider a resonator in space, or inside some large box, the resonator has almost no effect on the density of states of this larger system. In that view, what happens is that, for those modes of the overall system that happen to correspond to strong resonance within the resonator, the mode amplitude is strongly enhanced inside the resonator, which leads to a much larger $|d|^2$ for all such modes. In this case, it is the matrix element between the initial and final states that is enhanced, because the optical states of interest correspond to ones with much larger field concentration inside the resonator where the active material is. Now, in one view, the difference between these two pictures does not matter, at least for resonators; both will give the same answer if we come up with some supposed factor for the enhancement of the density of states by the resonator. However, once we consider other situations, such as the enhancement of optical field near some metallic tip, there is no obvious resonator, and no obvious way to define a true density of states that has been enhanced. Any increase in optical interaction for materials near such a tip is arguably physically from the increased optical field, not from any change in the density of optical states. Nonetheless, it is common to describe such enhanced interactions in terms of an effective "local density of states," even though, in this author's opinion, that terminology bears little or no relation to the actual physics. As a result, though, we will avoid using the term "local density of states" here, using the more physical idea of optical concentration. Incidentally, though the various terminologies might make this seem to be a confused topic where no clarity is possible, a direct quantum mechanical approach here is quite straightforward and will give unambiguous answers. For example, we could model the resonator system by putting it in a large box, and then evaluating all the electromagnetic modes of that large box, including the resonator. Then we could calculate a property like absorption or spontaneous or stimulated emission using those modes rather than plane waves, following a standard quantum optical approach, e.g., as in [84]; the result of such a calculation is quite independent of any of the definitions of terms like finesse, quality factor, or local density of states.

In fabrication, lithography might allow length precision ~ a few nanometers. Suppose that our device of interest has be on the scale of only a few microns in size so that the energy can be low enough and the density of devices high enough; then it would be difficult to set the operating wavelength of the device to better than ~ 1 part in 1000 directly in fabrication. Furthermore, device-by-device trimming to compensate for that lack of fabrication precision might not be feasible financially for the large numbers of devices we might need.

For light emitters, we could argue that the precise wavelength may not matter much, though that does mean that we cannot use other narrow band or wavelength-sensitive optics in the rest of the optical system; dense wavelength division multiplexed systems might therefore also not be possible with such lasers as sources without some further tuning.

For modulators, if they need high $Q$'s just to function sufficiently well, we could propose some active tuning stabilization for every device, but that raises two other issues: we would need additional detection and feedback loops for every device (as well as some wavelength reference), and we would need some physical resonator tuning mechanism for every device. There could be many different approaches to resonator tuning, but current approaches such as thermal tuning tend to consume significant energy; other microscopic mechanisms for changing refractive index can lead to loss (e.g., as in tuning by changing carrier density) and may also not be able to give large enough refractive index changes to tune a small device. One possible approach might be micromechanical tuning, which might not require any static power dissipation.

Even if we can devise an approach that allows such tuning of each resonator, the additional system complexity and power dissipation associated with such tuning could be prohibitive for any large number of modulator devices, so we should be cautious in proposing Q's beyond 1000 for any modulator device to be used in large numbers. As noted above, however, electroabsorptive devices can likely achieve low enough energies without such high Q's, so they remain an attractive modulator option.

One further important issue is that resonator wavelengths will in general drift with temperature. In real systems, we should expect that the entire system should be able to operate over some significant environmental temperature range, such as at least the commercial range of 0° − 70°C; local temperatures on a silicon chip can also vary substantially from position to position on the chip, possibly by as much as 40°C [220]. A typical order of magnitude for the change in refractive index with temperature is $dn / dT \sim 10^{-4} \mathrm{K}^{-1}$ in a semiconductor [221] and $\sim 10^{-5} \mathrm{K}^{-1}$ in glass [22]. One promising approach to such an issue is to compensate the refractive index change of one material with an opposite change in another [221], [222], [223].

If we consider only moderate $Q$ resonators, however, we may not need any tuning or compensation. For a semiconductor resonator with $dn / dT \sim 10^{-4} \mathrm{K}^{-1}$, then a 100°C temperature variation corresponds to a change in index of $\sim 10^{-2}$ and a corresponding fractional change in the resonant frequency or wavelength. For example, for a $Q \sim 30$ or smaller, such a fractional change would be significantly less than the fractional linewidth ($\sim 1 / Q$) of the resonator. For a $Q$ of this magnitude, it might also be possible to operate over most or all of the telecommunications C-band (1530 − 1565 nm wavelength) without tuning since that wavelength range corresponds to a fractional range of $\sim 1 / 44$. Hence, such a $Q \sim 30$ device could be quite a practical option.

## APPENDIX C − MATERIALS CRITERIA FOR MODULATORS

As mentioned in the main text, an important criterion for a modulator is the absolute difference $\Delta T$ in the transmission of the modulator in its two states [41]; this gives the fraction of the input optical power that is usefully available to drive the detector and receiving circuit. In general, when trying to maximize energy efficiency overall, optimizing $\Delta T$ is more important than optimizing contrast ratio itself [41]. As a result, a good device not only should have some significant contrast ratio between high and low transmission, but it should also have a high maximum transmission. Hence, background loss in modulators is particularly important. This leads to important consequences for the properties we require from electroabsorption and electrorefraction materials.

### a)     Criteria for electroabsorptive materials

For the moment, presume we have a device without any resonator. Suppose the background absorption coefficient of the material (i.e., the absorption coefficient in the "transmitting" or "on" state) is $\alpha_{trans}$. For an electroabsorptive modulator, suppose the absorption coefficient in the "absorbing" or "off" state is some larger amount $\alpha_{abs} = \rho \alpha_{trans}$, so that the ratio of the "off" to "on" absorption coefficients is $\rho$. For a length $L$, the "on" and "off" transmissions will be $T_{on} = \exp(-\alpha_{trans} L)$ and $T_{off} = \exp(-\alpha_{abs} L)$, respectively, with the difference being $\Delta T = T_{on} - T_{off}$.

An electroabsorptive material at a given wavelength and operating field will have some specific absorption coefficient ratio $\rho$. A simple maximization by differentiation shows that the largest $\Delta T$ is obtained for a length $L$ such that

$$\alpha_{off} L = \frac{\ln \rho}{\rho - 1} \tag{27}$$

with a resulting maximum $\Delta T$ of

$$\Delta T_{max} = \rho^{-1/(\rho-1)} - \rho^{-\rho/(\rho-1)} \tag{28}$$

This value of $\Delta T_{max}$ rises monotonically from 0 for $\rho = 1$ (so no contrast in absorption coefficients) through 3.5% (-14.5dB) for a low absorption coefficient contrast of $\rho = 1.1$, ~15% (-8.3dB) for $\rho = 1.5$, 25% (-6dB) for $\rho = 2$, ~50% (-3dB) for $\rho = 4.5$, and continuing to rise, but with progressively decreasing further benefits, for increasingly larger $\rho$ (e.g., ~ 70% (-1.6dB) for $\rho = 10$).

A reasonable approximate conclusion from this analysis is that we need an absorption contrast ratio

$$\rho = \frac{\alpha_{abs}}{\alpha_{trans}} \geq 2 \tag{29}$$

if we are to have a modulator that is reasonably (i.e., >25%) efficient in using the optical power. The penalty for lower absorption coefficient contrast increases steeply as $\rho$ reduces below about 2.

No matter how strong is the optical absorption in the material, we will have an optically very inefficient design unless we have at least about a factor of 2 or more contrast between the "off" and "on" absorption coefficients. This turns out to be quite a demanding criterion for electroabsorptive materials, and rules out several electroabsorptive mechanisms.

Such an example design using a material with $\rho = 2$ would have a length

$$L = \frac{\ln 2}{\alpha_{off}} \approx \frac{0.693}{\alpha_{off}} \tag{30}$$

so about 70% of an absorption length, and it would modulate from a high transmission of 50% to a low transmission of 25%.

### b)   Criteria for electrorefractive materials

For an electrorefractive modulator, to maximize $\Delta T$ we also want to avoid having too much loss. With a background optical absorption coefficient of $\alpha$, in a simple modulator without a resonator, we would therefore want to keep the length $L$ of the modulator less than about one absorption length, i.e., $L \leq 1/\alpha$. If we have no resonator, then we need to have sufficient refractive index change $\Delta n$ in the device length $L$ to give the desired $\sim \Delta n L \geq \lambda/2$ change in optical path. (Note that $\lambda$ here is the free-space wavelength, not the wavelength in the material.) Hence a desirable criterion for an electrorefractive material is

$$\frac{\Delta n}{\alpha} \geq \frac{\lambda}{2} \tag{31}$$

This can be a surprising difficult criterion to meet for many otherwise promising mechanisms for refractive index change, as we will discuss below. A key difficulty is that it can be difficult to find any high-speed mechanism that can in practice and under reasonable operating conditions give $\Delta n$ much greater than about $10^{-3}$ while still satisfying this criterion (6). That has been a long-standing problem in electrorefractive devices in general. As a result, electrorefractive devices without resonators tend to need to be quite long, e.g., $L \sim 750\,\mu\text{m}$ for $\Delta n \sim 10^{-3}$ and $\lambda = 1\,\mu\text{m}$. Organic polymer electrooptic materials have been projected to offer up to $\Delta n \sim 1\%$ at a field of $10^6\,\text{V/cm}$, and in a device in a 90 nm wide plasmonic waveguide with additional field concentration from group velocity effects has been able to operate with a 10 μm length [88], which may represent the shortest refractive modulator without a resonator. (This device operates at $\sim 25$ fJ/bit energy.)

### c)   Materials criteria and use of resonators

Both electroabsorptive and electrorefractive modulators can also exploit resonators. The use of resonators can allow us to work with smaller devices. Loosely, for a cavity finesse $\mathfrak{F}$, since the photon now makes $\sim \mathfrak{F}/\pi$ passes through the cavity (see Appendix B), we only need to pick up $\sim \pi/\mathfrak{F}$ as much path length change or absorption in each pass, so the device can be shorter by a factor $\sim \pi/\mathfrak{F}$.

Our analyses above for the case without a resonator lead to a device no longer than $\sim 1$ absorption length for the background or "on" state absorption. The amount of background loss we can tolerate per pass in the resonator case also has to go down by a similar factor $\sim \pi/\mathfrak{F}$, however. Then, this background absorptive loss per pass at most remains comparable to the loss through the mirrors per pass; that amount of loss is at a point where we are beginning to substantially affect the operation of the cavity because of this background absorption loss.

Hence, the material requirement (6) remains the same for electrorefractive modulators with resonators; essentially, we are dividing both sides of the equation by $\mathfrak{F}/\pi$, which leaves the material criterion the same.

It is also practically the case that to make substantial modulation in an absorptive device in a cavity, we will need at least roughly to double the amount of absorption; that would change the absorption per pass from being comparable to the mirror loss to being substantially greater than the mirror loss, thereby substantially changing the transmission of the resonator. So, we should expect the criterion (5) to remain approximately valid for electroabsorptive modulators with resonators.

These arguments are loose, and for any specific resonator design we should perform the actual analysis to get detailed answers for performance, but the basic conclusion is that changing to a resonator design does not substantially change the underlying requirements (5) and (6) on the materials. See, e.g., Refs. [75], [79] for recent example analysis and design of such devices.

Note, incidentally, that the use of asymmetric Fabry-Perot resonators – a useful trick to enhance the contrast ratio in absorptive modulators (as in [75], [79], for example) – in practice makes little difference to the total $\Delta T$ for the modulator, so it does not change the material requirements here.

### Appendix D – Example analysis of "near-receiverless" operation

We can make a simple estimate of how much energy we can tolerate to run a receiver amplifier so that we are benefitting overall in reducing the total energy to run the entire system.

Suppose, for example, that the effective optical loss[71] of the system from the optical power source to the receiving photodetector is some factor $L_{SP}$, that the "wall-plug" efficiency[72] of the optical power source is a factor $\eta_S$, and that, in a receiverless system, we need an optical energy $E_R$ per bit at the receiver. Then the corresponding transmitter "wall-plug" energy per bit for the receiverless system is

$$E_T = E_R \eta_S L_{SP} \qquad (32)$$

---

[71] Effective optical loss would include all actual loss factors together with a factor for the increased optical power required because of the limited difference in optical transmission between the "low" and "high" transmission states of a modulator (see Appendix C and Section IV D).

Adding in a receiver gain of some factor $g$ would reduce the required electrical "wall-plug" energy per bit by a factor $g$ because it would correspondingly reduce the required transmitter energy bit to $E_T / g$. So the transmitter energy saved would be an amount

$$\Delta E_T = E_T - \frac{E_T}{g} = E_T \left(1 - \frac{1}{g}\right) \qquad (33)$$

Presuming we are thinking of adding a receiver gain stage with $g$ significantly greater than 1 (e.g., $3 - 10$)[73], then the factor $1 - (1/g)$ is not far from 1, and the energy saved at the transmitter by adding the gain stage will be approximately $E_T$, i.e., $\Delta E_T \approx E_T$. So for any energy benefit in adding such a receiver amplifier, the energy per bit to run the additional receiver amplifier circuit, $E_{gain}$, should be at least somewhat less than the energy per bit $E_T$ currently being dissipated at the transmitter or there is no point in adding the gain stage. So

$$E_{gain} < E_R \eta_S L_{SP} \ (= E_T) \qquad (34)$$

Once we integrate photodetectors with very low total capacitance, the optical input energy required for receiverless operation ($E_R$) becomes small, and the energy $E_{gain}$ we can afford to spend on a receiver gain stage for any net energy benefit also becomes small. Nonetheless, $E_T$ may still be a significant number, such as 10's of fJ in such a hypothetical future system (see the discussion in Section IX A). So spending up to a few fJ per bit on a gain stage might make sense. Any such circuit would have to be quite simple, however, such as one CMOS stage of gain, to hit such an energy target, and would be unlikely to be designed as a noise-limited amplifier stage [111].

### Appendix E - Noise in low-capacitance and receiverless operation

One legitimate question is whether we truly can avoid problems of noise in receiverless or near-receiverless operation. We might seriously consider two potential sources of noise – Johnson (or thermal) noise, and shot (or Poissonian statistics) noise. The simplest answer, which is certainly valid for the receiverless case, is that, since these noise sources do not matter for ordinary electronic logic gates operating at logic voltage swings, then they do not matter when those same logic voltage swings are generated by photodetectors.

Note that, since a photon energy of ~0.8 eV is also equal to the energy of an electron at a logic voltage of 0.8 V, the numbers of photons and the numbers of electrons to drive a gate with an efficient detector are essentially the same, so if shot noise does not matter for the transistor, then it does not matter in the receiverless photodetector case.

In optical communications, analysis of the statistics of photons gives required minimum numbers of photons of 20 – 100 to avoid bit errors from photon statistics [224], depending on the specific statistical assumptions and the required bit error rate. For ~ 1 eV photons, a received optical energy of 100 aJ/bit corresponds to ~ 600 photons/bit, so at such a level we are likely far from shot noise being a significant problem. We might need to reconsider this, however, if we were to consider operating at ~ 10 aJ/bit levels.

For thermal noise, we can estimate this by considering what is sometimes referred to as "$kT / C$" noise. If we charge a capacitor $C$ through a resistor, and consider thermal noise in the resistor as a noise source, the resulting fluctuation of the voltage on the capacitor is essentially independent of the resistor value; this independence is because, though the thermal noise (voltage)$^2$ per unit bandwidth is proportional to the resistor value, the bandwidth of the $RC$ circuit is inversely proportional to the resistor value; so, the resistor value cancels out in the algebra. As a result, using standard Johnson noise analysis, the standard deviation of the voltage on the capacitor is $v_n = \sqrt{k_B T / C}$ where $k_B$ is Boltzmann's constant, regardless of the resistor used to charge it.

Such noise only appears if we have a resistor of some kind connected to the capacitor, which is not necessarily the case in an optical receiver. But, even assuming we have such a resistance to charge or discharge the photodetector capacitance, this noise is not likely to present much of a problem for a receiverless approach; for 1fF, $v_n \approx 2$ mV, and even for 10aF, $v_n \approx 20$ mV, both of which are much less than a logic swing.

If we do use some moderate signal amplification at the output of the photodetector in a "near-receiverless" approach, as long as that is only some small factor, such as $3 - 5$, such noise sources are still not likely to be much of a problem, though we should likely analyze the noise in such cases with amplification.

### Appendix F – Free space optical systems

#### I. Diffraction limit to the number of free-space channels

One important number we need to understand is the limiting number of possible separate channels we can have for communication between two surfaces, as determined from the laws of diffraction. For each polarization, we will not in practice be able to exceed this number, and any optical system will have to be designed so that it does not attempt to violate this limit. Fortunately, this problem is well understood, both intuitively and somewhat more rigorously (see, e.g., [126]).

We can think of free-space optical system in which we are communicating between one essentially plane "transmitting" surface and another (parallel) "receiving" surface, as sketched in Fig. 15. The area or "aperture" of the transmitting (receiving) surface is $A_T$ ($A_R$). The solid angle subtended by the transmitting surface at the receiving surface is $\Omega_T \approx A_T / L^2$, where $L$ is the separation of the surfaces, and we are taking a "paraxial" approximation, presuming $L$ is much greater than the linear dimensions of either area. Similarly, the solid angle subtended by the receiving surface at the position of the transmitting surface is $\Omega_R = A_R / L^2$ .

---

[72] By "wall-plug" efficiency we mean the ratio of useful optical power out from a light source to total electrical power in to the light source.

[73] There may not be much point in adding in gain much less than this, and just one CMOS inverter stage is likely to add gain of at least such an amount.

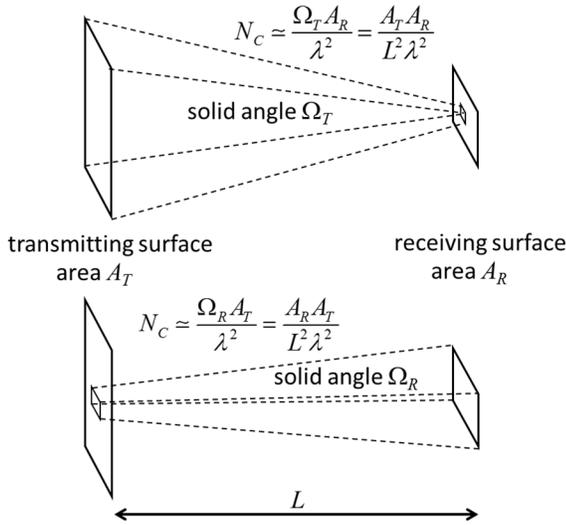

Fig. 15. Optical apertures and solid angles for calculating the number communication modes between surfaces.

For a wavelength $\lambda$ in the medium between the surfaces, the physics of diffraction sets the practical number of orthogonal (i.e., spatially separable) spatial channels or "communications modes" between the surfaces as [126]

$$N_C \simeq \frac{\Omega_R A_T}{\lambda^2} = \frac{\Omega_T A_R}{\lambda^2} = \frac{A_T A_R}{L^2 \lambda^2} \qquad (35)$$

We can if we want think of this as if we had some lens in the "transmitting" aperture focusing to the smallest spots allowed by diffraction at the position of the "receiving" aperture, with $N_C$ corresponding to the number of resolvable or approximately non-overlapping spots we could form or approximately non-overlapping positions we could focus a light beam on the receiving surface given the cross-sectional areas of both surfaces.

The minimum size of spot we can form is limited by diffraction; indeed, we could get intuitively to the result Eq. (35) by presuming that a spot of an area $A_S = m\lambda^2$ (for some number $m$) has a corresponding diffraction solid angle of $\Omega_S = 1/m$ steradians; that is essentially equivalent to saying that a spot of lateral dimension $d$ (e.g., in, say, the "vertical" direction) has a corresponding diffraction angle in radians (in that same "vertical" plane) of $\theta_d = \lambda / d$ , which is a standard type of result in diffraction theory: a small spot must have large diffraction angle, and equivalently it takes a large convergence angle to focus to a small spot[74].

Note this problem is symmetric – we could also consider this in terms of a lens in the receiving aperture capturing the light from multiple spots in the transmitting aperture, where those spots are as small as we can allow if their resulting diffracted beam just fits within the receiving aperture.

Of course such a counting is loose because it requires a choice of just how far apart we think spots have to be to count as "non-overlapping". More rigorously, we can formally solve such problems in a generalized fashion [126] to find the optimum best-coupled channels – the "communications modes", which we can do by performing the singular value decomposition of the coupling "diffraction" operator between the surfaces[75]. If we do so, we get the same result for the number here, so this result is quite rigorous[76]. So, for a given pair of such surfaces, we can state quite definitely the maximum number of orthogonal spatial channels we have for communication for a given polarization.

Recently, there has been some confusion about whether the use of different forms of beam can somehow increase the number of channels – that is, essentially violating Eq. (35). The fact that orbital angular momentum modes [225], [226] can be described in terms of an angular momentum "quantum number" could lead to the mistaken impression that this angular momentum is somehow an addition degree of freedom of the light field, and hence could increase the number of channels in the system beyond the result Eq. (35). In fact, this is not the case. Such angular momentum beams are merely a different choice of basis on which to represent spatial beams; they do *not* increase the number of available spatial channels as given by Eq. (35). They are also not necessarily the optimum modes for any given problem. Indeed, if we restrict ourselves to only using angular momentum beams that have a "ring"-like form, such beams use the available aperture of the optical system very inefficiently; instead, we would have to use all of the radial forms of beams with the same angular momentum to make good use of the available optical aperture. Specific analysis of information capacity of optical channels using angular momentum and other approaches [227] confirms such a conclusion. The true optimum choice of modes for a given power coupling linear optical problem (the communications modes) can be established by performing the singular value decomposition of the coupling operator, and that process does not violate Eq. (35); indeed, it actually proves Eq. (35) [126].

---

[74] We could work out an explicit example using Gaussian beam spots. As is conventional, we can define such a spot at its focus (e.g., on the receiving surface) to have with electric field amplitude of the form $\exp(-r^2/w_o^2)$ for some spot radius parameter $w_o$ and with $r$ being the distance from the center of the spot in the plane of the transmitting or receiving surface. As we move away from the focus, the beam stays Gaussian in shape, of a form $\exp(-r^2/w^2)$ but with $w$ growing with distance $z$ from the focus approximately as $w(z) \cong \lambda z / \pi w_o$, as the spot expands due to diffraction. If we take the effective area of the spots to be $\pi w_o^2$ on the surface where they are focused, and consider them to be focused from a transmitting surface of area $A_T = \pi [w(L)]^2$, then we will get exactly Eq. (35).

[75] The resulting optimal choice of "communications modes" functions for the case of rectangular or circular apertures are versions of so-called prolate

spheroidal functions, which are not generally spot-like functions on either surface; all such functions on both surfaces actually essentially fill the receiving aperture. See, e.g., [126].

[76] Technically, there is a sum rule for the sum of the squares of the "coupling strengths" between orthogonal source functions on one surface and resulting orthogonal wave functions on the other [126]. For plane parallel surfaces in the paraxial approximation, those couplings are strong up to a number given by the result Eq. (35), by which point the sum rule is essentially exhausted. Any other coupled sources and waves beyond this point have very small coupling, and can generally be neglected. This sum rule is the rigorous generalization of diffraction.

*J. Calculations of number of channels*

Suppose now we consider an optical system in which, for simplicity, the two areas are equal, i.e., $A_T = A_R \equiv A$, as we might use in connecting between chips as in Fig. 11 (h). Then from Eq. (35), the number of orthogonal spatial channels between the surfaces is limited by diffraction to

$$N_C \sim \frac{A^2}{L^2 \lambda^2} \qquad (36)$$

For example, consider some optics for communicating between $2 \times 2\,\mathrm{mm}^2$ arrays on chip to adjacent chips over 4 cm distance. For $\lambda \simeq 1.5\mu\mathrm{m}$, $L = 2\,\mathrm{cm}$ (the distance to an imaging lens) and $A \equiv 2 \times 2\,\mathrm{mm}^2$, then diffraction limits us to $N_C \sim 17{,}800$ channels. Hence, 1024 channels based on output couplers and lenslets [160] on 62.5 μm centers can readily be coupled through a free space channel of $2 \times 2\,\mathrm{mm}^2$ cross-section over centimeters with only a single imaging lens in the path; even increasing the density to 4096 channels on 31.25 μm centers should be viable optically. So, for example, a 1 cm focal length lens 2 cm from the "transmitting" lenslet plane would image to final "receiving" plane a total of 4 cm away from the transmitting plane, as sketched in Fig. 16. Of course, it is straightforward to add mirror surfaces, as in Figs. D2 (b) and XIA (h), in the regions between the lenses, to deflect the beam sideways as required.

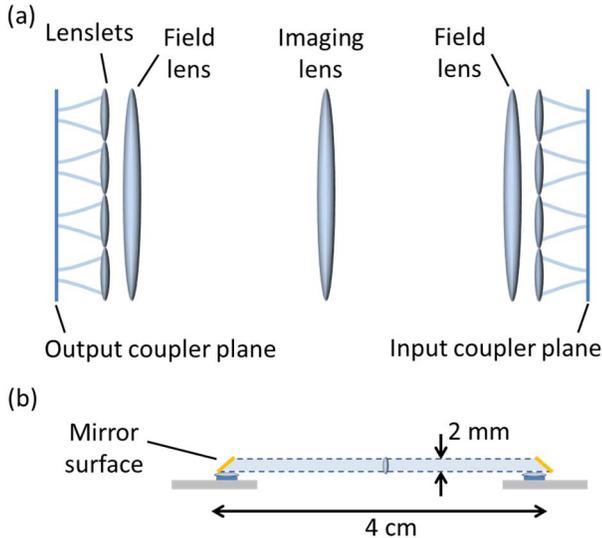

Fig. 16. (a) Sketch (not to scale) of an optical system from an output coupler plane through a lenslet array (only 4 lenslets are shown here for graphic clarity) and a field lens, an imaging lens, another field lens, and another lenslet array, onto an input coupler plane. (b) Optics shown "folded" by mirrors at the two ends for coupling to chip surfaces, at close to actual size for a $\sim 2 \times 2$ mm cross-section and a $\sim 4$ cm distance.

Here we have also included 2 cm focal length "field lenses" above each microlens array; the one in front of the "transmitting" lenslet plane effectively captures all the diverging light from the emitting microlenses so it passes through the imaging lens aperture, and the one at the final "receiving" lenslet plane effectively "straightens out" the light so that it is focused by the lenslets onto its optic axis. This makes the system from the initial output coupler plane to the final input coupler plane what is sometimes called "telecentric". These field lenses allow the whole system never to exceed $2 \times 2\,\mathrm{mm}^2$ in cross-section.

There are many ways such an optical system could be constructed, including substantially solid elements like gradient index (GRIN) optics, and we will not go into these here; our point here is just to illustrate the magnitudes of capacities of simple systems. There is also nothing special about the 4 cm distance illustrated here for such a 2 mm cross-section. Any shorter distance moves the optical system further away from any diffraction limits for the same number of channels. It is also possible to build "relaying" optical systems, with lenses spaced by twice their focal length, to extend to longer paths with the same number of channels.

Suppose we consider another example, this time hypothetically communicating through free space between two telephoto lenses, each with aperture of $A = 25\,\mathrm{cm}^2$, "staring" at each other over a separation distance of $L = 5\,\mathrm{m}$. Then we would calculate the maximum number of channels as limited by diffraction as $N_C \sim 110{,}000$. So such a hypothetical cabinet-to-cabinet link could readily carry 10's of thousands of channels.

*K. Wavelength dependence and Dammann grating spot array generators*

Since a spot array generator is a diffractive optical element, that overall size of the spot array scales with the operating wavelength, so that wavelength needs to be set to sufficient precision. For an array size of, say, $32 \times 32$ spots (so 1024 spots), in which we want the positions of the spots in the diagonal corners relative to those in the center to be correct to, say, 1/10 of the spot size, we need a relative precision of the wavelength of 1 part in $10 \times \sqrt{16^2 + 16^2} \simeq 226$. At 1550 nm wavelength, that corresponds to a wavelength precision of $\sim 7$ nm, or an optical frequency precision of $\sim 860$ GHz. This is a relatively slack tolerance for optical wavelength, especially if we are setting this in some single, centralized laser.

Incidentally, the fact that we have such a tolerance to the precise laser frequency means that we could also operate with pulsed light with pulse widths down to a few picoseconds without causing problems for such spot array generation. The usefulness of this will become apparent below when we discuss clocking and timing.

## Appendix G – Example optical requirements for modulo-synchronous systems

Here will illustrate the requirements and capabilities of optics for modulo-synchronous systems, in which propagation delays longer than one clock cycle are preset to match the clock cycle timing.

For example, suppose we run the entire system at a 2 GHz clock rate. Such a clock rate, which is in line with current practice for chips, means the chips can be run efficiently at relatively low power dissipations and with relatively full utilization of the chip's capability for information processing without exceeding thermal limits. That range of rates allows optical interconnect path lengths of up to ~15 cm in air or ~10 cm in glass or plastic path for a full clock cycle, or ~7.5 cm in air or 5 cm in glass for a half clock cycle. This is enough distance to consider groups of chips in a module within distances of several centimeters, all run with communications on a one-clock-cycle-or-less communication pattern.

Driving such a system with optical pulses so that the optics does not add substantial timing uncertainty (and could possibly reduce that uncertainty) would suggest that the pulses are some small fraction of a clock cycle, for example 10% or shorter. For 2 GHz clocks, this would suggest 50 ps pulses, or shorter. Such pulses can be generated by optical mode-locked sources or by direct modulation of a semiconductor power source laser.

Within a multiple-chip module, such on a scale of centimeters, such chip-to-chip connections can be done largely or even totally using free-space optics together with possibly some secondary optical waveguide layer for forming specific and moderately complex interconnection patterns (as discussed in Section IX).

Between more distant parts within the optically modulo-synchronous volume, we might use optical fiber connections. As discussed in Section VIIIB, distances of many meters are possible with only a few 10's of picoseconds variability in pulse arrival time from the variation of fiber refractive index with temperature. Since that temperature variation is not a significant problem, to get a specific delay from propagation in a fiber, we need to cut it to the correct length. To ensure propagation delay times in the fiber within, say 30 ps precision within a clock cycle, fibers lengths would have to be cut to specific clock-cycle lengths to within 6 mm precision, which is eminently feasible with simple techniques. Even 1 mm should be straightforward with simple cutting jigs even allowing for end polishing length loss and variation. Hence we could interconnect the larger units within the optically modulo-synchronous volume with fibers of lengths corresponding to integer numbers of clock cycle delays.

This modulo-synchronous approach would require that the clock frequency is specified and fixed for the entire system (and indeed for the fiber cable manufacture so they can cut to the correct length), but that in itself poses no substantial engineering challenge. For a maximum modulo-synchronous fiber cable length of, say, 10 m, which would correspond to ~100 clock cycles at ~2 GHz, and specifying that the timing precision is to be better than, say, 10 ps within a clock cycle even for the longest (~10 m) cable, or 1/50 of a clock cycle, then our clock frequency precision only has to be set to a precision of 1/10,000, which should represent no substantial engineering challenge at such frequencies.

Note also that, in such a modulo-synchronous system, we would deliver the clock itself optically from a centralize clock source to boards, modules, or even to chips, with the clock distribution itself being modulo-synchronous, thereby establishing a uniform, synchronous clock throughout the system.

## Acknowledgment

The author has benefitted from many conversations with a large number of individuals on these topics over many years, a list that would be too long to include and essentially impossible to construct reliably or completely. He would, however, particularly like to acknowledge stimulating and informative conversations with Tony Heinz, Joseph Kahn, Ashok Krishnamoorthy, and Jelena Vuckovic during the preparation of this review.

**David A. B. Miller** (M'83–F'95) received the Ph.D. degree in physics from Heriot-Watt University, Edinburgh, U.K., in 1979. He was with Bell Laboratories from 1981 to 1996, as a Department Head from 1987. He is currently the W. M. Keck Professor of Electrical Engineering, and a Co-Director of the Stanford Photonics Research Center at Stanford University, Stanford, CA. He was President of the IEEE Lasers and Electro-Optics Society (now Photonics Society) in 1995. His research interests include physics and devices in nanophotonics, nanometallics, and quantum-well optoelectronics, and fundamentals and applications of optics in information sensing, switching, and processing. He has published more than 260 scientific papers and the text Quantum Mechanics for Scientists and Engineers (Cambridge, U.K.: Cambridge Univ. Press, 2008), and holds 73 patents. Dr. Miller has received numerous awards. He is a Fellow of the Optical Society of America (OSA), the American Physical Society (APS), the Royal Society, and the Royal Society of Edinburgh, and a Member of the National Academy of Sciences and the National Academy of Engineering.